\documentclass{article}

\usepackage{arxiv}

\usepackage[T1]{fontenc}
\usepackage[utf8]{inputenc}
\usepackage{lmodern}
\usepackage{microtype}
\usepackage{amsmath,amssymb,bm,mathtools}
\usepackage{graphicx}
\usepackage{booktabs,longtable,tabularx,array,multirow,calc}
\usepackage{caption}
\usepackage{pdflscape}
\usepackage{enumitem}
\usepackage{xcolor}
\usepackage[numbers,sort&compress]{natbib}
\usepackage{xurl}
\usepackage[hidelinks]{hyperref}
\usepackage{etoolbox}

\newlength{\manuscripttablewidth}
\newcommand{\manuscripttablefont}{\footnotesize}
\AtBeginEnvironment{longtable}{\manuscripttablefont}

\newcommand{\secondbest}[1]{\underline{#1}}
\graphicspath{{./images/}}
\title{Local-Global Feature Mixer and Trend-Guided Consistent Learning for Remaining Useful Life Prediction of Rotating Machinery}

\author{
 Hanbyeol Park \\
  Department of Industrial Engineering\\
  Pusan National University\\
  Pusan, Republic of Korea\\
  \texttt{pb104@pusan.ac.kr} \\
  \And
 Hyerim Bae \\
  Department of Industrial Engineering\\
  Pusan National University\\
  Pusan, Republic of Korea\\
  \texttt{hrbae@pusan.ac.kr} \\
}

\begin{document}
\maketitle

\begin{abstract}
Degradation process (DP) modeling is widely used for remaining useful life (RUL) prediction, particularly when run-to-failure data are limited. Neural networks can represent complex degradation trajectories without prescribing a fixed degradation function; however, recursive HI forecasting is prone to error accumulation and may fail to preserve the irreversible degradation trend over long horizons. To address these limitations, this study proposes a local--global feature mixer (LGFM) and trend-guided rollout-consistent (TG-RC) loss. The LGFM combines the original HI sequence with statistical and degradation-related features to reduce sensitivity to high-frequency noise and capture both local changes and global degradation states. The TG-RC loss supplements the conventional one-step mean squared error with a recursive multi-step rollout loss and a soft dynamic time warping alignment term based on a global trend prior. Consequently, it reduces the discrepancy between training and recursive inference, while guiding the predicted trajectory toward a consistent degradation direction. Experiments on two public bearing datasets show that the proposed framework improves long-term HI extrapolation stability and RUL prediction performance across different inspection times. The LGFM also maintains low computational complexity, while the TG-RC loss can be incorporated into various DP models to improve their long-term forecasting performance.
\end{abstract}

\section{Introduction}
\label{sec:intro}

Rotating machinery is a critical component of industrial systems, including aerospace systems, wind turbines, and high-speed trains \citep{Zhou2025RULReview}. Because such machinery often operates under harsh conditions, it is susceptible to failure \citep{Bai2025SpatioTemporal}. Failure in a rotating component may propagate beyond the affected component and cause a system-level breakdown, threatening worker safety while reducing productivity and increasing maintenance costs \citep{Li2025InteractiveBiLSTM}. Therefore, prognostics and health management (PHM) has received considerable attention as a means of improving the reliability and operational safety of rotating machinery \citep{Guo2025WaveletFusion}. In particular, remaining useful life (RUL) prediction is a core PHM task because it enables planned maintenance while reducing unnecessary maintenance and operating costs \citep{He2025DigitalTwin}. Accurate RUL estimates are essential for effective maintenance decisions.

RUL prediction methods are generally classified as physics-based \citep{Gazizulin2015Physics,Qiu2002Damage,Gabrielli2024EquivalentDamagedVolume}, data-based \citep{Li2024GTFDAU,Lu2025LSTMTransformer,Park2025TrendFluctuation,Qin2021GDAU,Xiang2023CocktailLSTM,Zhou2023DTGRU}, and hybrid approaches \citep{Chen2025AdaptiveWiener,He2025PINNWiener,Qi2025HybridWiener}. Physics-based approaches can achieve high predictive accuracy when the degradation process is well characterized by equations derived from physical principles. However, they require extensive domain knowledge, and the degradation mechanisms in real environments are often too complex to describe physically \citep{Hu2022FeatureDisentanglement}. These requirements limit the scalability and applicability of these approaches across various conditions \citep{Park2025TrendFluctuation}. Data-driven approaches predict RUL from sensor data collected from machinery; these methods have been studied extensively alongside advances in data acquisition, storage, and computing infrastructure \citep{Wen2022PredictiveMaintenance}. Deep learning (DL), in particular, can automatically extract degradation-related features from sensor data through end-to-end learning \citep{Kumar2025NearFailure}. Thus, it can reduce the need for expert intervention and provide robust predictions under various operating conditions. Hybrid approaches combine physics-based and data-driven methods by embedding physical knowledge into data-driven models \citep{Li2024PhysicsInformedReview}. This integration can suppress physically implausible degradation trajectories under data-scarce and noisy conditions, thereby improving generalization. Nevertheless, hybrid approaches still require substantial physical knowledge and modeling expertise, which may restrict their general use in industrial applications \citep{Zhou2023DTGRU}. Consequently, developing DL-based data-driven methods that accurately predict RUL with minimal dependence on explicit physical knowledge remains an important research problem.

Data-driven RUL prediction methods can be broadly divided into direct mapping (DM) approaches \citep{Li2025GloballyOrdered,Cao2025TCNTransformer,Wang2025HighQualityRepresentation} and HI-based approaches \citep{Li2024GTFDAU,Lu2025LSTMTransformer,Park2025TrendFluctuation,Qin2023SupervisedMHAE,Peng2025HealthStates,Wang2008Similarity}. DM methods formulate prognostics as a supervised regression from sensor data to RUL, typically using linear or piecewise linear RUL targets derived from the observed failure time and, where applicable, an estimate of the degradation onset \citep{Lin2021LabelConstruction}. However, RUL depends on the underlying health states of the equipment. Therefore, time-based linear labels that do not consider these states may fail to represent real engineering conditions \citep{Fu2022MemoryAutoencoder}. Health indicator (HI)-based approaches are further divided into similarity matching (SM) \citep{Qin2023SupervisedMHAE} and degradation-process (DP) modeling \citep{Li2024GTFDAU,Lu2025LSTMTransformer,Park2025TrendFluctuation,Qin2021GDAU,Xiang2023CocktailLSTM,Zhou2023DTGRU}. SM estimates RUL by comparing the HI sequence of a target unit with the run-to-failure (RtF) HI trajectories of previously failed units and identifying similar trajectories as references \citep{Peng2025HealthStates}. SM methods require no explicit degradation model; however, their prediction reliability depends on the availability and representativeness of historical RtF trajectories \citep{Wang2008Similarity}. Therefore, DP has been used more widely for rotating machinery, for which sufficient RtF data cannot generally be assumed \citep{Escobar2006AcceleratedTest,Joseph2006ReliabilityImprovement}.

DP-based RUL prediction consists of two stages \citep{Li2024GTFDAU,Lu2025LSTMTransformer,Park2025TrendFluctuation,Qin2021GDAU,Xiang2023CocktailLSTM,Zhou2023DTGRU}: HI construction and HI extrapolation using a DP model. For HI extrapolation, an inspection time (IT) is first defined as the reference point for RUL estimation. The DP model learns the degradation pattern from the HI sequence observed up to the IT and extrapolates the sequence beyond that point. The first hitting time (FHT) at which the extrapolated HI reaches the failure threshold (FT) is then taken as the predicted failure time, from which the RUL is calculated. Because the performance of the resulting RUL estimate depends directly on the quality of the DP modeling, reliable HI extrapolation is essential.

Recurrent neural network (RNN)-based DP models have been widely studied for capturing the temporal dynamics of HI sequences \citep{Park2025TrendFluctuation}. Chen et al. \citep{Chen2021QuadraticDCAE} utilized a gated attention unit, which combines an attention mechanism with a gated recurrent unit (GRU) cell, for HI extrapolation within an RUL prediction framework. Qin et al. \citep{Qin2020MacroMicroAttention} proposed a macroscopic--microscopic attention long short-term memory (MMA-LSTM) network, which incorporates macroscopic attention to capture local details. MMA-LSTM produced more accurate HI predictions than the conventional GRU and LSTM models, leading to improved RUL prediction. Qin et al. \citep{Qin2021GDAU} also proposed a gated dual attention unit (GDAU), which applies attention between the input HI sequence and the previous hidden state, as well as between the reset and update gates. These two attention mechanisms improved prediction accuracy and convergence speed relative to conventional GRU-based models. Li et al. \citep{Li2021SelfAttentionConvLSTM} developed an SA-ConvLSTM cell by combining conventional LSTM with a self-attention mechanism. Its convolutional structure reduces information redundancy and improves the network's capacity to model nonlinear relationships. Zhou et al. \citep{Zhou2023DTGRU} proposed a dual-thread gated recurrent unit (DTGRU), which extracts stationary and nonstationary information from HI sequences using a multi-threaded learning strategy. The differences between hidden states at adjacent time steps are incorporated as complementary information. Experiments on several gearbox datasets demonstrated the effectiveness of the DTGRU for RUL prediction. Li et al. \citep{Li2024GTFDAU} emphasized the importance of transient fluctuations in HI sequences, and proposed a gated transient-fluctuation dual-attention unit (GTFDAU) to capture them automatically. The GTFDAU combines a transient-fluctuation gate with historical and current attention gates, allowing it to adaptively use past and current information. The model substantially improved HI forecasting and achieved particularly high accuracy in long-term recursive prediction. Collectively, these studies demonstrate that RNN-based DP models can learn temporal dependencies from observations without imposing a predefined functional form on the degradation trajectory.

Despite these advances, existing methods have two structural limitations that affect the stability of long-term recursive extrapolation: one related to the model architecture and the other to the loss function. First, most existing studies use many-to-one RNN architectures, which cannot explicitly represent temporal dependencies within the output sequence. Second, models are typically trained under teacher forcing, with ground-truth HI values supplied as inputs, whereas recursive inference relies on the model's own predictions. This mismatch between training and inference causes small one-step errors to accumulate over long horizons, and gradually shifts the input distribution away from that encountered during training. Specifically, the pointwise mean squared error (MSE) is commonly used for optimization; it measures only the magnitude of prediction errors and does not constrain the direction of degradation in the predicted sequence. Without a learning objective that reflects the irreversibility of degradation, a model may achieve a low MSE yet lose its directional tendency toward the FT during extrapolation. Consequently, the predicted trajectory may fail to reach the FT, making RUL estimation unreliable or even infeasible.

This study addresses these limitations by improving both the DP model and its training objective. The principal contributions are as follows:

\begin{enumerate}
\def\labelenumi{\arabic{enumi}.}
\item
  A local--global feature mixer (LGFM) is proposed for DP modeling. The LGFM is a many-to-many linear model that captures temporal dynamics while incorporating global and local statistical and degradation-related features into the prediction. This design supports stable recursive extrapolation. The LGFM comprises three components: 1) a global degradation feature (GDF) module, 2) a local degradation feature (LDF) module, and 3) a linear layer. The GDF module constructs a global difference vector between the current HI sequence and an initial HI sequence representing a healthy state. The LDF module constructs a local difference vector between the current sequence and its adjacent lagged sequence. Together, these components provide latent representations of the extent to which the current health state has deviated from both the healthy state and recent observations.
\item
  Trend-guided rollout-consistent (TG-RC) loss is proposed. The TG-RC loss combines three terms: 1) a conventional teacher-forcing MSE term \citep{Li2024GTFDAU,Lu2025LSTMTransformer,Park2025TrendFluctuation,Qin2021GDAU,Xiang2023CocktailLSTM,Zhou2023DTGRU}, 2) an MSE term that measures accumulated errors during recursive rollout without teacher forcing, and 3) a soft dynamic time warping (DTW) term that constrains the direction of the recursively predicted trajectory. The rollout MSE addresses error accumulation during recursive forecasting, while the soft-DTW term encourages a monotone degradation trajectory. By incorporating requirements specific to degradation modeling into the optimization objective, the TG-RC loss restricts the learned function space toward a physically consistent trajectory. Consequently, it guides predictions in a consistent direction, even when extrapolating unobserved HI values, thereby improving the stability of RUL estimation \citep{Li2024PhysicsInformedReview}.
\end{enumerate}

The proposed framework was evaluated on two publicly available datasets: the dataset collected using the PRONOSTIA platform for the PHM Challenge 2012 \citep{Nectoux2012PRONOSTIA} and the bearing dataset released by Xi'an Jiaotong University (XJTU) \citep{Wang2020HybridPrognostics}. The experiments show that the HI extrapolation and RUL prediction performance of existing DP models vary substantially with the IT. In contrast, the proposed framework provides accurate and robust predictions across different IT values. The LGFM has the lowest computational complexity among the compared models, making it suitable for resource-constrained applications such as embedded systems. Training existing DP models with the TG-RC loss also consistently improves their RUL prediction performance, indicating that the loss can be applied flexibly to different model architectures.

The remainder of this paper is organized as follows. Section \ref{sec:theory_back} presents the theoretical background relevant to this study. Section \ref{sec:methodology} describes the proposed methodology. Section \ref{sec:exp} presents the RUL prediction experiments and their results. Section \ref{sec:ablation} reports ablation studies that examine the individual contributions of the LGFM and TG-RC loss. Finally, Section \ref{sec:conclusion} concludes the paper and outlines directions for future research.

\section{Theoretical background}
\label{sec:theory_back}

\subsection{Spatiotemporal frequency feature fusion autoencoder (STF3AE)-based HI construction}
\label{sec:theory_STF3AE}

To estimate RUL using a DP model, the sensor signal acquired at each measurement time must be converted into a univariate HI \citep{Zhou2023DTGRU}. This study employed a spatiotemporal frequency feature fusion autoencoder (STF3AE) \citep{Park2026MultidomainSTF3AE}. The STF3AE consists of a time-domain encoder that processes the raw vibration signal and a time--frequency-domain encoder that processes its short-time Fourier transform (STFT) map. The representations obtained from the two encoders are fused into a single HI, from which the decoder reconstructs both inputs.

Let \(n = 1,\ldots,N\) index the training bearings, \(T_{n}\) denote the final measurement index in the RtF trajectory of the \(n\)-th bearing, and \(t = 1,\ldots,T_{n}\) denote the discrete measurement indices. The raw vibration signal and STFT map at time \(t\) are denoted by \(\mathbf{s}_{n,t} \in \mathbb{R}^{N_{s}}\) and \(\mathbf{x}_{n,t} \in \mathbb{R}^{F \times M}\), respectively. Here, \(N_{s}\) is the length of the raw vibration signal, and \(F\) and \(M\) denote the number of frequency bins and timeframes in the STFT map, respectively.

A shape constraint function (SCF) \citep{Li2024TransformerHI} is used as a pseudo-target to impose an increasing degradation direction on the HI produced by the STF3AE encoder. The SCF is defined to increase from 0 at the first measurement to 1 at the final RtF measurement \citep{Qin2023SupervisedMHAE,Chen2021QuadraticDCAE}:

\refstepcounter{equation}
\label{eq:1}
\begin{tabularx}{\textwidth}{@{}>{\raggedright\arraybackslash}X>{\raggedleft\arraybackslash}p{0.08\textwidth}@{}}

\begin{minipage}[b]{\linewidth}\raggedright
\[\xi_{n,t}^{(k)} = \left( \frac{t - 1}{T_{n} - 1} \right)^{k},\ \ t = 1,\ldots,T_{n},\ \ k > 0\]
\end{minipage} & \begin{minipage}[b]{\linewidth}\raggedright
(\theequation)
\end{minipage} \\
\end{tabularx}

where \(k\) controls the curvature of the SCF. When \(k = 1\), the function represents linear degradation. When \(k > 1\), it represents a pattern in which degradation proceeds slowly at first and accelerates as failure approaches. Equation~\eqref{eq:1} satisfies \(\xi_{n,1}^{(k)} = 0\), \(\xi_{n,T_{n}}^{(k)} = 1\), \(\xi_{n,t + 1}^{(k)} \geq \xi_{n,t}^{(k)}\). In this study, \(k = 1\) was used \citep{Park2026MultidomainSTF3AE}.

Let \(g_{\omega}^{e}\) and \(g_{\omega}^{d}\) denote the STF3AE encoder and decoder, respectively, where \(\omega\) represents all trainable model parameters. Let \(B_{HI}\) be a mini-batch in which each element \(b \in B_{HI}\) corresponds to a training-bearing and measurement-index pair \(\left( n_{b},t_{b} \right)\).

The HI predicted by the encoder is defined as \(y_{n_{b},t_{b}} = g_{\omega}^{e}\left( \mathbf{s}_{n_{b},t_{b}},\mathbf{x}_{n_{b},t_{b}} \right)\), and the reconstructions produced by the decoder are given by \(\left( {\widehat{\mathbf{s}}}_{n_{b},t_{b}},{\widehat{\mathbf{x}}}_{n_{b},t_{b}} \right) = g_{\omega}^{d}\left( y_{n_{b},t_{b}} \right)\). The STF3AE training objective is

\refstepcounter{equation}
\label{eq:2}
\begin{tabularx}{\textwidth}{@{}>{\raggedright\arraybackslash}X>{\raggedleft\arraybackslash}p{0.08\textwidth}@{}}

\begin{minipage}[b]{\linewidth}\raggedright
\[\mathcal{L}_{HI}(\omega) = \frac{1}{\left| B_{HI} \right|}\sum_{b \in B_{HI}}^{}\left\lbrack w_{1}\left| y_{n_{b},t_{b}} - \xi_{n_{b},t_{b}}^{(k)} \right|^{2} + w_{2}\left\| {\widehat{\mathbf{s}}}_{n_{b},t_{b}} - \mathbf{s}_{n_{b},t_{b}} \right\|_{2}^{2} + w_{3}\left\| {\widehat{\mathbf{x}}}_{n_{b},t_{b}} - \mathbf{x}_{n_{b},t_{b}} \right\|_{2}^{2} \right\rbrack\]
\end{minipage} & \begin{minipage}[b]{\linewidth}\raggedright
(\theequation)
\end{minipage} \\

\end{tabularx}

where \(w_{1},w_{2},w_{3} > 0\) control the relative contributions of the HI regression loss, time-domain reconstruction loss, and time--frequency-domain reconstruction loss, respectively. These weights are updated dynamically during training using dynamic weight averaging (DWA) \citep{Liu2019MultitaskAttention}.

After training, the HI of a target bearing \(m\) at time \(t\) is computed using only the inputs observed at that time: \(y_{m,t} = g_{\omega}^{e}\left( \mathbf{s}_{m,t},\mathbf{x}_{m,t} \right)\). Accordingly, at IT \(i\), the HI prefix available to the DP model is \(\mathbf{y}_{m,1:i} = \left\lbrack y_{m,1},\ldots,y_{m,i} \right\rbrack^{T} \in \mathbb{R}^{i}\).

\subsection{Long-term time series forecasting linear model}
\label{sec:theory_LSTFLinear}

The long-term time-series forecasting linear model (LTSF-Linear) maps an input sequence of length \(L\) directly onto a future sequence of length \(H\) using a single linear transformation \citep{Zeng2023TransformersForecasting}. Let \(\mathbf{y}_{in} \in \mathbb{R}^{L}\) and \({\widehat{\mathbf{y}}}_{out} \in \mathbb{R}^{H}\) denote the input and predicted output sequences of a univariate HI, respectively. The model is defined as

\refstepcounter{equation}
\label{eq:3}
\begin{tabularx}{\textwidth}{@{}>{\raggedright\arraybackslash}X>{\raggedleft\arraybackslash}p{0.08\textwidth}@{}}

\begin{minipage}[b]{\linewidth}\raggedright
\[{\widehat{\mathbf{y}}}_{out} = \mathbf{W}{\widehat{\mathbf{y}}}_{out}\mathbf{,\ \ W} \in \mathbb{R}^{H \times L}\]
\end{minipage} & \begin{minipage}[b]{\linewidth}\raggedright
(\theequation)
\end{minipage} \\

\end{tabularx}

This architecture follows a direct multi-step (DMS) forecasting scheme \citep{Zeng2023TransformersForecasting}, in which the values at all \(H\) future time steps are predicted simultaneously. Unlike RNN-based one-step DP models, which recursively feed each prediction task into the model to generate subsequent values, DMS allows the entire predicted sequence to be included directly in the training objective. This structure reduces the number of recursive forecasting operations required to reach the FT and enables the model to learn temporal patterns across the predicted sequence.

The proposed LGFM builds on this DMS-based linear projection. Unlike the standard LTSF-Linear model, which uses only the current HI observations as input, LGFM additionally considers (i) the global degradation difference between the current HI sequence and a healthy reference sequence and (ii) the local degradation difference between the current HI sequence and the sequence observed \(L\) lags earlier. Therefore, the LGFM captures not only the recent HI level, but also the accumulated change from the initial healthy state and the local progression of degradation.

\section{Methodology}
\label{sec:methodology}

\begin{figure}[!htbp]
\centering
\includegraphics[scale=0.6]{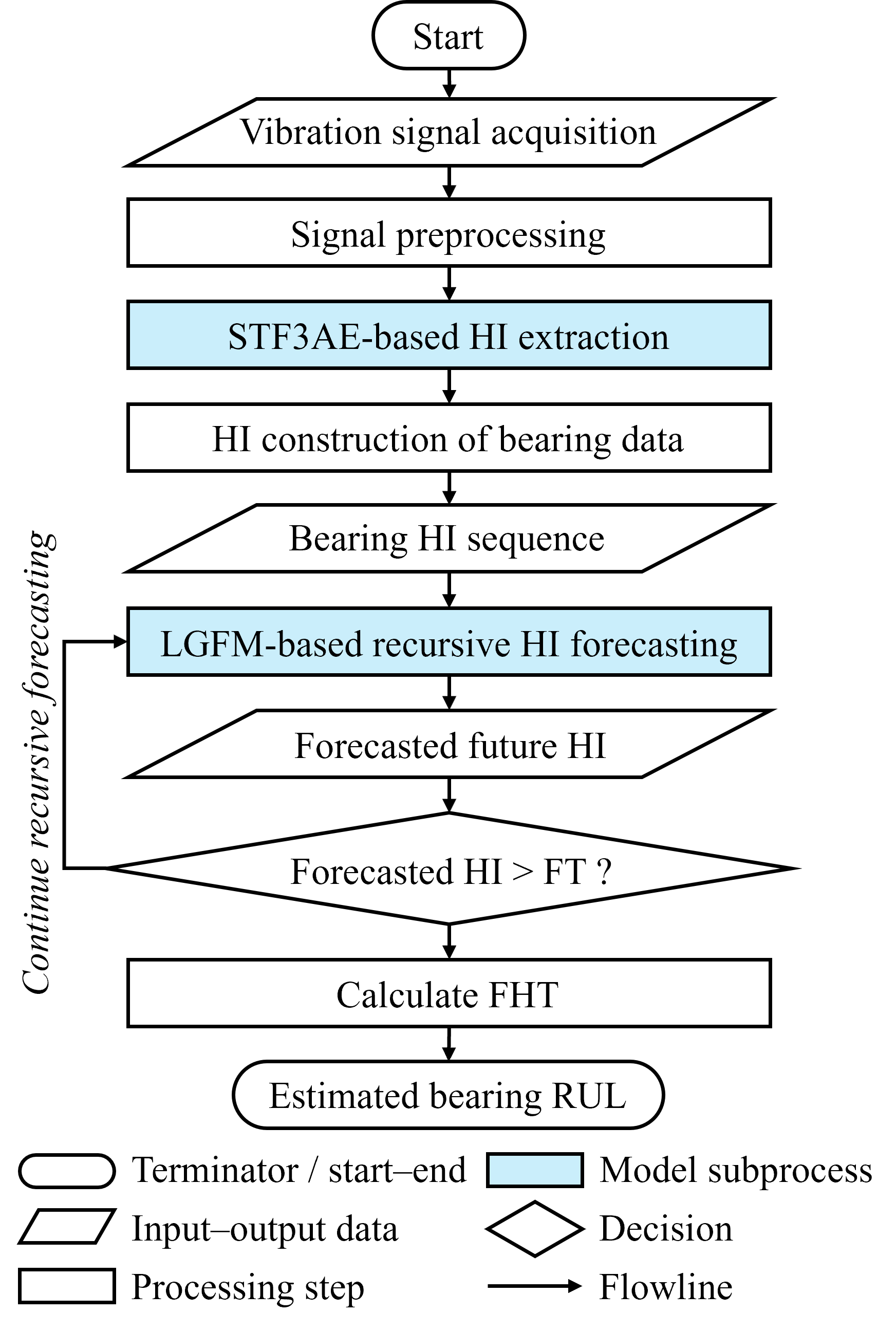}
\caption{Procedure of STF3AE-based HI construction, LGFM training, recursive HI extrapolation, and RUL estimation.}
\label{fig:1}
\end{figure}

This section describes the procedure for training the LGFM using only the HI prefix of a test bearing observed up to IT index \(i\), recursively extending the trained model until the predicted HI reaches the FT, and estimating the resulting RUL. The overall procedure is illustrated in Figure \ref{fig:1}, and the principal notation is summarized in Table \ref{tab:1}.

\begin{longtable}[]{@{}
  >{\raggedright\arraybackslash}p{(\manuscripttablewidth - 2\tabcolsep) * \real{0.1692}}
  >{\raggedright\arraybackslash}p{(\manuscripttablewidth - 2\tabcolsep) * \real{0.8308}}@{}}
\caption{Definitions of the principal notation.}\label{tab:1}\\
\toprule\noalign{}
\begin{minipage}[b]{\linewidth}\raggedright
Notation
\end{minipage} & \begin{minipage}[b]{\linewidth}\raggedright
Description
\end{minipage} \\
\midrule\noalign{}
\endfirsthead

\toprule\noalign{}
\begin{minipage}[b]{\linewidth}\raggedright
Notation
\end{minipage} & \begin{minipage}[b]{\linewidth}\raggedright
Description
\end{minipage} \\
\midrule\noalign{}
\endhead
\bottomrule\noalign{}
\endlastfoot
\(n\) & Index of a bearing used to train STF3AE. \\
\(m\) & Index of a target bearing used for online RUL inference. \\
\(u\) & General bearing index encompassing both cases. \\
\(t\) & Forecast origin or general discrete measurement index. \\
\(i\) & Absolute inspection index at which HI extrapolation begins for the target bearing. \\
\(T_{u}\) & Index of the final physical RtF measurement for bearing \(u\). \\
\(N_{s}\) & Number of samples in a raw vibration record collected at one measurement time. \\
\(F,M\) & Number of frequency bins and time frames, respectively, in the STFT. \\
\(\mathbf{s}_{u,t}\) & Raw vibration record at measurement index \(t\), where \(\mathbf{s}_{u,t} \in \mathbb{R}^{N_{s}}\). \\
\(\mathbf{x}_{u,t}\) & STFT feature map at measurement index \(t\), where \(\mathbf{x}_{u,t} \in \mathbb{R}^{F \times M}\). \\
\(y_{u,t}\) & Scalar HI at measurement index \(t\). \\
\(L\) & Common length of the healthy, previous, and current HI windows. \\
\(H\) & Number of future HI values produced by the LGFM in a single forward pass. \\
\(R\) & Number of recursive rollout blocks used during training. \\
\(P\) & Total length of the training rollout, where \(P = RH\). \\
\(C\) & Representation dimension of each the LGFM branch. \\
\(\mathbf{h}_{u},\mathbf{p}_{u,t},\mathbf{c}_{u,t}\) & Healthy, previous, and current HI windows, respectively. \\
\(\Phi( \cdot )\) & Function that returns five statistical and physical degradation features. \\
\(f_{\theta}( \cdot )\) & LGFM multi-step predictor parameterized by \(\theta\). \\
\(\gamma\) & Soft-DTW smoothing parameter. \\
\(\tau_{DWA}\) & Temperature parameter used in dynamic weight averaging. \\
\(FT\) & HI threshold used to determine predicted failure. \\
\(K_{\max}\) & Maximum allowable HI extrapolation length after the inspection index. \\
\(\mathrm{\Delta}t\) & Physical time interval between two consecutive HI measurements. \\
\end{longtable}

\subsection{Problem formulation and causal window construction}
\label{sec:methodology_problem}

Let \(\left\{ y_{u,t} \right\}_{t = 1}^{T_{u}}\) denote the HI sequence of bearing \(u\). For any endpoint \(t \geq L\), the window containing the most recent \(L\) HI values and the subsequent target vector of length \(Q\) are defined as

\begin{tabularx}{\textwidth}
	{@{}>{\raggedright\arraybackslash}X
		>{\raggedleft\arraybackslash}p{0.08\textwidth}@{}}
	
	\begin{minipage}[b]{\linewidth}\raggedright
		\[
		\mathbf{w}_{u,t}
		=
		\left[
		y_{u,t-L+1},\ldots,y_{u,t}
		\right]^{T}
		\in\mathbb{R}^{L}
		\]
	\end{minipage}
	&
	\begin{minipage}[b]{\linewidth}\raggedright
		\refstepcounter{equation}
		\label{eq:4}
		(\theequation)
	\end{minipage}
	\\
	
	\begin{minipage}[b]{\linewidth}\raggedright
		\[
		\mathbf{y}_{u,t}^{+}(Q)
		=
		\left[
		y_{u,t+1},\ldots,y_{u,t+Q}
		\right]^{T}
		\in\mathbb{R}^{Q}
		\]
	\end{minipage}
	&
	\begin{minipage}[b]{\linewidth}\raggedright
		\refstepcounter{equation}
		\label{eq:5}
		(\theequation)
	\end{minipage}
	\\
	
\end{tabularx}

The healthy reference used by the LGFM is fixed as the first \(L\) HI observations of the corresponding bearing:

\refstepcounter{equation}
\label{eq:6}
\begin{tabularx}{\textwidth}{@{}>{\raggedright\arraybackslash}X>{\raggedleft\arraybackslash}p{0.08\textwidth}@{}}

\begin{minipage}[b]{\linewidth}\raggedright
\[\mathbf{h}_{u} = \mathbf{w}_{u,L} = \left\lbrack y_{u,1},\ldots,y_{u,L} \right\rbrack^{T} \in \mathbb{R}^{L}\]
\end{minipage} & \begin{minipage}[b]{\linewidth}\raggedright
(\theequation)
\end{minipage} \\

\end{tabularx}

For a forecast origin \(t \geq 2L\), the previous window is defined as the non-overlapping window of length \(L\) immediately preceding the current window:

\refstepcounter{equation}
\label{eq:7}
\begin{tabularx}{\textwidth}{@{}>{\raggedright\arraybackslash}X>{\raggedleft\arraybackslash}p{0.08\textwidth}@{}}

\begin{minipage}[b]{\linewidth}\raggedright
\[\mathbf{p}_{u,t} = \mathbf{w}_{u,t - L} = \left\lbrack y_{u,t - 2L + 1},\ldots,y_{u,t - L} \right\rbrack^{T} \in \mathbb{R}^{L}\]
\end{minipage} & \begin{minipage}[b]{\linewidth}\raggedright
(\theequation)
\end{minipage} \\

\end{tabularx}

The current window contains the most recent \(L\) HI observations available at forecast origin \(t\):

\refstepcounter{equation}
\label{eq:8}
\begin{tabularx}{\textwidth}{@{}>{\raggedright\arraybackslash}X>{\raggedleft\arraybackslash}p{0.08\textwidth}@{}}

\begin{minipage}[b]{\linewidth}\raggedright
\[\mathbf{c}_{u,t} = \mathbf{w}_{u,t} = \left\lbrack y_{u,t - L + 1},\ldots,y_{u,t} \right\rbrack^{T} \in \mathbb{R}^{L}\]
\end{minipage} & \begin{minipage}[b]{\linewidth}\raggedright
(\theequation)
\end{minipage} \\

\end{tabularx}

Therefore, each LGFM sample consists of the healthy reference \(\mathbf{h}_{u}\), previous window \(\mathbf{p}_{u,t}\), current window \(\mathbf{c}_{u,t}\), and a target sequence beginning after \(t\). All model inputs contain only HI values available at or before the forecast origin, thereby preserving temporal causality.

\subsection{Local-global feature mixer}
\label{sec:methodology_LGFM}

\begin{figure}[!htbp]
\centering
\includegraphics[width=\textwidth]{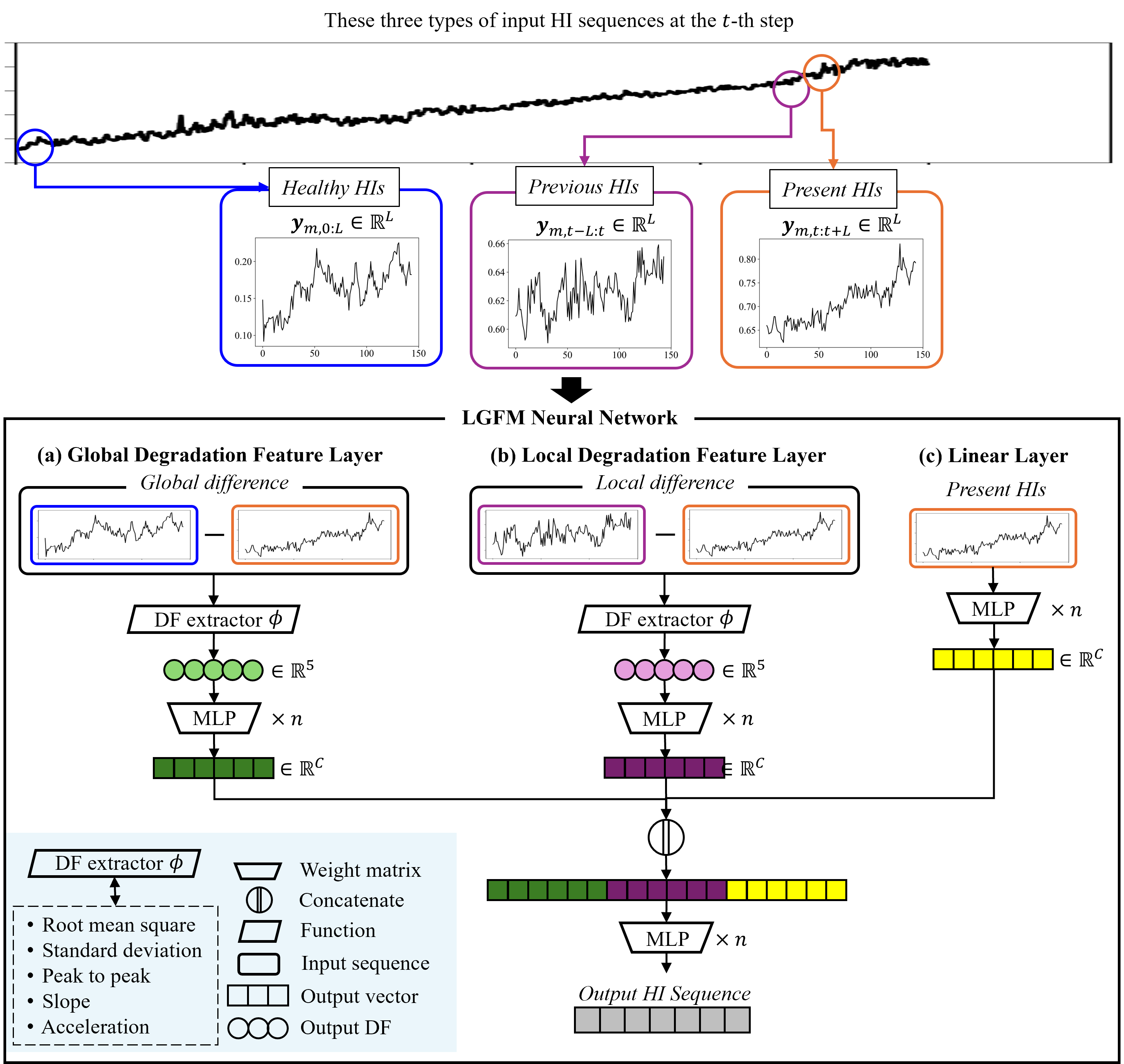}
\caption{Architecture of the LGFM neural network.}
\label{fig:2}
\end{figure}

This section presents the mathematical formulation of the proposed LGFM. Its overall architecture is illustrated in Figure \ref{fig:2}.

\subsubsection{Statistical and physical degradation feature extractor}
\label{sec:methodology_DF}

For an arbitrary input window \(\mathbf{v} = \left\lbrack v_{1},\ldots,v_{L} \right\rbrack^{T} \in \mathbb{R}^{L}\), LGFM computes five degradation features (DFs): the root mean square (RMS), standard deviation (STD), peak-to-peak amplitude (P2P), mean absolute first difference (SLOPE), and mean absolute change in consecutive first-difference magnitudes (ACC). These DFs are calculated as follows:

\begin{tabularx}{\textwidth}
	{@{}>{\raggedright\arraybackslash}X
		>{\raggedleft\arraybackslash}p{0.08\textwidth}@{}}
	
	\begin{minipage}[b]{\linewidth}\raggedright
		\[
		RMS\left(\mathbf{v}\right)
		=
		\sqrt{\frac{1}{L}\sum_{j=1}^{L}v_j^2}
		\]
	\end{minipage}
	&
	\begin{minipage}[b]{\linewidth}\raggedright
		\refstepcounter{equation}
		\label{eq:9}
		(\theequation)
	\end{minipage}
	\\
	
	\begin{minipage}[b]{\linewidth}\raggedright
		\[
		STD\left(\mathbf{v}\right)
		=
		\sqrt{
			\frac{1}{L}
			\sum_{j=1}^{L}
			\left(v_j-\overline{v}\right)^2
		},
		\quad
		\overline{v}
		=
		\frac{1}{L}\sum_{j=1}^{L}v_j,
		\]
	\end{minipage}
	&
	\begin{minipage}[b]{\linewidth}\raggedright
		\refstepcounter{equation}
		\label{eq:10}
		(\theequation)
	\end{minipage}
	\\
	
	\begin{minipage}[b]{\linewidth}\raggedright
		\[
		P2P\left(\mathbf{v}\right)
		=
		\max_{1\leq j\leq L}v_j
		-
		\min_{1\leq j\leq L}v_j
		\]
	\end{minipage}
	&
	\begin{minipage}[b]{\linewidth}\raggedright
		\refstepcounter{equation}
		\label{eq:11}
		(\theequation)
	\end{minipage}
	\\
	
	\begin{minipage}[b]{\linewidth}\raggedright
		\[
		SLOPE\left(\mathbf{v}\right)
		=
		\frac{1}{L-1}
		\sum_{j=1}^{L-1}
		d_j\left(\mathbf{v}\right),
		\quad
		d_j\left(\mathbf{v}\right)
		=
		\left|v_{j+1}-v_j\right|
		\]
	\end{minipage}
	&
	\begin{minipage}[b]{\linewidth}\raggedright
		\refstepcounter{equation}
		\label{eq:12}
		(\theequation)
	\end{minipage}
	\\
	
	\begin{minipage}[b]{\linewidth}\raggedright
		\[
		ACC\left(\mathbf{v}\right)
		=
		\frac{1}{L-2}
		\sum_{j=1}^{L-2}
		\left|
		d_{j+1}\left(\mathbf{v}\right)
		-
		d_j\left(\mathbf{v}\right)
		\right|
		\]
	\end{minipage}
	&
	\begin{minipage}[b]{\linewidth}\raggedright
		\refstepcounter{equation}
		\label{eq:13}
		(\theequation)
	\end{minipage}
	\\
	
\end{tabularx}

The resulting DF extractor is defined as

\refstepcounter{equation}
\label{eq:14}
\begin{tabularx}{\textwidth}{@{}>{\raggedright\arraybackslash}X>{\raggedleft\arraybackslash}p{0.08\textwidth}@{}}

\begin{minipage}[b]{\linewidth}\raggedright
\[\phi\left( \mathbf{v} \right) = \left\lbrack \begin{array}{r}
RMS\left( \mathbf{v} \right) \\
STD\left( \mathbf{v} \right) \\
P2P\left( \mathbf{v} \right) \\
SLOPE\left( \mathbf{v} \right) \\
ACC\left( \mathbf{v} \right)
\end{array} \right\rbrack \in \mathbb{R}^{5}\]
\end{minipage} & \begin{minipage}[b]{\linewidth}\raggedright
(\theequation)
\end{minipage} \\

\end{tabularx}

RMS, STD, and P2P summarize the level and dispersion of HI values within the window, whereas SLOPE and ACC describe the magnitude and variability of local changes, respectively.

\subsubsection{Global, local, and current-state branches}
\label{sec:methodology_GDFLDF}

The LGFM consists of global, local, and current-state branches. The global branch represents the extent to which the current window has changed from the initial healthy state. Its input is an elementwise difference vector, defined as

\refstepcounter{equation}
\label{eq:15}
\begin{tabularx}{\textwidth}{@{}>{\raggedright\arraybackslash}X>{\raggedleft\arraybackslash}p{0.08\textwidth}@{}}

\begin{minipage}[b]{\linewidth}\raggedright
\[\Delta\mathbf{g}_{u,t} = \mathbf{c}_{u,t} - \mathbf{h}_{u} \in \mathbb{R}^{L}\]
\end{minipage} & \begin{minipage}[b]{\linewidth}\raggedright
(\theequation)
\end{minipage} \\

\end{tabularx}

The local branch uses the elementwise difference between the current window and the immediately preceding window:

\refstepcounter{equation}
\label{eq:16}
\begin{tabularx}{\textwidth}{@{}>{\raggedright\arraybackslash}X>{\raggedleft\arraybackslash}p{0.08\textwidth}@{}}

\begin{minipage}[b]{\linewidth}\raggedright
\[\Delta\mathbf{l}_{u,t} = \mathbf{c}_{u,t} - \mathbf{p}_{u,t} \in \mathbb{R}^{L}\]
\end{minipage} & \begin{minipage}[b]{\linewidth}\raggedright
(\theequation)
\end{minipage} \\

\end{tabularx}

The GDFs are mapped to a \(C\)-dimensional representation through an affine projection:

\refstepcounter{equation}
\label{eq:17}
\begin{tabularx}{\textwidth}{@{}>{\raggedright\arraybackslash}X>{\raggedleft\arraybackslash}p{0.08\textwidth}@{}}

\begin{minipage}[b]{\linewidth}\raggedright
\[\mathbf{a}_{u,t}^{G} = \mathbf{W}_{G}\phi\left( \Delta\mathbf{g}_{u,t} \right) + \mathbf{b}_{G} \in \mathbb{R}^{C},\ \ \mathbf{W}_{G} \in \mathbb{R}^{C \times 5},\ \ \mathbf{b}_{G} \in \mathbb{R}^{C}\]
\end{minipage} & \begin{minipage}[b]{\linewidth}\raggedright
(\theequation)
\end{minipage} \\

\end{tabularx}

The LDFs are projected into a representation of the same dimension:

\refstepcounter{equation}
\label{eq:18}
\begin{tabularx}{\textwidth}{@{}>{\raggedright\arraybackslash}X>{\raggedleft\arraybackslash}p{0.08\textwidth}@{}}

\begin{minipage}[b]{\linewidth}\raggedright
\[\mathbf{a}_{u,t}^{L} = \mathbf{W}_{L}\phi\left( \Delta\mathbf{l}_{u,t} \right) + \mathbf{b}_{L} \in \mathbb{R}^{C},\ \ \mathbf{W}_{L} \in \mathbb{R}^{C \times 5},\ \ \mathbf{b}_{L} \in \mathbb{R}^{C}\]
\end{minipage} & \begin{minipage}[b]{\linewidth}\raggedright
(\theequation)
\end{minipage} \\

\end{tabularx}

The current state branch directly projects the current HI window, thereby retaining information regarding its absolute HI level:

\refstepcounter{equation}
\label{eq:19}
\begin{tabularx}{\textwidth}{@{}>{\raggedright\arraybackslash}X>{\raggedleft\arraybackslash}p{0.08\textwidth}@{}}

\begin{minipage}[b]{\linewidth}\raggedright
\[\mathbf{a}_{u,t}^{C} = \mathbf{W}_{C}\phi\left( \mathbf{c}_{u,t} \right) + \mathbf{b}_{C} \in \mathbb{R}^{C},\ \ \mathbf{W}_{C} \in \mathbb{R}^{C \times L},\ \ \mathbf{b}_{L} \in \mathbb{R}^{C}\]
\end{minipage} & \begin{minipage}[b]{\linewidth}\raggedright
(\theequation)
\end{minipage} \\

\end{tabularx}

The outputs of the three branches are concatenated along the feature dimension:

\refstepcounter{equation}
\label{eq:20}
\begin{tabularx}{\textwidth}{@{}>{\raggedright\arraybackslash}X>{\raggedleft\arraybackslash}p{0.08\textwidth}@{}}

\begin{minipage}[b]{\linewidth}\raggedright
\[\mathbf{a}_{u,t} = \left\lbrack \mathbf{a}_{u,t}^{G}\|\mathbf{a}_{u,t}^{L}\|\mathbf{a}_{u,t}^{C} \right\rbrack \in \mathbb{R}^{3C}\]
\end{minipage} & \begin{minipage}[b]{\linewidth}\raggedright
(\theequation)
\end{minipage} \\

\end{tabularx}

The resulting representation is projected onto an output block of length \(H\):

\refstepcounter{equation}
\label{eq:21}
\begin{tabularx}{\textwidth}{@{}>{\raggedright\arraybackslash}X>{\raggedleft\arraybackslash}p{0.08\textwidth}@{}}

\begin{minipage}[b]{\linewidth}\raggedright
\[{\widehat{\mathbf{f}}}_{u,t}^{(0)} = \psi_{\theta}\left( \mathbf{h}_{u},\mathbf{p}_{u,t},\mathbf{c}_{u,t} \right) = \mathbf{W}_{O}\mathbf{a}_{u,t} + \mathbf{b}_{O} \in \mathbb{R}^{H},\ \ \mathbf{W}_{O} \in \mathbb{R}^{H \times 3C},\ \ \mathbf{b}_{O} \in \mathbb{R}^{H}\]
\end{minipage} & \begin{minipage}[b]{\linewidth}\raggedright
(\theequation)
\end{minipage} \\

\end{tabularx}

Here, \(\theta\) denotes all learnable weights and biases in LGFM. The elements of \({\widehat{\mathbf{f}}}_{u,t}^{(0)}\) correspond to the future time indices \(t + 1,\ldots,t + H\). The one-shot MSE loss is defined as

\refstepcounter{equation}
\label{eq:22}
\begin{tabularx}{\textwidth}{@{}>{\raggedright\arraybackslash}X>{\raggedleft\arraybackslash}p{0.08\textwidth}@{}}

\begin{minipage}[b]{\linewidth}\raggedright
\[\mathcal{L}_{OS}(u,t) = \frac{1}{H}\left\| {\widehat{\mathbf{f}}}_{u,t}^{(0)} - \mathbf{y}_{u,t}^{+}(H) \right\|_{2}^{2}\]
\end{minipage} & \begin{minipage}[b]{\linewidth}\raggedright
(\theequation)
\end{minipage} \\

\end{tabularx}

\subsection{Trend-guided rollout-consistent loss}
\label{sec:methodology_TGRC}

The TG-RC loss jointly optimizes the one-shot error, differentiable recursive-rollout error, and shape alignment with a linear trend prior extrapolated from the complete observation history available at the forecast origin. Given an input length \(L\), an output-block length \(H\), and \(R\) rollout blocks, the total training-rollout horizon is \(P = RH\). For the forecast origin \(t\), the supervised target over the complete rollout horizon is defined as

\refstepcounter{equation}
\label{eq:23}
\begin{tabularx}{\textwidth}{@{}>{\raggedright\arraybackslash}X>{\raggedleft\arraybackslash}p{0.08\textwidth}@{}}

\begin{minipage}[b]{\linewidth}\raggedright
\[\mathbf{v}_{u,t} = \mathbf{y}_{u,t}^{+}(P) = \left\lbrack y_{u,t + 1},\ldots,y_{u,t + P} \right\rbrack^{T} \in \mathbb{R}^{P}\]
\end{minipage} & \begin{minipage}[b]{\linewidth}\raggedright
(\theequation)
\end{minipage} \\

\end{tabularx}

\subsubsection{Linear trend-prior reference}
\label{sec:methodology_TGRC_ref}

The linear trend prior is constructed by fitting a straight line to the complete HI history from the first observation through the RUL IT \(i\) and extrapolating the fitted line over the subsequent \(P\) time steps. The complete observation history and its corresponding design matrix are defined as follows:

\begin{tabularx}{\textwidth}
	{@{}>{\raggedright\arraybackslash}X
		>{\raggedleft\arraybackslash}p{0.08\textwidth}@{}}
	
	\begin{minipage}[b]{\linewidth}\raggedright
		\[
		\mathbf{y}_{u,i}^{hist}
		=
		\left[
		y_{u,1},\ldots,y_{u,i}
		\right]^{T}
		\in \mathbb{R}^{i}
		\]
	\end{minipage}
	&
	\begin{minipage}[b]{\linewidth}\raggedright
		\refstepcounter{equation}
		\label{eq:24}
		(\theequation)
	\end{minipage}
	\\
	
	\begin{minipage}[b]{\linewidth}\raggedright
		\[
		\mathbf{X}_{i}
		=
		\begin{bmatrix}
			1      & 1      \\
			\vdots & \vdots \\
			1      & i
		\end{bmatrix}
		\in \mathbb{R}^{i\times 2}
		\]
	\end{minipage}
	&
	\begin{minipage}[b]{\linewidth}\raggedright
		\refstepcounter{equation}
		\label{eq:25}
		(\theequation)
	\end{minipage}
	\\
	
\end{tabularx}

The linear trend at measurement index \(k\) is expressed as

\begin{tabularx}{\textwidth}{@{}>{\raggedright\arraybackslash}X>{\raggedleft\arraybackslash}p{0.08\textwidth}@{}}

\refstepcounter{equation}
\label{eq:26}
\begin{minipage}[b]{\linewidth}\raggedright
\[y_{u,k} \approx \beta_{u,i,0} + \beta_{u,i,1}k,\ \ k = 1,\ldots,i\]
\end{minipage} & \begin{minipage}[b]{\linewidth}\raggedright
(\theequation)
\end{minipage} \\

\end{tabularx}

where \(\mathbf{\beta}_{u,i} = \left\lbrack \beta_{u,i,0},\beta_{u,i,1} \right\rbrack^{T} \in \mathbb{R}^{2}\) is the vector of linear coefficients fitted to the complete HI history available up to IT \(i\). The coefficients \(\beta_{u,i,0}\) and \(\beta_{u,i,1}\) represent the intercept and slope, respectively.

The coefficient vector is estimated by solving the following least-squares problem:

\refstepcounter{equation}
\label{eq:27}
\begin{tabularx}{\textwidth}{@{}>{\raggedright\arraybackslash}X>{\raggedleft\arraybackslash}p{0.08\textwidth}@{}}

\begin{minipage}[b]{\linewidth}\raggedright
\[{\widehat{\mathbf{\beta}}}_{u,i} = \underset{\mathbf{\beta} \in \mathbb{R}^{2}}{\arg\min}\left\| \mathbf{y}_{u,i}^{hist} - \mathbf{X}_{i}\mathbf{\beta} \right\|_{2}^{2} = \mathbf{X}_{i}^{\dagger}\mathbf{y}_{u,i}^{hist}\]
\end{minipage} & \begin{minipage}[b]{\linewidth}\raggedright
(\theequation)
\end{minipage} \\

\end{tabularx}

where \(\mathbf{X}_{i}^{\dagger}\) denotes the Moore--Penrose pseudoinverse of \(\mathbf{X}_{i}\).

By extrapolating the fitted line to future indices \(i + 1,\ldots,i + P\), the corresponding design matrix is defined as

\refstepcounter{equation}
\label{eq:28}
\begin{tabularx}{\textwidth}{@{}>{\raggedright\arraybackslash}X>{\raggedleft\arraybackslash}p{0.08\textwidth}@{}}

\begin{minipage}[b]{\linewidth}\raggedright
\[\mathbf{X}_{i,P}^{ext} = \begin{bmatrix}
1 & i + 1 \\
 \vdots & \vdots \\
1 & i + P
\end{bmatrix} \in \mathbb{R}^{P \times 2}\]
\end{minipage} & \begin{minipage}[b]{\linewidth}\raggedright
(\theequation)
\end{minipage} \\

\end{tabularx}

The \(P\)-step linear trend-prior reference is then given by

\refstepcounter{equation}
\label{eq:29}
\begin{tabularx}{\textwidth}{@{}>{\raggedright\arraybackslash}X>{\raggedleft\arraybackslash}p{0.08\textwidth}@{}}

\begin{minipage}[b]{\linewidth}\raggedright
\[{\widetilde{\mathbf{v}}}_{u,i} = \mathbf{X}_{i,P}^{ext}{\widehat{\mathbf{\beta}}}_{u,i} \in \mathbb{R}^{P}\]
\end{minipage} & \begin{minipage}[b]{\linewidth}\raggedright
(\theequation)
\end{minipage} \\

\end{tabularx}

Accordingly, the trend-prior value at each future index is

\refstepcounter{equation}
\label{eq:30}
\begin{tabularx}{\textwidth}{@{}>{\raggedright\arraybackslash}X>{\raggedleft\arraybackslash}p{0.08\textwidth}@{}}

\begin{minipage}[b]{\linewidth}\raggedright
\[\left\lbrack {\widetilde{\mathbf{v}}}_{u,i} \right\rbrack_{j} = {\widehat{\beta}}_{u,i,0} + {\widehat{\beta}}_{u,i,1}(i + j),\ \ j = 1,\ldots,P\]
\end{minipage} & \begin{minipage}[b]{\linewidth}\raggedright
(\theequation)
\end{minipage} \\

\end{tabularx}

The soft-DTW loss associated with this reference does not mathematically enforce pointwise monotonicity. Instead, it encourages the predicted trajectory to align with the dominant linear direction estimated from the complete observation history.

\subsubsection{Differentiable recursive rollout}
\label{sec:methodology_recursive_rollout}

Because the LGFM requires both previous and current windows, its rollout state retains the most recent \(2L\) HI values. The initial state at forecast origin \(t\) is defined as

\refstepcounter{equation}
\label{eq:31}
\begin{tabularx}{\textwidth}{@{}>{\raggedright\arraybackslash}X>{\raggedleft\arraybackslash}p{0.08\textwidth}@{}}

\begin{minipage}[b]{\linewidth}\raggedright
\[\mathbf{q}_{u,t}^{(0)} = \left\lbrack \begin{array}{r}
\mathbf{p}_{u,t} \\
\mathbf{c}_{u,t}
\end{array} \right\rbrack \in \mathbb{R}^{2L}\]
\end{minipage} & \begin{minipage}[b]{\linewidth}\raggedright
(\theequation)
\end{minipage} \\

\end{tabularx}

For an arbitrary state \(\mathbf{q} = \left\lbrack q_{1},\ldots,q_{2L} \right\rbrack^{T}\), define \(Prev\left( \mathbf{q} \right) = \left\lbrack q_{1},\ldots,q_{L} \right\rbrack^{T}\) and \(Curr\left( \mathbf{q} \right) = \left\lbrack q_{L + 1},\ldots,q_{2L} \right\rbrack^{T}\). For all rollout blocks \(r = 0,\ldots,R - 1\), the previous and current windows are reconstructed from the current state, after which the next \(H\) HI values are predicted:

\refstepcounter{equation}
\label{eq:32}
\begin{tabularx}{\textwidth}{@{}>{\raggedright\arraybackslash}X>{\raggedleft\arraybackslash}p{0.08\textwidth}@{}}

\begin{minipage}[b]{\linewidth}\raggedright
\[\mathbf{p}_{u,t}^{(r)} = prev\left( \mathbf{q}_{u,t}^{(r)} \right),\ \ \mathbf{c}_{u,t}^{(r)} = Curr\left( \mathbf{q}_{u,t}^{(r)} \right),\ \ {\widehat{\mathbf{f}}}_{u,t}^{(r)} = \psi_{\theta}\left( \mathbf{h}_{u},\mathbf{p}_{u,t}^{(r)},\mathbf{c}_{u,t}^{(r)} \right) \in \mathbb{R}^{H}\]
\end{minipage} & \begin{minipage}[b]{\linewidth}\raggedright
(\theequation)
\end{minipage} \\

\end{tabularx}

The block \({\widehat{\mathbf{f}}}_{u,t}^{(r)}\) contains predictions for indices \(t + rH + 1,\ldots,t + (r + 1)H\). After appending the predicted block to the current state, the most recent \(2L\) values are retained as the state for the next rollout block:

\refstepcounter{equation}
\label{eq:33}
\begin{tabularx}{\textwidth}{@{}>{\raggedright\arraybackslash}X>{\raggedleft\arraybackslash}p{0.08\textwidth}@{}}

\begin{minipage}[b]{\linewidth}\raggedright
\[\mathbf{q}_{u,t}^{(r + 1)} = {Last}_{2L}\left( \left\lbrack \begin{array}{r}
\mathbf{p}_{u,t}^{(r)} \\
{\widehat{\mathbf{f}}}_{u,t}^{(r)}
\end{array} \right\rbrack \right) \in \mathbb{R}^{2L}\]
\end{minipage} & \begin{minipage}[b]{\linewidth}\raggedright
(\theequation)
\end{minipage} \\

\end{tabularx}

During training, the computation graph is retained across all \(R\) rollout blocks. Consequently, the gradients from losses in the later blocks propagate through the predictions generated in the earlier blocks. The complete rollout prediction is formed by concatenating the blockwise outputs:

\refstepcounter{equation}
\label{eq:34}
\begin{tabularx}{\textwidth}{@{}>{\raggedright\arraybackslash}X>{\raggedleft\arraybackslash}p{0.08\textwidth}@{}}

\begin{minipage}[b]{\linewidth}\raggedright
\[{\widehat{\mathbf{v}}}_{u,t} = \left\lbrack \begin{array}{r}
{\widehat{\mathbf{f}}}_{u,t}^{(0)} \\
{\widehat{\mathbf{f}}}_{u,t}^{(1)} \\
 \vdots \\
{\widehat{\mathbf{f}}}_{u,t}^{(R - 1)}
\end{array} \right\rbrack \in \mathbb{R}^{P},\ \ P = RH\]
\end{minipage} & \begin{minipage}[b]{\linewidth}\raggedright
(\theequation)
\end{minipage} \\

\end{tabularx}

The recursive-rollout MSE is defined as the MSE between the complete rollout prediction and its ground-truth target:

\refstepcounter{equation}
\label{eq:35}
\begin{tabularx}{\textwidth}{@{}>{\raggedright\arraybackslash}X>{\raggedleft\arraybackslash}p{0.08\textwidth}@{}}

\begin{minipage}[b]{\linewidth}\raggedright
\[\mathcal{L}_{RO}(u,t) = \frac{1}{P}\left\| {\widehat{\mathbf{v}}}_{u,t} - \mathbf{v}_{u,t} \right\|_{2}^{2}\]
\end{minipage} & \begin{minipage}[b]{\linewidth}\raggedright
(\theequation)
\end{minipage} \\

\end{tabularx}

Let \(D_{\gamma}( \bullet , \bullet )\) denote the soft-DTW divergence defined in \hyperref[sec:appendixC]{Appendix C}. The trend-guided loss is

\refstepcounter{equation}
\label{eq:36}
\begin{tabularx}{\textwidth}{@{}>{\raggedright\arraybackslash}X>{\raggedleft\arraybackslash}p{0.08\textwidth}@{}}

\begin{minipage}[b]{\linewidth}\raggedright
\[\mathcal{L}_{TG}(u,t) = D_{\gamma}\left( {\widehat{\mathbf{v}}}_{u,t},{\widetilde{\mathbf{v}}}_{u,t} \right),\ \ \gamma > 0\]
\end{minipage} & \begin{minipage}[b]{\linewidth}\raggedright
(\theequation)
\end{minipage} \\

\end{tabularx}

The TG-RC loss is defined as the weighted sum of the three loss terms:

\refstepcounter{equation}
\label{eq:37}
\begin{tabularx}{\textwidth}{@{}>{\raggedright\arraybackslash}X>{\raggedleft\arraybackslash}p{0.08\textwidth}@{}}

\begin{minipage}[b]{\linewidth}\raggedright
\[\mathcal{L}_{TG - RC}^{(e)}(u,t) = \lambda_{1}^{(e)}\mathcal{L}_{OS}(u,t) + \lambda_{2}^{(e)}\mathcal{L}_{RO}(u,t) + \lambda_{3}^{(e)}\mathcal{L}_{TG}(u,t)\]
\end{minipage} & \begin{minipage}[b]{\linewidth}\raggedright
(\theequation)
\end{minipage} \\

\end{tabularx}

where \(e\) denotes the training epoch and is used to compute the dynamic weights. Let \({\overline{\mathcal{L}}}_{k}^{(e)}\) denote the epochwise mean of loss component \(k \in \left\{ 1,2,3 \right\}\) at epoch \(e\), and let \(\varepsilon > 0\) be a numerical-stability constant. The recent relative rate of change and corresponding DWA weight are calculated as

\refstepcounter{equation}
\label{eq:38}
\begin{tabularx}{\textwidth}{@{}>{\raggedright\arraybackslash}X>{\raggedleft\arraybackslash}p{0.08\textwidth}@{}}

\begin{minipage}[b]{\linewidth}\raggedright
\[r_{k}^{(e - 1)} = \frac{{\overline{\mathcal{L}}}_{k}^{(e - 1)}}{{\overline{\mathcal{L}}}_{k}^{(e - 2)} + \varepsilon},\ \ \lambda_{k}^{(e)} = \frac{|k|\exp\left( \frac{r_{k}^{(e - 1)}}{\tau_{DWA}} \right)}{\sum_{j = 1}^{3}{\exp\left( \frac{r_{j}^{(e - 1)}}{\tau_{DWA}} \right)}}\]
\end{minipage} & \begin{minipage}[b]{\linewidth}\raggedright
(\theequation)
\end{minipage} \\

\end{tabularx}

\subsection{Online training and RUL estimation}
\label{sec:methodology_training_rul}

At IT index \(i\), only the HI prefix \(\left\{ y_{m,t} \right\}_{t = 1}^{i}\) of target bearing \(m\) is available. The set of valid forecast origins for which the previous window, current window, and rollout target of length \(P\) are all contained within the observed prefix is defined as

\refstepcounter{equation}
\label{eq:39}
\begin{tabularx}{\textwidth}{@{}>{\raggedright\arraybackslash}X>{\raggedleft\arraybackslash}p{0.08\textwidth}@{}}

\begin{minipage}[b]{\linewidth}\raggedright
\[\mathcal{J}_{m,i} = \left\{ t \in \mathbb{Z}: 2L \leq t \leq i - P \right\}\]
\end{minipage} & \begin{minipage}[b]{\linewidth}\raggedright
(\theequation)
\end{minipage} \\

\end{tabularx}

The resulting number of training samples is

\refstepcounter{equation}
\label{eq:40}
\begin{tabularx}{\textwidth}{@{}>{\raggedright\arraybackslash}X>{\raggedleft\arraybackslash}p{0.08\textwidth}@{}}

\begin{minipage}[b]{\linewidth}\raggedright
\[N_{m,i}^{samp} = \max(0,i - P - 2L + 1)\]
\end{minipage} & \begin{minipage}[b]{\linewidth}\raggedright
(\theequation)
\end{minipage} \\

\end{tabularx}

For a mini-batch \(B \subseteq \mathcal{J}_{m,i}\), training objective at epoch \(e\) is

\refstepcounter{equation}
\label{eq:41}
\begin{tabularx}{\textwidth}{@{}>{\raggedright\arraybackslash}X>{\raggedleft\arraybackslash}p{0.08\textwidth}@{}}

\begin{minipage}[b]{\linewidth}\raggedright
\[\mathcal{J}_{B}^{(e)} = \frac{1}{|B|}\sum_{t \in B}^{}{\mathcal{L}_{TG - RC}^{(e)}(m,t)},\ \ B \subseteq \mathcal{J}_{m,i}\]
\end{minipage} & \begin{minipage}[b]{\linewidth}\raggedright
(\theequation)
\end{minipage} \\

\end{tabularx}

After training, the forecast origin is set to the IT index \(i\), and the \(H\)-step prediction and state-update operations given in ~\eqref{eq:31}--~\eqref{eq:33} are applied recursively from the initial state \(\mathbf{q}_{m,i}^{(0)}\). Unlike the finite \(R\)-block rollout used during training, the inference continues the same \(H\)-step state update until the predicted HI reaches the FT. Only the recursive predictions generated by the trained LGFM are used for RUL estimation.

Given the recursively extrapolated HI sequence and the FT, the predicted FHT is defined as

\refstepcounter{equation}
\label{eq:42}
\begin{tabularx}{\textwidth}{@{}>{\raggedright\arraybackslash}X>{\raggedleft\arraybackslash}p{0.08\textwidth}@{}}

\begin{minipage}[b]{\linewidth}\raggedright
\[{\widehat{\tau}}_{F}(i) = \min\left( t \in \left\{ i + 1,\ldots,i + \rho \right\}:{\widehat{y}}_{m,t} \geq FT \right)\]
\end{minipage} & \begin{minipage}[b]{\linewidth}\raggedright
(\theequation)
\end{minipage} \\

\end{tabularx}

The predicted RUL is then calculated as follows.

\refstepcounter{equation}
\label{eq:43}
\begin{tabularx}{\textwidth}{@{}>{\raggedright\arraybackslash}X>{\raggedleft\arraybackslash}p{0.08\textwidth}@{}}

\begin{minipage}[b]{\linewidth}\raggedright
\[\widehat{RUL}(i) = \left\lbrack {\widehat{\tau}}_{F}(i) - i \right\rbrack\Delta t\]
\end{minipage} & \begin{minipage}[b]{\linewidth}\raggedright
(\theequation)
\end{minipage} \\

\end{tabularx}

\section{Experiments}
\label{sec:exp}

This section presents the experimental evaluation. Section \ref{sec:exp4.1} describes the datasets, Section \ref{sec:exp4.2} introduces the evaluation metrics, Section \ref{sec:exp4.3} details the experimental setup, and Section \ref{sec:exp4.4} reports the results.

\subsection{Dataset}
\label{sec:exp4.1}

\begin{figure}[!htbp]
\centering
\includegraphics[width=\textwidth]{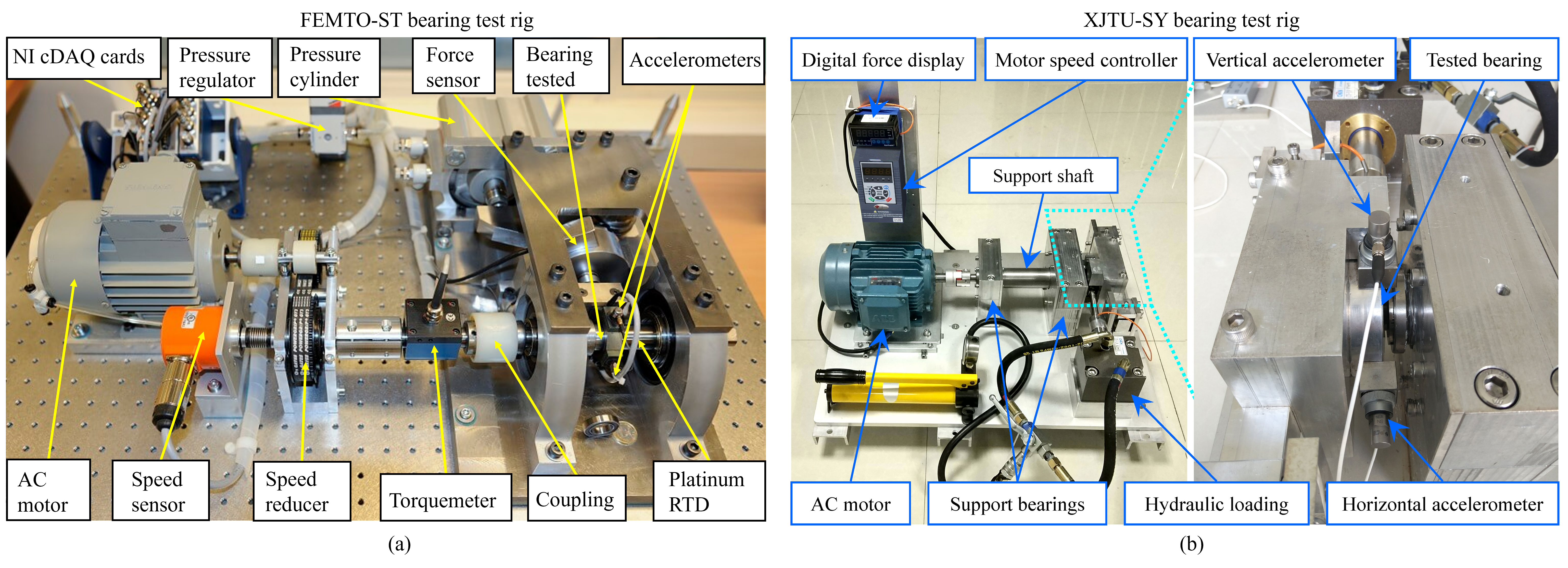}
\caption{Bearing test rigs used for evaluation: FEMTO-ST and XJTU-SY.}
\label{fig:3}
\end{figure}

The RUL prediction performance of the proposed framework was evaluated using two publicly available bearing datasets: the FEMTO-ST bearing dataset collected on the PRONOSTIA platform, and the XJTU-SY bearing dataset. The corresponding test rigs are illustrated in Figures \ref{fig:3}(a) and \ref{fig:3}(b), and their operating and data-acquisition conditions are described below.

\textbf{FEMTO-ST bearing (FB) dataset}

The FB dataset, released as part of the 2012 PHM challenge, was collected using the PRONOSTIA test platform, which was designed for fault diagnosis and prognostics of rolling bearings \citep{Nectoux2012PRONOSTIA}. It contains 17 complete RtF trajectories obtained under three operating conditions. Each bearing was monitored using accelerometers mounted in the horizontal and vertical directions. The operating conditions were as follows: Condition 1, 1,800 rpm and 4,000 N; Condition 2, 1,650 rpm and 4,200 N; and Condition 3, 1,500 rpm and 5,000 N. Vibration signals were sampled at 25.6 kHz for 0.1 s every 10 s.

\textbf{XJTU-SY bearing (XB) dataset}

The XB dataset is provided by the Institute of Design Science and Basic Components at XJTU, Shaanxi, P.R. China, and Changxing Sumyoung Technology Co., Ltd. (SY), Zhejiang, P. R. China \citep{Wang2020HybridPrognostics}. It contains complete RtF trajectories for 15 rolling element bearings obtained through accelerated degradation experiments. The experiments were conducted under three operating conditions: Condition 1, 2,100 rpm and 12,000 N; Condition 2, 2,250 rpm and 11,000 N; and Condition 3, 2,400 rpm and 10,000 N. Each vibration record was sampled at 25.6 kHz for 1.28 s, yielding 32,768 samples, and the interval between consecutive records was 1 min.

\subsection{Evaluation metrics}
\label{sec:exp4.2}

RUL prediction performance was evaluated using established metrics---the mean absolute error (MAE), normalized root mean squared error (NRMSE), and Score---together with the proposed non-crossing rate (NCR).

The MAE and RMSE are defined as \citep{Park2025TrendFluctuation}

\begin{tabularx}{\textwidth}
	{@{}>{\raggedright\arraybackslash}X
		>{\raggedleft\arraybackslash}p{0.08\textwidth}@{}}
	
	\begin{minipage}[b]{\linewidth}\raggedright
		\[
		MAE(i)
		=
		\frac{1}{10}
		\sum_{o=1}^{10}
		\left|
		RUL(i)-\widehat{RUL}_{o}(i)
		\right|
		\]
	\end{minipage}
	&
	\begin{minipage}[b]{\linewidth}\raggedright
		\refstepcounter{equation}
		\label{eq:44}
		(\theequation)
	\end{minipage}
	\\
	
	\begin{minipage}[b]{\linewidth}\raggedright
		\[
		RMSE(i)
		=
		\sqrt{
			\frac{1}{10}
			\sum_{o=1}^{10}
			\left[
			RUL(i)-\widehat{RUL}_{o}(i)
			\right]^2
		}
		\]
	\end{minipage}
	&
	\begin{minipage}[b]{\linewidth}\raggedright
		\refstepcounter{equation}
		\label{eq:45}
		(\theequation)
	\end{minipage}
	\\
	
\end{tabularx}

where \(RUL(i)\) is the actual RUL at IT \(i\), and \({\widehat{RUL}}_{o}(i)\) is its estimate from the \(o\)-th run. The NRMSE is calculated as

\refstepcounter{equation}
\label{eq:46}
\begin{tabularx}{\textwidth}{@{}>{\raggedright\arraybackslash}X>{\raggedleft\arraybackslash}p{0.08\textwidth}@{}}

\begin{minipage}[b]{\linewidth}\raggedright
\[NRMSE(i) = \frac{RMSE(i)}{\frac{1}{10}\sum_{o = 1}^{10}\left| {\widehat{RUL}}_{o}(i) \right|}\]
\end{minipage} & \begin{minipage}[b]{\linewidth}\raggedright
(\theequation)
\end{minipage} \\

\end{tabularx}

where \(o\) indexes repeated prediction runs. In this study, each prediction at IT \(i\) was repeated ten times \citep{Li2024GTFDAU,Lu2025LSTMTransformer,Park2025TrendFluctuation,Qin2021GDAU,Xiang2023CocktailLSTM,Zhou2023DTGRU}. Therefore, the MAE represents the mean absolute RUL prediction error across ten runs, whereas the RMSE places greater weight on large errors by taking the RMS of the differences between the actual and predicted RUL values. The NRMSE normalizes the RMSE by the mean absolute predicted RUL across the ten runs.

Second, the Score metric assigns asymmetric penalties to relative RUL prediction errors depending on whether the actual RUL is overestimated or underestimated \citep{Nectoux2012PRONOSTIA}. For the \(o\)-th repeated experiment, the relative error is defined as

\refstepcounter{equation}
\label{eq:47}
\begin{tabularx}{\textwidth}{@{}>{\raggedright\arraybackslash}X>{\raggedleft\arraybackslash}p{0.08\textwidth}@{}}

\begin{minipage}[b]{\linewidth}\raggedright
\[{Er}_{o}(i) = \frac{RUL(i) - {\widehat{RUL}}_{o}(i)}{RUL(i)}\]
\end{minipage} & \begin{minipage}[b]{\linewidth}\raggedright
(\theequation)
\end{minipage} \\

\end{tabularx}

If \({Er}_{o}(i) < 0\), then \({\widehat{RUL}}_{o}(i) > RUL(i)\), which indicates RUL overestimation. In this case, the model predicts a longer remaining lifetime than is actually available and may consequently issue a late failure warning. Conversely, \({Er}_{o}(i) > 0\) indicates RUL underestimation, which corresponds to a conservative prediction that anticipates failure earlier than it occurs.

The scoring term \(A_{o}(i)\) is defined by an exponential function whose scaling factor depends on the sign of \({Er}_{o}(i)\). A larger penalty is assigned to RUL overestimation:

\begin{tabularx}{\textwidth}
	{@{}>{\raggedright\arraybackslash}X
		>{\raggedleft\arraybackslash}p{0.08\textwidth}@{}}
	
	\begin{minipage}[b]{\linewidth}\raggedright
		\[
		A_o(i)
		=
		\begin{cases}
			\displaystyle
			\exp\left[
			-\ln(0.5)
			\left(\frac{{Er}_o(i)}{5}\right)
			\right],
			& \text{if } {Er}_o(i)\leq 0,
			\\[6pt]
			\displaystyle
			\exp\left[
			\ln(0.5)
			\left(\frac{{Er}_o(i)}{20}\right)
			\right],
			& \text{if } {Er}_o(i)>0.
		\end{cases}
		\]
	\end{minipage}
	&
	\begin{minipage}[b]{\linewidth}\raggedright
		\refstepcounter{equation}
		\label{eq:48}
		(\theequation)
	\end{minipage}
	\\
	
\end{tabularx}

The final Score is the mean of \(A_{o}(i)\) over the ten repeated experiments:

\refstepcounter{equation}
\label{eq:49}
\begin{tabularx}{\textwidth}{@{}>{\raggedright\arraybackslash}X>{\raggedleft\arraybackslash}p{0.08\textwidth}@{}}

\begin{minipage}[b]{\linewidth}\raggedright
\[Score(i) = \frac{1}{10}\sum_{o = 1}^{10}{A_{o}(i)}\]
\end{minipage} & \begin{minipage}[b]{\linewidth}\raggedright
(\theequation)
\end{minipage} \\

\end{tabularx}

Finally, previous studies have reported that neural-network-based recursive extrapolation may fail to reach the FT within a prescribed forecasting horizon \citep{Park2025TrendFluctuation,Li2023GRUDeepAR}. To evaluate extrapolation stability, this study introduces the NCR. For the \(o\)-th repeated experiment, a prediction is classified as non-crossing if the recursively extrapolated HI fails to reach the FT within \(\rho\) future steps. The corresponding indicator is

\begin{tabularx}{\textwidth}
	{@{}>{\raggedright\arraybackslash}X
		>{\raggedleft\arraybackslash}p{0.08\textwidth}@{}}
	
	\begin{minipage}[b]{\linewidth}\raggedright
		\[
		\mathrm{NC}_{o}(i;\mathrm{FT},\rho)
		=
		\begin{cases}
			1,
			&
			\begin{aligned}
				&\text{if the recursive prediction initiated at IT } i \\
				&\text{does not reach the FT within } \rho \text{ steps},
			\end{aligned}
			\\[4pt]
			0,
			&
			\text{otherwise}.
		\end{cases}
		\]
	\end{minipage}
	&
	\begin{minipage}[b]{\linewidth}\raggedright
		\refstepcounter{equation}
		\label{eq:50}
		(\theequation)
	\end{minipage}
	\\
	
\end{tabularx}

Here, \(\rho\) should be set sufficiently large to distinguish failure to reach the FT from an ordinarily long extrapolation. In this study, \(\rho = 500\).

Accordingly, \(NCR(i;FT,\rho)\) is defined as

\refstepcounter{equation}
\label{eq:51}
\begin{tabularx}{\textwidth}{@{}>{\raggedright\arraybackslash}X>{\raggedleft\arraybackslash}p{0.08\textwidth}@{}}

\begin{minipage}[b]{\linewidth}\raggedright
\[NCR(i;FT,\rho) = \frac{1}{10}\sum_{o = 1}^{10}{{NC}_{o}(i;FT,\rho)}\]
\end{minipage} & \begin{minipage}[b]{\linewidth}\raggedright
(\theequation)
\end{minipage} \\

\end{tabularx}

A lower \(NCR(i;FT,\rho)\) indicates that the HI trajectories extrapolated from IT \(i\) reach the FT more consistently within the specified horizon.

\subsection{Experimental settings}
\label{sec:exp4.3}

This section describes the experimental setup. It first specifies the training and test RtF trajectories used for each bearing dataset, then introduces the comparison methods for RUL prediction, and finally explains the experimental protocol, including the IT at which the inference was performed.

\begin{longtable}[]{@{}
  >{\raggedright\arraybackslash}p{(\manuscripttablewidth - 6\tabcolsep) * \real{0.2500}}
  >{\raggedright\arraybackslash}p{(\manuscripttablewidth - 6\tabcolsep) * \real{0.2500}}
  >{\raggedright\arraybackslash}p{(\manuscripttablewidth - 6\tabcolsep) * \real{0.2500}}
  >{\raggedright\arraybackslash}p{(\manuscripttablewidth - 6\tabcolsep) * \real{0.2500}}@{}}
\caption{FB dataset classified into three operating conditions.}\label{tab:2}\\
\toprule\noalign{}
\multirow{2}{*}{Subsets} & \multicolumn{3}{c}{Operating conditions} \\
& \begin{minipage}[b]{\linewidth}\raggedright
Condition 1
\end{minipage} & \begin{minipage}[b]{\linewidth}\raggedright
Condition 2
\end{minipage} & \begin{minipage}[b]{\linewidth}\raggedright
Condition 3
\end{minipage} \\
\midrule\noalign{}
\endfirsthead

\toprule\noalign{}
\multirow{2}{*}{Subsets} & \multicolumn{3}{c}{Operating conditions} \\
& \begin{minipage}[b]{\linewidth}\raggedright
Condition 1
\end{minipage} & \begin{minipage}[b]{\linewidth}\raggedright
Condition 2
\end{minipage} & \begin{minipage}[b]{\linewidth}\raggedright
Condition 3
\end{minipage} \\
\midrule\noalign{}
\endhead
\bottomrule\noalign{}
\endlastfoot
Train set & Bearing 1-2 (FB1-2) & Bearing 2-1 (FB2-1) & Bearing 3-1 (FB3-1) \\
& Bearing 1-4 (FB1-4) & Bearing 2-2 (FB2-2) & Bearing 3-2 (FB3-2) \\
& Bearing 1-5 (FB1-5) & Bearing 2-3 (FB2-3) & \\
& Bearing 1-6 (FB1-6) & Bearing 2-5 (FB2-5) & \\
& Bearing 1-7 (FB1-7) & Bearing 2-6 (FB2-6) & \\
\midrule\noalign{}
Test set & Bearing 1-1 (FB1-1) & Bearing 2-4 (FB2-4) & Bearing 3-3 (FB3-3) \\
& Bearing 1-3 (FB1-3) & Bearing 2-7 (FB2-7) & \\
\end{longtable}

\begin{longtable}[]{@{}
  >{\raggedright\arraybackslash}p{(\manuscripttablewidth - 6\tabcolsep) * \real{0.2500}}
  >{\raggedright\arraybackslash}p{(\manuscripttablewidth - 6\tabcolsep) * \real{0.2500}}
  >{\raggedright\arraybackslash}p{(\manuscripttablewidth - 6\tabcolsep) * \real{0.2500}}
  >{\raggedright\arraybackslash}p{(\manuscripttablewidth - 6\tabcolsep) * \real{0.2500}}@{}}
\caption{XB dataset classified into three operating conditions.}\label{tab:3}\\
\toprule\noalign{}
\multirow{2}{*}{Subsets} & \multicolumn{3}{c}{Operating conditions} \\
& \begin{minipage}[b]{\linewidth}\raggedright
Condition 1
\end{minipage} & \begin{minipage}[b]{\linewidth}\raggedright
Condition 2
\end{minipage} & \begin{minipage}[b]{\linewidth}\raggedright
Condition 3
\end{minipage} \\
\midrule\noalign{}
\endfirsthead

\toprule\noalign{}
\multirow{2}{*}{Subsets} & \multicolumn{3}{c}{Operating conditions} \\
& \begin{minipage}[b]{\linewidth}\raggedright
Condition 1
\end{minipage} & \begin{minipage}[b]{\linewidth}\raggedright
Condition 2
\end{minipage} & \begin{minipage}[b]{\linewidth}\raggedright
Condition 3
\end{minipage} \\
\midrule\noalign{}
\endhead
\bottomrule\noalign{}
\endlastfoot
Train set & Bearing 1-2 (XB1-2) & Bearing 2-1 (XB2-1) & Bearing 3-1 (XB3-1) \\
& Bearing 1-3 (XB1-3) & Bearing 2-2 (XB2-2) & Bearing 3-2 (XB3-2) \\
& Bearing 1-4 (XB1-4) & Bearing 2-3 (XB2-3) & Bearing 3-4 (XB3-4) \\
& & Bearing 2-4 (XB2-4) & Bearing 3-5 (XB3-5) \\
\midrule\noalign{}
Test set & Bearing 1-1 (XB1-1) & Bearing 2-5 (XB2-5) & Bearing 3-3 (XB3-3) \\
& Bearing 1-5 (XB1-5) & & \\
\end{longtable}

Tables \ref{tab:2} and \ref{tab:3} summarize the training and test sets of the FB and XB datasets, respectively. The DP-based RUL prediction framework combines HI construction with DP modeling. The HI construction model was trained using the training set, whereas the DP modeling used only the HI sequence observed before the IT of each test bearing. This protocol follows those of previous studies: the FB training and test sets were divided according to \citep{Li2024GTFDAU,Qin2021GDAU,Xiang2023CocktailLSTM,Zhou2023DTGRU}, whereas the XB datasets were divided according to \citep{Jiang2023DualChannelTransformer}.

Following \citep{Tefera2025ConstraintGuided}, the training set used for HI construction was divided into the following three degradation stages. This stagewise division was adopted to obtain balanced samples across the life cycle because HI trajectories are nonstationary.

\begin{enumerate}
\def\labelenumi{\arabic{enumi}.}
\item
  Healthy state: The first 10\% of the RtF trajectory, assumed to represent fully healthy operation.
\item
  Slight degradation phase: The interval from 10\% to 95\% of the RtF trajectory, during which degradation gradually develops.
\item
  Sharp degradation phase: The interval after 95\% of the RtF trajectory, characterized by rapid degradation leading to clear failure.
\end{enumerate}

Samples were drawn randomly from these three stages to construct each batch. Healthy-state, slight-degradation, and sharp-degradation samples accounted for 20\%, 70\%, and 10\% of each batch, respectively \citep{Tefera2025ConstraintGuided}. Of the combined samples from the three stages, 75\% were used for training and 25\% were used for validation. HI construction was terminated by early stopping when the validation loss failed to improve by at least 0.0001 for seven consecutive epochs.

\begin{figure}[!htbp]
\centering
\includegraphics[width=\textwidth,height=0.78\textheight,keepaspectratio]{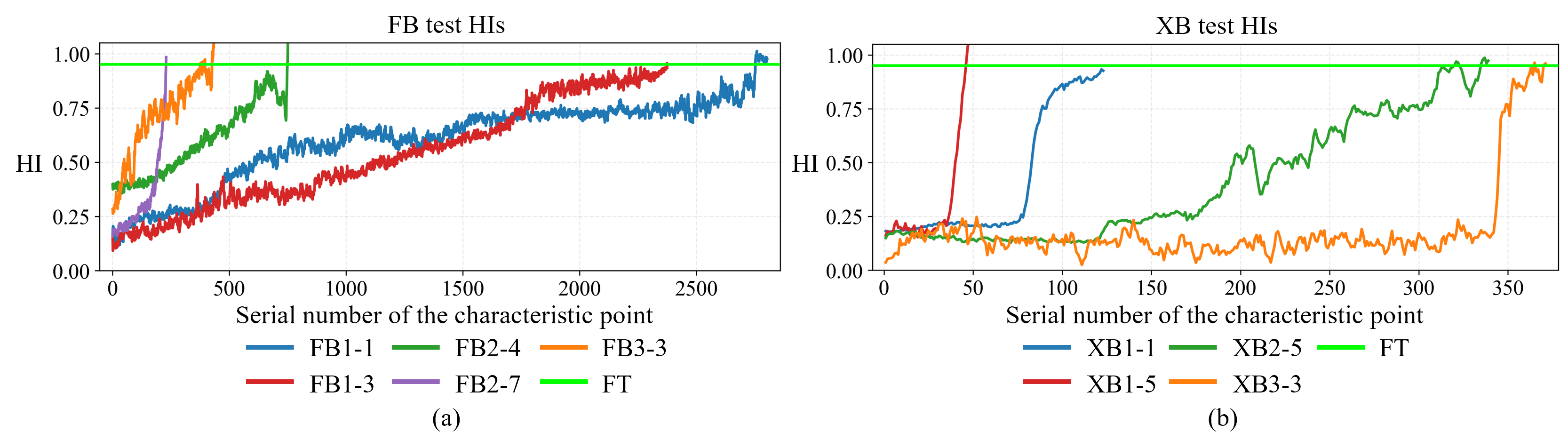}
\caption{HIs constructed for the test bearings in the (a) FB and (b) XB test sets.}
\label{fig:4}
\end{figure}

Figures \ref{fig:4}(a) and \ref{fig:4}(b) show the HI sequences constructed using the STF3AE for the FB and XB test sets, respectively. The resulting sequences exhibit substantial variation in both RtF duration and degradation patterns. Because RUL prediction performance may depend on the inference point, the models were evaluated under multiple IT conditions. For reproducibility, the FB experiments used IT settings comparable to or broader than those adopted in previous studies \citep{Li2024GTFDAU,Qin2021GDAU,Xiang2023CocktailLSTM,Zhou2023DTGRU}. For the XB dataset, ITs were determined with reference to the first prediction time (FPT) identified from each constructed HI sequence \citep{Lv2026AdaptiveFBM}. The FPT denotes the estimated onset of degradation at which the RUL inference begins. Specifically, the first 20\% of each XB trajectory was assumed to represent a healthy state. The mean \(\widehat{\mu}\) and standard deviation \(\widehat{\sigma}\) were estimated from this initial segment, and the FPT was defined as the first measurement index at which the HI departed from the healthy-state interval \(\widehat{\mu} \pm 4\widehat{\sigma}\) \citep{Lv2026AdaptiveFBM}. This index was treated as the onset of degradation and the earliest point from which RUL inference was performed. The RtF lengths and ITs of the FB and XB test bearings are summarized in Table \ref{tab:4}, and the FPT identified for each XB test bearing is reported in \hyperref[sec:appendixB]{Appendix B}.

\begin{longtable}[]{@{}
  >{\raggedright\arraybackslash}p{(\manuscripttablewidth - 6\tabcolsep) * \real{0.2505}}
  >{\raggedright\arraybackslash}p{(\manuscripttablewidth - 6\tabcolsep) * \real{0.2500}}
  >{\raggedright\arraybackslash}p{(\manuscripttablewidth - 6\tabcolsep) * \real{0.2500}}
  >{\raggedright\arraybackslash}p{(\manuscripttablewidth - 6\tabcolsep) * \real{0.2496}}@{}}
\caption{Experimental settings for each test bearing according to its RtF length and ITs.}\label{tab:4}\\
\toprule\noalign{}
\begin{minipage}[b]{\linewidth}\raggedright
Bearing
\end{minipage} & \begin{minipage}[b]{\linewidth}\raggedright
Test set
\end{minipage} & \begin{minipage}[b]{\linewidth}\raggedright
RtF
\end{minipage} & \begin{minipage}[b]{\linewidth}\raggedright
ITs
\end{minipage} \\
\midrule\noalign{}
\endfirsthead

\toprule\noalign{}
\begin{minipage}[b]{\linewidth}\raggedright
Bearing
\end{minipage} & \begin{minipage}[b]{\linewidth}\raggedright
Test set
\end{minipage} & \begin{minipage}[b]{\linewidth}\raggedright
RtF
\end{minipage} & \begin{minipage}[b]{\linewidth}\raggedright
ITs
\end{minipage} \\
\midrule\noalign{}
\endhead
\bottomrule\noalign{}
\endlastfoot
FB & FB11 & 2,803 & 50, 100, 150, 200 \\
& FB13 & 2,075 & 50, 100, 150, 200 \\
& FB24 & 751 & 50, 100, 150, 200 \\
& FB27 & 230 & 15, 30, 45, 60 \\
& FB33 & 434 & 15, 30, 45, 60 \\
\midrule\noalign{}
XB & XB11 & 123 & 5, 10, 15, 20 \\
& XB15 & 52 & 3, 6, 9, 12 \\
& XB25 & 339 & 50, 100, 150, 200 \\
& XB33 & 371 & 6, 12, 18, 24 \\
\end{longtable}

\begin{longtable}[]{@{}
  >{\raggedright\arraybackslash}p{(\manuscripttablewidth - 6\tabcolsep) * \real{0.2317}}
  >{\raggedright\arraybackslash}p{(\manuscripttablewidth - 6\tabcolsep) * \real{0.2620}}
  >{\raggedright\arraybackslash}p{(\manuscripttablewidth - 6\tabcolsep) * \real{0.2481}}
  >{\raggedright\arraybackslash}p{(\manuscripttablewidth - 6\tabcolsep) * \real{0.2582}}@{}}
\caption{Existing DP models used to evaluate RUL prediction performance.}\label{tab:5}\\
\toprule\noalign{}
\begin{minipage}[b]{\linewidth}\raggedright
Base model
\end{minipage} & \begin{minipage}[b]{\linewidth}\raggedright
Model
\end{minipage} & \begin{minipage}[b]{\linewidth}\raggedright
Year
\end{minipage} & \begin{minipage}[b]{\linewidth}\raggedright
Authors
\end{minipage} \\
\midrule\noalign{}
\endfirsthead

\toprule\noalign{}
\begin{minipage}[b]{\linewidth}\raggedright
Base model
\end{minipage} & \begin{minipage}[b]{\linewidth}\raggedright
Model
\end{minipage} & \begin{minipage}[b]{\linewidth}\raggedright
Year
\end{minipage} & \begin{minipage}[b]{\linewidth}\raggedright
Authors
\end{minipage} \\
\midrule\noalign{}
\endhead
\bottomrule\noalign{}
\endlastfoot
\multirow{4}{*}{RNN-based} & GDAU \citep{Qin2021GDAU} & 2021 & Qin et al. \\
& DTGRU \citep{Zhou2023DTGRU} & 2023 & Zhou et al. \\
& CLSTM \citep{Xiang2023CocktailLSTM} & 2023 & Xiang et al. \\
& GTFDAU \citep{Li2024GTFDAU} & 2024 & Li et al. \\
Transformer & LSTM-TF \citep{Lu2025LSTMTransformer} & 2025 & Lu et al. \\
\end{longtable}

Table \ref{tab:5} summarizes the comparison models used to evaluate the proposed DP model. The baselines were selected from existing DP-based RUL prediction methods. Specifically, four RNN-based many-to-one DP models developed for RUL prediction in rotating machinery were included: GDAU \citep{Qin2021GDAU}, DTGRU \citep{Zhou2023DTGRU}, CLSTM \citep{Xiang2023CocktailLSTM}, and GTFDAU \citep{Li2024GTFDAU}. To examine whether the proposed method remains effective across models developed for other types of equipment, LSTM-Transformer (TF) \citep{Lu2025LSTMTransformer}, originally proposed for RUL prediction of proton exchange membrane fuel cells, was also included. All DP models were evaluated using the same STF3AE-based HI sequences to ensure a consistent comparison.

\begin{longtable}[]{@{}
  >{\raggedright\arraybackslash}p{(\manuscripttablewidth - 8\tabcolsep) * \real{0.0847}}
  >{\raggedright\arraybackslash}p{(\manuscripttablewidth - 8\tabcolsep) * \real{0.1838}}
  >{\raggedright\arraybackslash}p{(\manuscripttablewidth - 8\tabcolsep) * \real{0.1130}}
  >{\raggedright\arraybackslash}p{(\manuscripttablewidth - 8\tabcolsep) * \real{0.3249}}
  >{\raggedright\arraybackslash}p{(\manuscripttablewidth - 8\tabcolsep) * \real{0.2936}}@{}}
\caption{Hyperparameter search spaces for the DP models.}\label{tab:6}\\
\toprule\noalign{}
\begin{minipage}[b]{\linewidth}\raggedright
Bearing
\end{minipage} & \begin{minipage}[b]{\linewidth}\raggedright
Architecture
\end{minipage} & \begin{minipage}[b]{\linewidth}\raggedright
Models
\end{minipage} & \begin{minipage}[b]{\linewidth}\raggedright
Hyperparameter search space
\end{minipage} & \begin{minipage}[b]{\linewidth}\raggedright
\end{minipage} \\
& & & Non-fixed & Fixed \\
\midrule\noalign{}
\endfirsthead

\toprule\noalign{}
\begin{minipage}[b]{\linewidth}\raggedright
Bearing
\end{minipage} & \begin{minipage}[b]{\linewidth}\raggedright
Architecture
\end{minipage} & \begin{minipage}[b]{\linewidth}\raggedright
Models
\end{minipage} & \begin{minipage}[b]{\linewidth}\raggedright
Hyperparameter search space
\end{minipage} & \begin{minipage}[b]{\linewidth}\raggedright
\end{minipage} \\
& & & Non-fixed & Fixed \\
\midrule\noalign{}
\endhead
\bottomrule\noalign{}
\endlastfoot
\multirow{2}{*}{FB} & Many-to-one & GDAU, DTGRU, CLSTM, GTFDAU & \(H_{m2o}^{FB} = \left\{ 1 \right\}\)

\(\eta_{m2o} = \left\{ 0.03,0.05,0.07 \right\}\) & \multirow{2}{=}{\centering\tiny\(\begin{aligned}L^{FB} &= \{81,100,121,144,169\},\\h_d &= \{25,27,29\}\end{aligned}\)} \\
& Many-to-many & LSTM-TF, LGFM & \(H_{m2m}^{FB} = \left\{ 10,30,50 \right\}\)

\(\eta_{m2m} = \left\{ {5e}^{- 3},2e^{- 3},e^{- 3} \right\}\) \\
\midrule\noalign{}
\multirow{2}{*}{XB} & Many-to-one & GDAU, DTGRU, CLSTM, GTFDAU & \(q_{m2o}^{XB} = \left\{ 1 \right\},\)

\(\eta_{m2o} = \left\{ 0.03,0.05,0.07 \right\}\) & \multirow{2}{=}{\centering\tiny\(\begin{aligned}L^{XB} &= \{25,36,49,64,81\},\\h_d &= \{25,27,29\}\end{aligned}\)} \\
& Many-to-many & LSTM-TF, LGFM & \(H_{m2m}^{XB} = \left\{ 3,5,7 \right\},\)

\(\eta_{m2m} = \left\{ {5e}^{- 3},2e^{- 3},e^{- 3} \right\}\) \\
\end{longtable}

Table \ref{tab:6} summarizes the hyperparameter search spaces used for the DP models. The search spaces were specified according to the forecasting structure of each model, namely many-to-one or many-to-many, and candidate values were selected based on previous studies \citep{Li2024GTFDAU,Qin2021GDAU,Xiang2023CocktailLSTM,Zhou2023DTGRU}. Here, \(L^{FB}\) and \(L^{XB}\) denote the input sequence lengths of the FB and XB test sets, respectively. \(H_{m2o}^{FB}\) and \(H_{m2m}^{FB}\) denote the output sequence lengths of the many-to-one and many-to-many structures, respectively, in the FB test set. Similarly, \(H_{m2o}^{XB}\) and \(H_{m2m}^{XB}\) represent the output sequence lengths of the XB test set. \(h_{d}\) denotes the hidden state dimension, and \(\eta\) the learning rate. Because the FB and XB datasets differ substantially in terms of RtF length, different search ranges were used for the two datasets. The smoothing parameter of the TG-RC loss was fixed at \(\gamma = 0.1\). For each DP model, the reported results were obtained using the best-performing hyperparameter configuration among the candidate combinations. The selection procedure follows the experimental protocols adopted in previous studies \citep{Li2024GTFDAU,Qin2021GDAU,Xiang2023CocktailLSTM,Zhou2023DTGRU}.

\subsection{Experimental results}
\label{sec:exp4.4}

This section presents the experimental results in four parts. First, the quantitative RUL prediction results and extrapolated HI trajectories are compared for the FB and XB test sets. Second, the sensitivity of the proposed method to its hyperparameters is examined. Third, the computational complexity of the models is compared. Finally, the interpretability of the proposed method is analyzed.

\begin{longtable}[]{@{}
  >{\raggedright\arraybackslash}p{(\manuscripttablewidth - 10\tabcolsep) * \real{0.1516}}
  >{\raggedright\arraybackslash}p{(\manuscripttablewidth - 10\tabcolsep) * \real{0.1748}}
  >{\raggedright\arraybackslash}p{(\manuscripttablewidth - 10\tabcolsep) * \real{0.1954}}
  >{\raggedright\arraybackslash}p{(\manuscripttablewidth - 10\tabcolsep) * \real{0.1704}}
  >{\raggedright\arraybackslash}p{(\manuscripttablewidth - 10\tabcolsep) * \real{0.1610}}
  >{\raggedright\arraybackslash}p{(\manuscripttablewidth - 10\tabcolsep) * \real{0.1468}}@{}}
\caption{RUL prediction performance on FB1-1 at different ITs.}\label{tab:7}\\
\toprule\noalign{}
\begin{minipage}[b]{\linewidth}\raggedright
IT
\end{minipage} & \begin{minipage}[b]{\linewidth}\raggedright
Model
\end{minipage} & \begin{minipage}[b]{\linewidth}\raggedright
Performance
\end{minipage} & \begin{minipage}[b]{\linewidth}\raggedright
\end{minipage} & \begin{minipage}[b]{\linewidth}\raggedright
\end{minipage} & \begin{minipage}[b]{\linewidth}\raggedright
\end{minipage} \\
& & MAE & NRMSE & Score & NCR \\
\midrule\noalign{}
\endfirsthead

\toprule\noalign{}
\begin{minipage}[b]{\linewidth}\raggedright
IT
\end{minipage} & \begin{minipage}[b]{\linewidth}\raggedright
Model
\end{minipage} & \begin{minipage}[b]{\linewidth}\raggedright
Performance
\end{minipage} & \begin{minipage}[b]{\linewidth}\raggedright
\end{minipage} & \begin{minipage}[b]{\linewidth}\raggedright
\end{minipage} & \begin{minipage}[b]{\linewidth}\raggedright
\end{minipage} \\
& & MAE & NRMSE & Score & NCR \\
\midrule\noalign{}
\endhead
\bottomrule\noalign{}
\endlastfoot

50 & GDAU & 41.5 & 0.832 & 0.058 & 0.0 \\
& \secondbest{DTGRU} & \secondbest{26.3} & 0.702 & \secondbest{0.140} & 0.0 \\
& CLSTM & 33.6 & \secondbest{0.701} & 0.133 & 0.0 \\
& GTFDAU & 40.4 & 0.811 & 0.062 & 0.0 \\
& LSTM-TF & 39.2 & 0.807 & 0.070 & 0.0 \\
& \textbf{LGFM} & \textbf{10.7} & \textbf{0.258} & \textbf{0.527} & 0.0 \\
\midrule\noalign{}
100 & GDAU & 64.3 & 0.672 & 0.139 & 0.0 \\
& \secondbest{DTGRU} & \secondbest{31.9} & \secondbest{0.333} & 0.023 & 0.0 \\
& CLSTM & 47.4 & 0.506 & \secondbest{0.140} & 0.0 \\
& GTFDAU & 49.5 & 0.512 & 0.009 & 0.0 \\
& LSTM-TF & 111.1 & 1.483 & 0.082 & 0.1 \\
& \textbf{LGFM} & \textbf{28.6} & \textbf{0.304} & \textbf{0.396} & 0.0 \\
\midrule\noalign{}
150 & GDAU & 123.3 & 0.829 & 0.062 & 0.0 \\
& DTGRU & 243.5 & 1.655 & 0.001 & 0.1 \\
& \secondbest{CLSTM} & \secondbest{61.9} & \secondbest{0.513} & \textbf{0.285} & 0.0 \\
& GTFDAU & 254.6 & 1.721 & 0.001 & 0.1 \\
& LSTM-TF & 171.3 & 1.386 & 0.072 & 0.3 \\
& \textbf{LGFM} & \textbf{56.7} & \textbf{0.382} & \secondbest{0.274} & 0.0 \\
\midrule\noalign{}
200 & GDAU & N/A & N/A & N/A & 1.0 \\
& DTGRU & N/A & N/A & N/A & 1.0 \\
& CLSTM & N/A & N/A & N/A & 1.0 \\
& GTFDAU & N/A & N/A & N/A & 1.0 \\
& LSTM-TF & N/A & N/A & N/A & 1.0 \\
& \textbf{LGFM} & \textbf{40.4} & \textbf{0.205} & \textbf{0.500} & \textbf{0.0} \\
\end{longtable}

Tables \ref{tab:7} and \ref{tab:8} present the RUL prediction performance at four ITs for FB1-1 and XB1-1, respectively. The results for the remaining test bearings are provided in \hyperref[sec:appendixA]{Appendix A}. The best result for each metric is shown in bold, and the second-best result is underlined. Lower values indicate better performance for MAE, NRMSE, and NCR, whereas a higher Score is preferable. If none of the ten recursive HI extrapolations reached the FT within the maximum extrapolation horizon, RUL could not be estimated; consequently, MAE, NRMSE, and Score are reported as N/A in these cases, with NCR set to 1.0.

Across the four ITs in Table \ref{tab:7}, the DTGRU and CLSTM generally performed better than the other baseline models. At IT=50, the MAE values of the GDAU, DTGRU, CLSTM, GTFDAU, and LSTM-TF were 41.5, 26.3, 33.6, 40.4, and 39.2, respectively, with the DTGRU having the lowest error among all baselines. The DTGRU also maintained a relatively low MAE at IT=100. However, its error increased markedly at IT=150, as did those of several other baselines. These variations indicate that the baseline models were sensitive to the location of the IT and the corresponding observed HI prefix.

CLSTM provided comparatively consistent predictions from IT=50 to IT=150. However, at IT=200, none of the baseline models produced an HI trajectory that reached the FT within the specified extrapolation horizon, thus preventing RUL estimation. This failure should not be attributed to a longer prediction horizon because IT=200 is later than the other ITs, and therefore corresponds to a shorter actual RUL. Instead, it indicates that the extrapolation direction learned from the prefix available at this IT was insufficient to drive the predicted HI toward the FT.

The LGFM achieved MAE values of 10.7, 28.6, 56.7, and 40.4 at the four ITs, respectively, yielding the lowest MAE in every case. Although its MAE increased at IT=150, it remained lower than those of all comparison models. Moreover, the LGFM maintained an NCR of 0 across all four ITs, whereas every baseline failed to reach the FT in all ten runs at IT=200. These results indicate that the LGFM was less sensitive to the selected IT and produced more reliable FT-crossing behavior on FB1-1.

\begin{longtable}[]{@{}
  >{\raggedright\arraybackslash}p{(\manuscripttablewidth - 10\tabcolsep) * \real{0.1516}}
  >{\raggedright\arraybackslash}p{(\manuscripttablewidth - 10\tabcolsep) * \real{0.1748}}
  >{\raggedright\arraybackslash}p{(\manuscripttablewidth - 10\tabcolsep) * \real{0.1954}}
  >{\raggedright\arraybackslash}p{(\manuscripttablewidth - 10\tabcolsep) * \real{0.1704}}
  >{\raggedright\arraybackslash}p{(\manuscripttablewidth - 10\tabcolsep) * \real{0.1610}}
  >{\raggedright\arraybackslash}p{(\manuscripttablewidth - 10\tabcolsep) * \real{0.1468}}@{}}
\caption{RUL prediction performance on XB1-1 at different ITs.}\label{tab:8}\\
\toprule\noalign{}
\begin{minipage}[b]{\linewidth}\raggedright
IT
\end{minipage} & \begin{minipage}[b]{\linewidth}\raggedright
Model
\end{minipage} & \begin{minipage}[b]{\linewidth}\raggedright
Performance
\end{minipage} & \begin{minipage}[b]{\linewidth}\raggedright
\end{minipage} & \begin{minipage}[b]{\linewidth}\raggedright
\end{minipage} & \begin{minipage}[b]{\linewidth}\raggedright
\end{minipage} \\
& & MAE & NRMSE & Score & NCR \\
\midrule\noalign{}
\endfirsthead

\toprule\noalign{}
\begin{minipage}[b]{\linewidth}\raggedright
IT
\end{minipage} & \begin{minipage}[b]{\linewidth}\raggedright
Model
\end{minipage} & \begin{minipage}[b]{\linewidth}\raggedright
Performance
\end{minipage} & \begin{minipage}[b]{\linewidth}\raggedright
\end{minipage} & \begin{minipage}[b]{\linewidth}\raggedright
\end{minipage} & \begin{minipage}[b]{\linewidth}\raggedright
\end{minipage} \\
& & MAE & NRMSE & Score & NCR \\
\midrule\noalign{}
\endhead
\bottomrule\noalign{}
\endlastfoot

5 & GDAU & 4.0 & 0.800 & 0.062 & 0.0 \\
& \textbf{DTGRU} & \textbf{1.0} & \textbf{0.268} & \textbf{0.375} & 0.0 \\
& CLSTM & 25.1 & 1.106 & 0.249 & 0.0 \\
& \secondbest{GTFDAU} & \secondbest{1.6} & \secondbest{0.456} & \secondbest{0.363} & 0.0 \\
& LSTM-TF & 4.4 & 1.229 & 0.131 & 0.0 \\
& LGFM & 2.4 & 0.529 & 0.256 & 0.0 \\
\midrule\noalign{}
10 & GDAU & 9.0 & 0.900 & 0.044 & 0.0 \\
& \textbf{DTGRU} & \textbf{2.7} & \textbf{0.304} & \secondbest{0.400} & 0.0 \\
& CLSTM & 10.7 & 0.728 & 0.282 & 0.0 \\
& \secondbest{GTFDAU} & \secondbest{2.9} & \textbf{0.486} & \textbf{0.432} & 0.0 \\
& LSTM-TF & 7.4 & 0.857 & 0.164 & 0.0 \\
& LGFM & 3.9 & \secondbest{0.457} & 0.216 & 0.0 \\
\midrule\noalign{}
15 & GDAU & 14.0 & 0.933 & 0.039 & 0.0 \\
& \textbf{DTGRU} & \textbf{2.5} & \textbf{0.211} & \textbf{0.369} & 0.0 \\
& \secondbest{CLSTM} & \secondbest{6.3} & \secondbest{0.477} & 0.243 & 0.0 \\
& GTFDAU & 55.1 & 10.243 & 0.221 & 0.1 \\
& LSTM-TF & 11.1 & 1.123 & \secondbest{0.269} & 0.0 \\
& LGFM & 8.6 & 0.707 & 0.194 & 0.0 \\
\midrule\noalign{}
20 & GDAU & 17.9 & 0.895 & 0.045 & 0.0 \\
& \secondbest{DTGRU} & \secondbest{11.0} & \secondbest{0.563} & \secondbest{0.164} & 0.0 \\
& CLSTM & 16.1 & 0.806 & 0.062 & 0.0 \\
& GTFDAU & 13.4 & 0.673 & 0.100 & 0.0 \\
& LSTM-TF & 13.9 & 0.818 & 0.146 & 0.0 \\
& \textbf{LGFM} & \textbf{3.4} & \textbf{0.240} & \textbf{0.644} & 0.0 \\
\end{longtable}

Table \ref{tab:8} presents a different pattern. At the earlier ITs, several baseline models outperformed the LGFM in terms of MAE. The DTGRU achieved the lowest MAE at IT=5, IT=10, and IT=15, with values of 1.0, 2.7, and 2.5, respectively. The GTFDAU was also competitive at IT=5 and IT=10, although its performance deteriorated sharply at IT=15. These results show that the existing methods can provide accurate RUL estimates under certain small-IT conditions. At IT=20, the LGFM attained the lowest MAE and NRMSE (3.4 and 0.240, respectively) and achieved the highest Score of 0.644. The baseline errors generally increased at this IT, whereas the LGFM improved relative to its result at IT=15. Because a later IT provides a longer observed HI prefix and a shorter remaining extrapolation interval, this result suggests that the LGFM used the available additional degradation information more effectively by IT=20. The XB1-1 results do not establish that the LGFM is uniformly superior over longer extrapolation horizons; rather, they show that its relative advantage depends on the quantity and characteristics of the observed degradation history data.

\begin{figure}[!htbp]
\centering
\includegraphics[width=\textwidth,height=0.78\textheight,keepaspectratio]{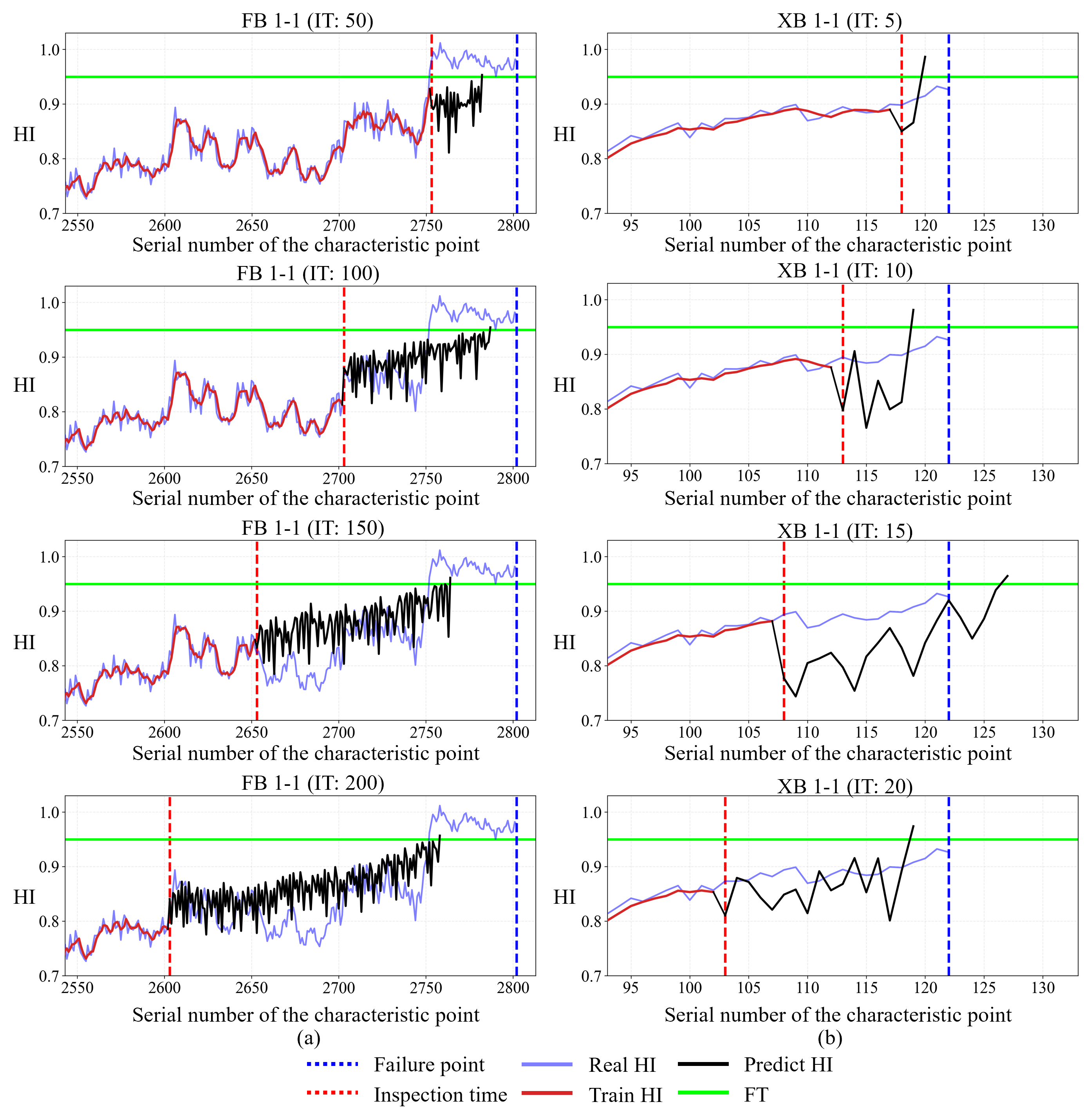}
\caption{HI extrapolation results obtained by LGFM for FB1-1 and XB1-1 at four ITs.}
\label{fig:5}
\end{figure}

Figures \ref{fig:5}(a) and \ref{fig:5}(b) show the recursively extrapolated HI trajectories of FB1-1 and XB1-1, respectively, under four IT conditions. In each panel, the horizontal and vertical axes represent the measurement index and HI, respectively. The red vertical dashed line indicates the IT at which recursive forecasting begins, and the blue vertical dashed line indicates the actual failure time. The blue solid line represents the original HI sequence, and the red solid line represents the smooth HI sequence using a window size of three; this sequence was used to train the LGFM model. The black solid line denotes the HI trajectory recursively extrapolated using the LGFM.

The proposed framework produced stable extrapolations across the considered ITs. For FB1-1 in Figure \ref{fig:5}(a), the predicted HI exhibited a nonlinear drift accompanied by recurrent fluctuations while continuing to progress toward the FT. The dominant global degradation direction was retained even over an extended recursive rollout. For XB1-1 in Figure \ref{fig:5}(b), the predicted segment showed larger fluctuations, but its overall trajectory continued toward the FT. These results indicate that the LGFM trained with the TG-RC loss can retain the direction of degradation under markedly different trajectory patterns. This interpretation concerns directional stability rather than pointwise accuracy, which must be assessed separately using quantitative metrics.

\begin{figure}[!htbp]
\centering
\includegraphics[width=\textwidth,height=0.78\textheight,keepaspectratio]{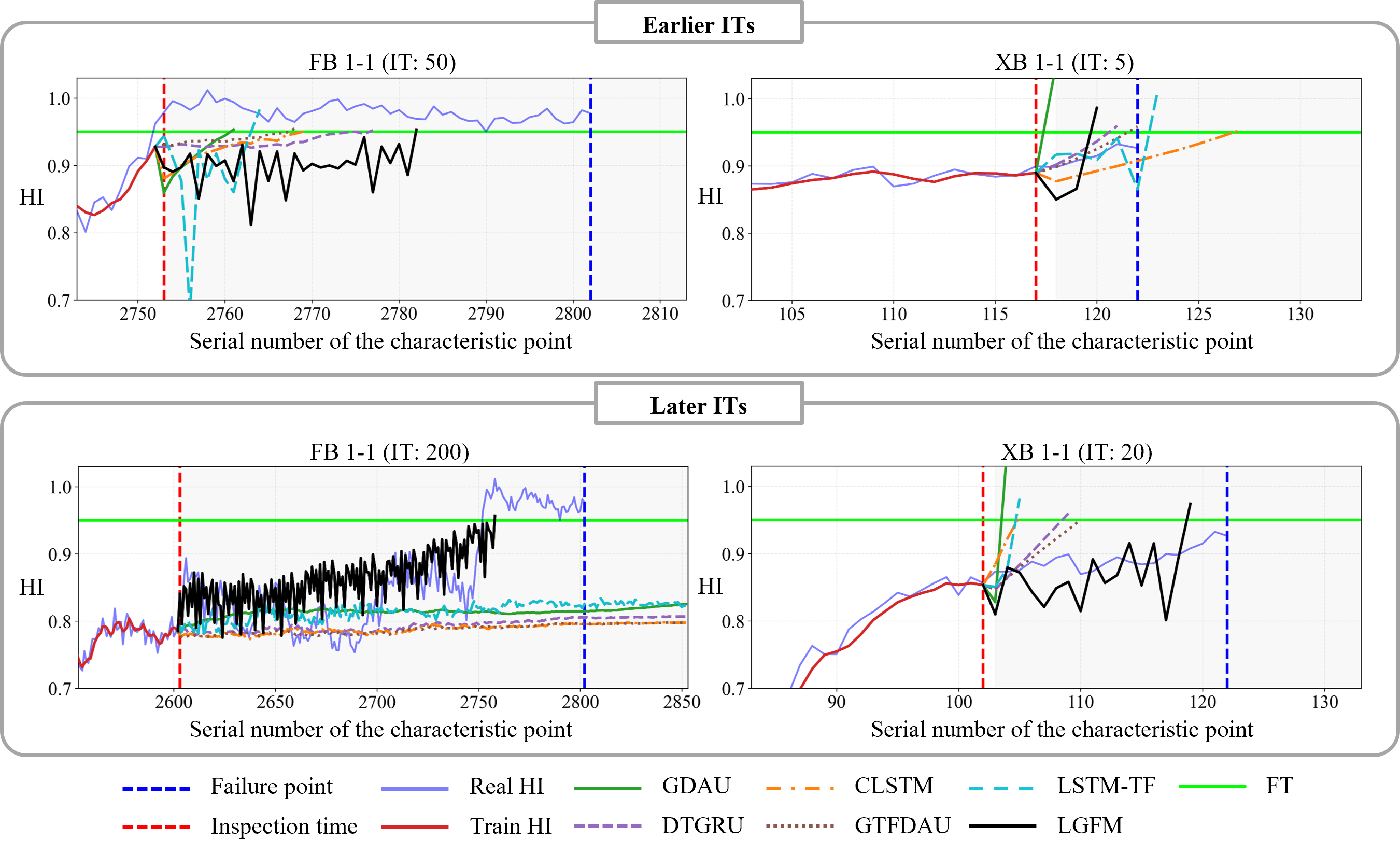}
\caption{HI extrapolation results of the six models for FB1-1 and XB1-1 with earlier and later ITs selected.}
\label{fig:6}
\end{figure}

Figure \ref{fig:6} compares the HI extrapolation behavior of the LGFM and the five baseline models for FB1-1 and XB1-1 at selected earlier and later ITs. At the earlier ITs shown in the figure, the RNN-based models (GDAU, DTGRU, CLSTM, and GTFDAU) generally retained information from the observed HI prefix and produced comparatively consistent extrapolations. Some baselines also achieved lower RUL prediction errors than the LGFM under specific conditions, including XB1-1 at IT=5. However, LSTM-TF exhibited pronounced fluctuations even at an earlier IT, suggesting greater sensitivity to recursively generated inputs and the risk of prematurely crossing the FT.

A different pattern was observed for FB1-1 at IT=200. None of the baseline trajectories progressed consistently toward the FT within the maximum extrapolation horizon. This behavior is consistent with the NCR of 1.0 reported for all five baselines at IT=200 in Table \ref{tab:7}. Because none of their extrapolated HI trajectories reached the FT, an RUL estimate could not be obtained. In contrast, the proposed LGFM DP and TG-RC loss-function framework initially produced a gradual change and subsequently maintained a trajectory directed toward the FT, allowing for RUL estimation at this IT.

This difference cannot be attributed solely to the LGFM architecture. The baseline models were primarily optimized using pointwise prediction losses, which do not explicitly regulate the direction of a recursively extrapolated trajectory. By contrast, the TG-RC loss incorporates recursive-rollout error and alignment with a trend prior into training, thereby encouraging the prediction to retain a consistent degradation direction. An ablation study examining the respective contributions of the TG-RC loss is presented in Section \ref{sec:ablation3}.

\begin{figure}[!htbp]
\centering
\includegraphics[width=\textwidth,height=0.78\textheight,keepaspectratio]{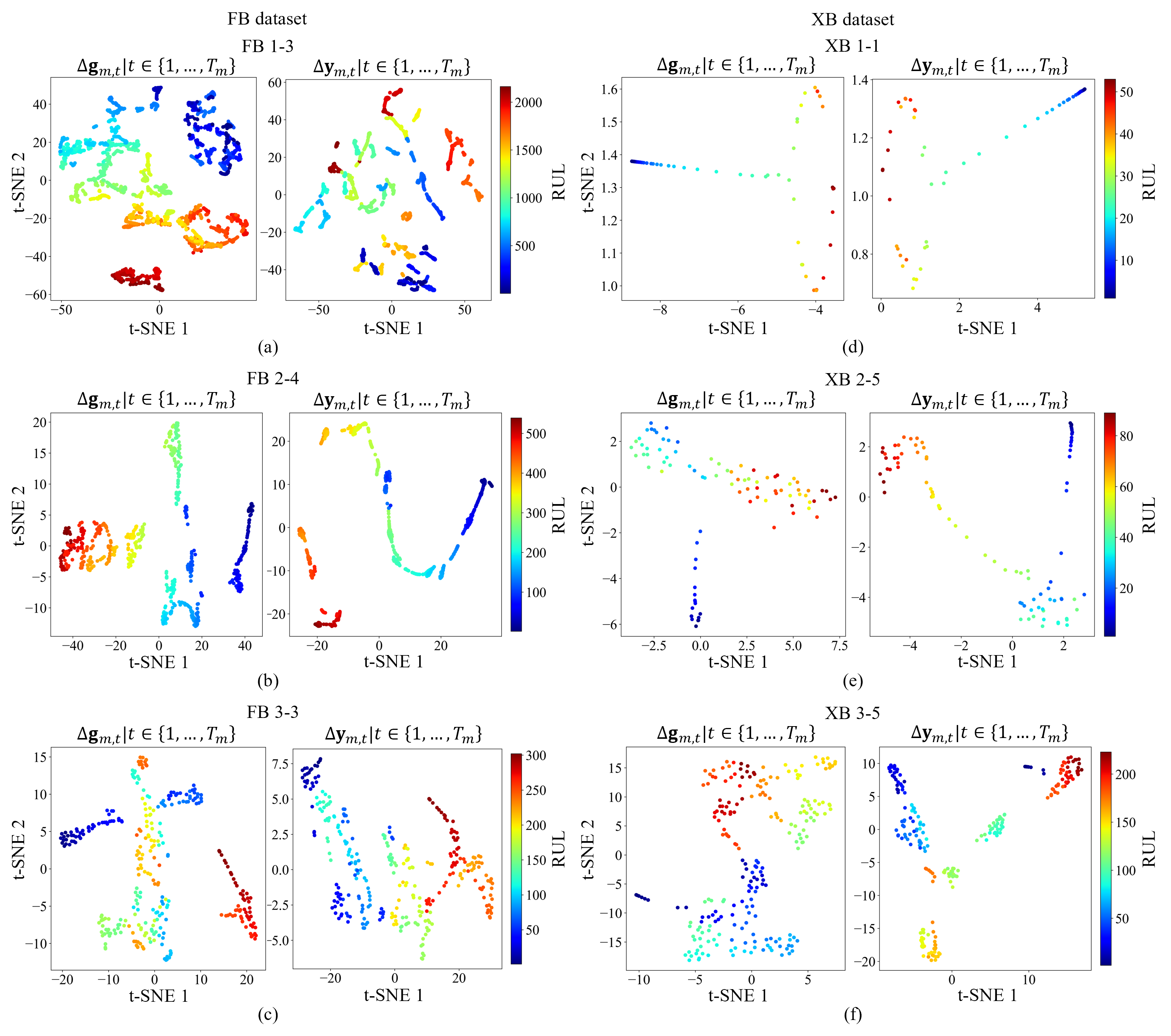}
\caption{t-SNE visualization of DF vector with respect to changes in RUL.}
\label{fig:7}
\end{figure}

Figure \ref{fig:7} presents two-dimensional t-distributed stochastic neighbor embedding (t-SNE) visualizations of the five-dimensional DF vectors extracted from the RtF trajectories of the FB and XB test bearings. Figures \ref{fig:7}(a)--(c) correspond to the FB dataset and Figures \ref{fig:7}(d)--(f) correspond to the XB dataset. In each subfigure, the horizontal and vertical axes represent the two t-SNE embedding dimensions, and each point represents one DF vector. The point color indicates the corresponding RUL: red denotes a healthy state with a large RUL, whereas blue denotes a state closer to failure with a small RUL.

The GDF is computed from the difference between the HI window at a given IT and the reference HI window, representing the initial healthy state. It is therefore expressed as \(\phi\left( \Delta\mathbf{g}_{\tau} \right)\) and describes global changes from the healthy condition. In contrast, the LDF is computed from the difference between the HI window at the IT and the window located \(L\) lags earlier. It is expressed as \(\phi\left( \Delta\mathbf{l}_{\tau} \right)\) and describes local changes in the HI trajectory.

For the FB dataset shown in Figures \ref{fig:7}(a)--(c), the locations of the embedded DF vectors generally varied with RUL, although the consistency of this relationship differed across the test bearings. In the GDF plot shown in Figure \ref{fig:7}(a), for example, the point colors progressively change from red to blue across the embedding space. This pattern indicates a clear association between the GDF vectors and the RUL. The corresponding LDF plot exhibits a less consistent color transition, suggesting that short-term difference features are not necessarily aligned with the degradation stage throughout the entire trajectory. Conversely, Figure \ref{fig:7}(c) shows a clearer displacement of the LDF vectors as RUL decreases, indicating that local changes may represent degradation progression more directly under certain conditions. For the XB dataset in Figures \ref{fig:7}(d)--(f), changes in the embedding location generally correspond to changes in RUL for both the global and local DFs. This result suggests that for the XB RtF trajectories considered here, both global shifts in the health state and local variations are associated with degradation progression.

Taken together, the FB and XB results indicate that global and local DFs play different roles across bearings and operating conditions. In some cases, the global differences provided a more consistent separation of RUL levels, whereas in others, local differences responded more sensitively to degradation progression. Both DFs may also exhibit an ordered relationship with RUL, as observed for parts of the XB dataset. Therefore, jointly using global and local DFs can provide complementary information and reduce the loss of degradation-related information that may occur when only one DF type is used.

\begin{figure}[!htbp]
\centering
\includegraphics[width=\textwidth,height=0.78\textheight,keepaspectratio]{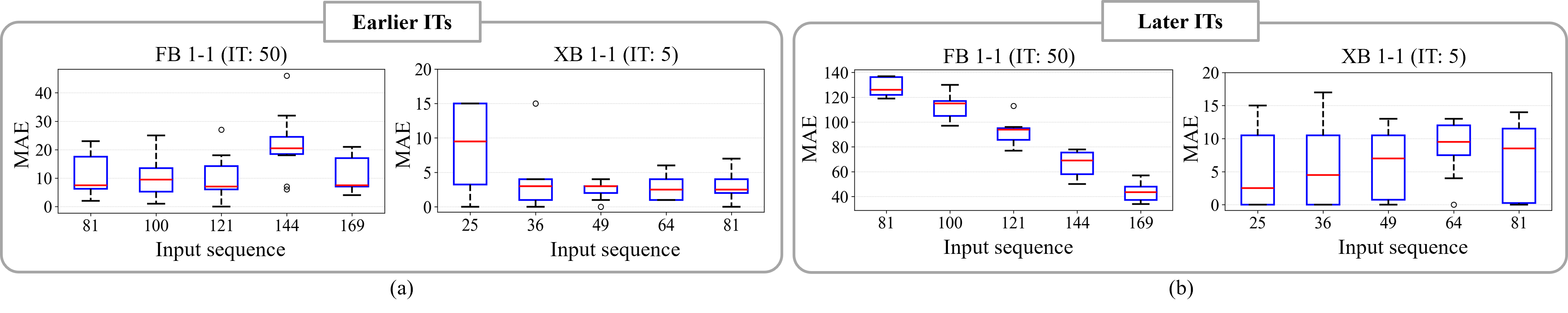}
\caption{Sensitivity of LGFM performance to the input-sequence length. Fixed parameters: output-sequence length, 20 for FB1-1 and 5 for XB1-1; learning rate, 0.002; hidden dimension, 27.}
\label{fig:8}
\end{figure}

\begin{figure}[!htbp]
\centering
\includegraphics[width=\textwidth,height=0.78\textheight,keepaspectratio]{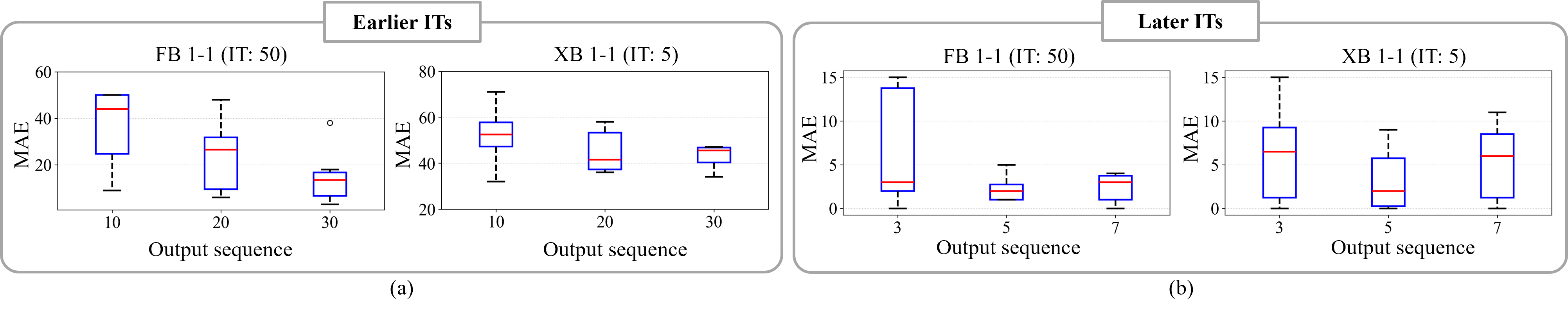}
\caption{Sensitivity of LGFM performance to the output-sequence length. Fixed parameters: input-sequence length, 169 for FB1-1 and 81 for XB1-1; learning rate, 0.002; hidden dimension, 27.}
\label{fig:9}
\end{figure}

Figures \ref{fig:8} and \ref{fig:9} illustrate the sensitivity of the LGFM to the input- and output-sequence lengths, respectively. Each box plot summarizes the results of ten repeated experiments, and the title of each panel indicates the IT used for evaluation. As shown in Figure \ref{fig:8}(a), the LGFM showed similar performance across the input lengths at IT=50 for FB1-1. However, at IT=200, the performance varied more substantially with the input length. The mean error was approximately 120 with an input length of 81 and decreased to approximately 40 when the input length was increased to 169. This suggests that using a longer input sequence may be beneficial for long-horizon recursive forecasting at IT=200. For XB1-1, as shown in Figure \ref{fig:8}(b), input length did not greatly affect performance at either IT. Thus, the benefit of increasing the input length was inconsistent across the two bearings and depended on the characteristics of the observed HI trajectory.

Figures \ref{fig:9}(a) and \ref{fig:9}(b) show no systematic difference in the mean performance across the considered output-sequence lengths. However, as shown in Figure \ref{fig:9}(b), an output length of three produced greater variability across repeated experiments, indicating reduced prediction stability under this setting. These results do not support the approach of selecting the longest input and output sequences in all cases. Instead, the sequence lengths should be chosen by considering both prediction stability and the amount of observed HI data available for training. Although increasing either length adds a relatively small computational burden to LGFM, longer sequences reduce the number of valid training windows, requiring a longer observed prefix.

\begin{figure}[!htbp]
\centering
\includegraphics[width=\textwidth]{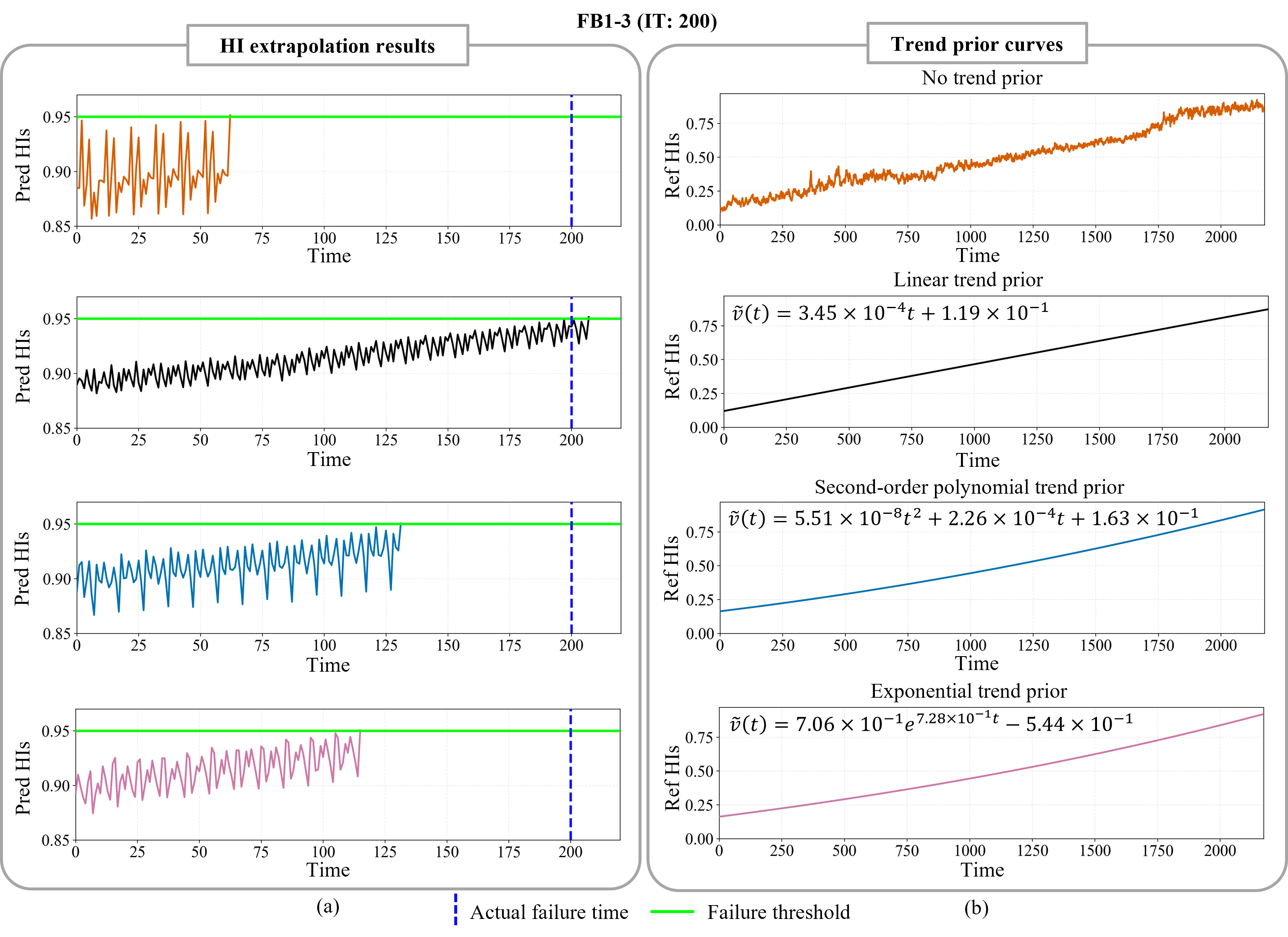}
\caption{Comparison of HI extrapolation results under different constructions of the reference sequence \(\widetilde{\mathbf{v}}\) used in the soft-DTW term of the TG-RC loss: (a) HI extrapolation under each reference construction and (b) the original, unfitted HI sequence and the fitted linear, polynomial, and exponential trend-prior curves.}
\label{fig:10}
\end{figure}

Figure \ref{fig:10} illustrates how the construction of the reference sequence \(\widetilde{v}\) in the soft-DTW component of the TG-RC loss affects HI extrapolation. Rather than directly using the original HI sequence as the alignment reference, the proposed method constructs a trend prior from the observed HI history. Equation~\eqref{eq:30} defines \(\widetilde{v}\) using a linear trend. Figure \ref{fig:10} compares four reference constructions: (i) the original HI sequence without a fitted trend prior, (ii) a linear trend prior, (iii) a polynomial trend prior, and (iv) an exponential trend prior. This comparison evaluates the effect of the assumed functional family of the prior on the direction and stability of the extrapolated trajectory.

The ``No trend prior'' case shown in Figure \ref{fig:10}(a) uses the original HI sequences as the soft-DTW reference rather than fitting a linear, polynomial, or exponential trend. This reference contains both the global degradation trend and local fluctuations. Consequently, soft-DTW may align the prediction with local variations, rather than primarily reflecting agreement in the overall degradation direction. As illustrated by the orange curve in Figure \ref{fig:10}(a), this setting produced a more oscillatory trajectory and caused a premature FT crossing.

Linear, polynomial, and exponential trend priors smooth the observed HI history using low-dimensional functional forms. This suppresses local fluctuations and emphasizes the global trend, allowing the soft-DTW term to respond more directly to discrepancies in the overall degradation direction. The resulting trajectories retain a more consistent direction over the extrapolation interval. The linear trend prior was adopted in this study because it provides a simple reference without imposing excessive curvature. The results shown in Figure \ref{fig:10} indicate that polynomial and exponential priors may also be viable alternatives for the examined trajectory. However, this finding should not be interpreted as evidence that one functional family is universally appropriate. The trend prior serves only to provide auxiliary directional guidance and does not mathematically enforce pointwise monotonicity.

\begin{longtable}[]{@{}
  >{\raggedright\arraybackslash}p{(\manuscripttablewidth - 8\tabcolsep) * \real{0.1227}}
  >{\raggedright\arraybackslash}p{(\manuscripttablewidth - 8\tabcolsep) * \real{0.1674}}
  >{\raggedright\arraybackslash}p{(\manuscripttablewidth - 8\tabcolsep) * \real{0.1298}}
  >{\raggedright\arraybackslash}p{(\manuscripttablewidth - 8\tabcolsep) * \real{0.1846}}
  >{\raggedright\arraybackslash}p{(\manuscripttablewidth - 8\tabcolsep) * \real{0.3956}}@{}}
\caption{Comparison of model complexity. Recursive RUL extrapolation times were averaged over 1,000 runs.}\label{tab:9}\\
\toprule\noalign{}
\begin{minipage}[b]{\linewidth}\raggedright
Models
\end{minipage} & \begin{minipage}[b]{\linewidth}\raggedright
Parameters (K)
\end{minipage} & \begin{minipage}[b]{\linewidth}\raggedright
FLOPs (K)
\end{minipage} & \begin{minipage}[b]{\linewidth}\raggedright
Time complexity
\end{minipage} & \begin{minipage}[b]{\linewidth}\raggedright
Recursive RUL extrapolation time (ms)
\end{minipage} \\
\midrule\noalign{}
\endfirsthead

\toprule\noalign{}
\begin{minipage}[b]{\linewidth}\raggedright
Models
\end{minipage} & \begin{minipage}[b]{\linewidth}\raggedright
Parameters (K)
\end{minipage} & \begin{minipage}[b]{\linewidth}\raggedright
FLOPs (K)
\end{minipage} & \begin{minipage}[b]{\linewidth}\raggedright
Time complexity
\end{minipage} & \begin{minipage}[b]{\linewidth}\raggedright
Recursive RUL extrapolation time (ms)
\end{minipage} \\
\midrule\noalign{}
\endhead
\bottomrule\noalign{}
\endlastfoot
GDAU & 24.62 & 49.00 & \(O(N)\) & 17.45 \\
DTGRU & 32.35 & 64.04 & \(O(N)\) & 19.28 \\
CLSTM & 44.28 & 87.97 & \(O(N)\) & 37.54 \\
GTFDAU & 18.85 & 37.39 & \(O(N)\) & 16.12 \\
LSTM-TF & 94.06 & 4588.28 & \(O\left( N^{2} \right)\) & 19.85 \\
LGFM & 9.13 & 11.48 & \(O(N)\) & 0.40 \\
\end{longtable}

Table \ref{tab:9} compares the computational complexity of the DP models. Because the models differ in their output structures, recursive extrapolation time was measured over a common 50-step forecasting horizon. The RNN-based models employ a many-to-one structure and therefore require 50 successive model evaluations to produce a 50-step forecast. LSTM-TF and the LGFM use a many-to-many structure; with an output length of 10, they require five successive evaluations to cover the same horizon. Under this protocol, each model was evaluated 1,000 times, and the mean elapsed time was reported. The timing measures only recursive HI extrapolation after model fitting. The experiments were conducted using a system equipped with an Intel Core i9-14900KF CPU operating at 3.20 GHz and an NVIDIA GeForce RTX 4080 GPU. The models were implemented in Python 3.8 using PyTorch. CUDA 12.6 and NVIDIA driver version 560.94 were used for GPU acceleration.

The RNN-based baselines had relatively modest parameter counts and FLOPs. Among them, the GTFDAU had the fewest parameters and FLOPs, whereas CLSTM had the largest values. LSTM-TF had the highest parameter count and substantially more FLOPs because of its combined LSTM--Transformer architecture. Its self-attention operation also has quadratic complexity with respect to the sequence length. Its many-to-many output reduces the number of recursive model evaluations, although its measured extrapolation time was the second highest among the baselines. The LGFM had the lowest computational cost, with 9.13 K parameters, 11.48 K FLOPs, and linear complexity with respect to the input-sequence length. Its mean 50-step extrapolation time was 0.40 ms. This corresponds to a reduction of 15.72 ms relative to the GTFDAU, the fastest baseline, and 37.14 ms relative to CLSTM, the slowest model in this comparison. These measurements indicate that the LGFM has a low forward-computation cost and may therefore be suitable for deployment on resource-constrained platforms.

\begin{longtable}[]{@{}
  >{\raggedright\arraybackslash}p{(\manuscripttablewidth - 8\tabcolsep) * \real{0.2091}}
  >{\raggedright\arraybackslash}p{(\manuscripttablewidth - 8\tabcolsep) * \real{0.2092}}
  >{\raggedright\arraybackslash}p{(\manuscripttablewidth - 8\tabcolsep) * \real{0.1940}}
  >{\raggedright\arraybackslash}p{(\manuscripttablewidth - 8\tabcolsep) * \real{0.1940}}
  >{\raggedright\arraybackslash}p{(\manuscripttablewidth - 8\tabcolsep) * \real{0.1936}}@{}}
\caption{Per-epoch training time of the individual loss components. Statistics were calculated from 10,000 repeated measurements; each measurement used 1,000 samples and a batch size of 16.}\label{tab:10}\\
\toprule\noalign{}
\begin{minipage}[b]{\linewidth}\raggedright
Loss function
\end{minipage} & \multicolumn{4}{c}{Computational time (ms)} \\
& Mean & Std & Min & Max \\
\midrule\noalign{}
\endfirsthead

\toprule\noalign{}
\begin{minipage}[b]{\linewidth}\raggedright
Loss function
\end{minipage} & \multicolumn{4}{c}{Computational time (ms)} \\
& Mean & Std & Min & Max \\
\midrule\noalign{}
\endhead
\bottomrule\noalign{}
\endlastfoot

\(\mathcal{L}_{OS}\) & 11.5 & 1.1 & 10.0 & 21.4 \\
\(\mathcal{L}_{RO}\) & 11.3 & 0.7 & 10.2 & 19.1 \\
\(\mathcal{L}_{TG}\) & 58.9 & 5.5 & 50.8 & 203.4 \\
\end{longtable}

Table \ref{tab:10} reports the per-epoch training time measured when each component of the proposed objective was evaluated separately. The one-shot MSE term \(\mathcal{L}_{OS}\) and recursive-rollout MSE term \(\mathcal{L}_{RO}\) had similar mean computation times, differing by only 0.2 ms under the specified experimental setting. In contrast, the soft-DTW-based trend-guided term \(\mathcal{L}_{TG}\) required 58.9 ms on average, approximately 47.4 ms more than \(\mathcal{L}_{OS}\). Summing the independently measured means gives 81.7 ms for the three components, which is 70.2 ms greater than the 11.5 ms required by \(\mathcal{L}_{OS}\) alone. This sum should be interpreted only as an approximate componentwise estimate. The actual per-epoch cost of the complete TG-RC objective must be measured directly because its components may share forward computations and intermediate representations.

The additional loss terms do not affect the recursive forward extrapolation once training has been completed. However, they increase the total latency of the proposed online RUL estimation procedure because the LGFM is trained using the target-bearing prefix at each IT. The sampling intervals of the FB and XB datasets were 10 s and 1 min, respectively; therefore, the measured training cost may remain small relative to the interval between successive measurements. Nevertheless, the complete online training time, rather than the loss-evaluation time alone, should be compared with these sampling intervals before real-time applicability is claimed.

\section{Ablation study}
\label{sec:ablation}

The proposed RUL prediction framework differs from existing methods in terms of its model architecture and loss formulation. This section presents ablation experiments designed to quantify the contribution of each component. Section \ref{sec:ablation1} compares RUL prediction performance across different combinations of the local, global, and current-state branches. Section \ref{sec:ablation2} examines the contribution of each DF to the LGFM. Section \ref{sec:ablation3} evaluates how the TG-RC loss affects the existing DP models. All ablation experiments were conducted on FB1-1 as a representative test bearing at an earlier IT of 50 and a later IT of 200. Each experiment was repeated ten times, and the tables report the mean performance.

\subsection{Contributions of the local, global, and current-state branches}
\label{sec:ablation1}

Seven architectural configurations were evaluated: three single-branch variants, three two-branch combinations, and the complete LGFM comprising all three branches. The same TG-RC loss was used to train each configuration.

\begin{longtable}[]{@{}
  >{\raggedright\arraybackslash}p{(\manuscripttablewidth - 12\tabcolsep) * \real{0.0928}}
  >{\raggedright\arraybackslash}p{(\manuscripttablewidth - 12\tabcolsep) * \real{0.1672}}
  >{\raggedright\arraybackslash}p{(\manuscripttablewidth - 12\tabcolsep) * \real{0.2296}}
  >{\raggedright\arraybackslash}p{(\manuscripttablewidth - 12\tabcolsep) * \real{0.1860}}
  >{\raggedright\arraybackslash}p{(\manuscripttablewidth - 12\tabcolsep) * \real{0.1350}}
  >{\raggedright\arraybackslash}p{(\manuscripttablewidth - 12\tabcolsep) * \real{0.0990}}
  >{\raggedright\arraybackslash}p{(\manuscripttablewidth - 12\tabcolsep) * \real{0.0904}}@{}}
\caption{RUL prediction performance of the seven LGFM configurations.}\label{tab:11}\\
\toprule\noalign{}
\begin{minipage}[b]{\linewidth}\raggedright
ITs
\end{minipage} & \begin{minipage}[b]{\linewidth}\raggedright
Model type
\end{minipage} & \begin{minipage}[b]{\linewidth}\raggedright
Architecture
\end{minipage} & \begin{minipage}[b]{\linewidth}\raggedright
Performance
\end{minipage} & \begin{minipage}[b]{\linewidth}\raggedright
\end{minipage} & \begin{minipage}[b]{\linewidth}\raggedright
\end{minipage} & \begin{minipage}[b]{\linewidth}\raggedright
\end{minipage} \\
& & & MAE & NRMSE & Score & NCR \\
\midrule\noalign{}
\endfirsthead

\toprule\noalign{}
\begin{minipage}[b]{\linewidth}\raggedright
ITs
\end{minipage} & \begin{minipage}[b]{\linewidth}\raggedright
Model type
\end{minipage} & \begin{minipage}[b]{\linewidth}\raggedright
Architecture
\end{minipage} & \begin{minipage}[b]{\linewidth}\raggedright
Performance
\end{minipage} & \begin{minipage}[b]{\linewidth}\raggedright
\end{minipage} & \begin{minipage}[b]{\linewidth}\raggedright
\end{minipage} & \begin{minipage}[b]{\linewidth}\raggedright
\end{minipage} \\
& & & MAE & NRMSE & Score & NCR \\
\midrule\noalign{}
\endhead
\bottomrule\noalign{}
\endlastfoot

\multirow{7}{*}{50} & BC1 & Local & N/A & N/A & N/A & 1.0 \\
& BC2 & Global & 62.3 & 1.329 & 0.026 & 0.0 \\
& BC3 & Current & 22.9 & 0.498 & 0.247 & 0.0 \\
& BC4 & Local + Global & 57.3 & 1.258 & 0.036 & 0.0 \\
& BC5 & Local + Current & 17.9 & \secondbest{0.426} & 0.361 & 0.0 \\
& BC6 & Global + Current & \secondbest{17.7} & 0.473 & \secondbest{0.441} & 0.0 \\
& \textbf{BC7} & \textbf{LGFM (proposed)} & \textbf{10.7} & \textbf{0.258} & \textbf{0.527} & 0.0 \\
\midrule\noalign{}
\multirow{7}{*}{200} & BC1 & Local & N/A & N/A & N/A & 1.0 \\
& BC2 & Global & 149.1 & 0.940 & 0.207 & 0.3 \\
& BC3 & Current & 44.5 & 0.225 & 0.466 & 0.0 \\
& BC4 & Local + Global & 254.0 & 1.343 & 0.010 & 0.7 \\
& BC5 & Local + Current & 43.2 & 0.220 & 0.477 & 0.0 \\
& BC6 & Global + Current & \secondbest{42.9} & \secondbest{0.219} & \secondbest{0.482} & 0.0 \\
& \textbf{BC7} & \textbf{LGFM (proposed)} & \textbf{40.4} & \textbf{0.205} & \textbf{0.500} & 0.0 \\
\end{longtable}

Table \ref{tab:11} summarizes the RUL prediction performance of the seven architectural configurations. Among the single-branch variants (BC1--BC3), BC1, which used only the local branch, failed to produce an extrapolated HI trajectory that reached the FT at either IT. Consequently, an RUL estimate could not be obtained. BC2, which used only the global branch, yielded MAEs of 62.3 and 149.1 at IT=50 and IT=200, respectively. BC3, which directly retained the current HI level, achieved substantially lower MAEs of 22.9 and 44.5 and was the best-performing single-branch configuration. These results indicate that the current absolute HI level provides important information for RUL estimation. The local and global branches operate on the differences between windows and therefore do not directly preserve the absolute level of the current HI sequence. Although the global difference represents accumulated departure from the healthy reference, its compressed DF alone appears to be insufficient to characterize the current degradation state accurately in these experiments.

Among the two-branch variants (BC4--BC6), BC4, which combined the local and global branches but lacked the current-state branch, produced MAEs of 57.3 and 254.0 at IT=50 and IT=200, respectively. It also failed to reach the FT in seven of the ten runs at IT=200, as indicated by the NCR of 0.7. Thus, combining the two difference-based branches alone did not provide stable RUL estimates. Adding either difference-based branch to the current-state branch improved the mean performance over BC3. Relative to BC3, BC5 reduced the MAE by 5.0 at IT=50 and 1.3 at IT=200, whereas BC6 reduced it by 5.2 and 1.6, respectively. These reductions suggest that local and global changes provide useful information beyond the absolute HI level, although their individual effects were modest at the later IT.

BC7, which is the complete LGFM, achieved the best performance at both ITs. Compared with BC6, the strongest two-branch configuration, BC7 further reduced MAE by 7.0 at IT=50 and 2.5 at IT=200. Additionally, it maintained an NCR of 0 under both conditions. Therefore, the results support the complementary roles of the three branches: the current-state branch retains the absolute HI level, whereas the local and global branches describe recent changes and accumulated deviation from the healthy state, respectively.

\subsection{Comparison of RUL prediction performance across DF components}
\label{sec:ablation2}

This section evaluates the individual contributions of the five descriptors used to construct the DFs: RMS, STD, P2P, SLOPE, and ACC. An exhaustive evaluation of all multivariate combinations would require numerous experiments. Therefore, for clarity, the analysis was restricted to single-descriptor variants. The complete LGFM using all five DFs was also included as the reference configuration. For each single DF descriptor variant, the selected descriptor replaced the five-dimensional DF vector in both the local and global branches, while the remaining architecture and training procedure remained unchanged.

\begin{longtable}[]{@{}
  >{\raggedright\arraybackslash}p{(\manuscripttablewidth - 10\tabcolsep) * \real{0.1223}}
  >{\raggedright\arraybackslash}p{(\manuscripttablewidth - 10\tabcolsep) * \real{0.2124}}
  >{\raggedright\arraybackslash}p{(\manuscripttablewidth - 10\tabcolsep) * \real{0.2392}}
  >{\raggedright\arraybackslash}p{(\manuscripttablewidth - 10\tabcolsep) * \real{0.1750}}
  >{\raggedright\arraybackslash}p{(\manuscripttablewidth - 10\tabcolsep) * \real{0.1298}}
  >{\raggedright\arraybackslash}p{(\manuscripttablewidth - 10\tabcolsep) * \real{0.1214}}@{}}
\caption{RUL prediction performance of the five single-descriptor variants.}\label{tab:12}\\
\toprule\noalign{}
\begin{minipage}[b]{\linewidth}\raggedright
ITs
\end{minipage} & \begin{minipage}[b]{\linewidth}\raggedright
DF
\end{minipage} & \begin{minipage}[b]{\linewidth}\raggedright
Performance
\end{minipage} & \begin{minipage}[b]{\linewidth}\raggedright
\end{minipage} & \begin{minipage}[b]{\linewidth}\raggedright
\end{minipage} & \begin{minipage}[b]{\linewidth}\raggedright
\end{minipage} \\
& & MAE & NRMSE & Score & NCR \\
\midrule\noalign{}
\endfirsthead

\toprule\noalign{}
\begin{minipage}[b]{\linewidth}\raggedright
ITs
\end{minipage} & \begin{minipage}[b]{\linewidth}\raggedright
DF
\end{minipage} & \begin{minipage}[b]{\linewidth}\raggedright
Performance
\end{minipage} & \begin{minipage}[b]{\linewidth}\raggedright
\end{minipage} & \begin{minipage}[b]{\linewidth}\raggedright
\end{minipage} & \begin{minipage}[b]{\linewidth}\raggedright
\end{minipage} \\
& & MAE & NRMSE & Score & NCR \\
\midrule\noalign{}
\endhead
\bottomrule\noalign{}
\endlastfoot

\multirow{6}{*}{50} & RMS & 23.2 & 0.517 & 0.085 & 0.0 \\
& STD & 25.4 & 0.560 & 0.129 & 0.0 \\
& P2P & 24.7 & 0.548 & 0.070 & 0.0 \\
& \secondbest{SLOPE} & \secondbest{21.3} & \secondbest{0.471} & \secondbest{0.150} & 0.0 \\
& ACC & 27.9 & 0.609 & 0.101 & 0.0 \\
& \textbf{LGFM (Proposed)} & \textbf{10.7} & \textbf{0.258} & \textbf{0.527} & 0.0 \\
\midrule\noalign{}
\multirow{6}{*}{200} & RMS & 47.5 & 0.240 & 0.443 & 0.0 \\
& STD & 43.7 & 0.233 & 0.487 & 0.0 \\
& P2P & 46.8 & 0.238 & 0.450 & 0.0 \\
& \textbf{SLOPE} & \textbf{38.8} & \textbf{0.195} & \textbf{0.512} & 0.0 \\
& ACC & 40.8 & 0.207 & 0.497 & 0.0 \\
& \secondbest{LGFM (Proposed)} & \secondbest{40.4} & \secondbest{0.205} & \secondbest{0.500} & 0.0 \\
\end{longtable}

Table \ref{tab:12} compares the single DF descriptor variants with the complete LGFM. At IT=50, the complete model achieved the lowest MAE of 10.7. Among the single DF descriptor variants, SLOPE performed the best, with an MAE of 21.3. At IT=200, SLOPE achieved the lowest MAE and NRMSE of 38.8 and 0.195, respectively, and the highest Score of 0.512. The complete LGFM produced similar but slightly weaker results at this IT, with an MAE of 40.4, an NRMSE of 0.205, and a Score of 0.500. The results show that no single DF descriptor was uniformly optimal relative to the complete model across both ITs. Combining all five DFs was clearly beneficial at IT=50, whereas SLOPE alone performed slightly better at IT=200. Therefore, the usefulness of a descriptor may depend on the HI pattern observed at the selected IT.

The differences among descriptor variants can be interpreted from their definitions. STD and P2P quantify the dispersion and amplitude range and may be informative when these quantities increase with degradation. RMS summarizes the overall magnitude of the window. SLOPE, as defined in this study, measures the mean magnitude of consecutive HI changes rather than the slope of a fitted trend. ACC describes the variation in these consecutive change magnitudes. Consequently, the relative usefulness of these descriptor variants may depend on whether degradation is expressed primarily through changes in the level, dispersion, local increment magnitude, or variation in those increments.

\subsection{RUL prediction performance under different TG-RC loss configurations}
\label{sec:ablation3}

This section examines the contribution of the loss components used in the TG-RC loss. Section \ref{sec:ablation3.1} compares seven combinations of the three constituent loss terms. Section \ref{sec:ablation3.2} the TG-RC loss to other DP models and quantifies the resulting changes in performance.

\subsubsection{Ablation of the TG-RC loss in LGFM}
\label{sec:ablation3.1}

Seven loss configurations were evaluated: three variants using a single loss term (LC1--LC3), three variants combining two terms (LC4--LC6), and the complete TG-RC objective containing all three terms (LC7).

\begin{longtable}[]{@{}
  >{\raggedright\arraybackslash}p{(\manuscripttablewidth - 12\tabcolsep) * \real{0.0874}}
  >{\raggedright\arraybackslash}p{(\manuscripttablewidth - 12\tabcolsep) * \real{0.1744}}
  >{\raggedright\arraybackslash}p{(\manuscripttablewidth - 12\tabcolsep) * \real{0.2616}}
  >{\raggedright\arraybackslash}p{(\manuscripttablewidth - 12\tabcolsep) * \real{0.1714}}
  >{\raggedright\arraybackslash}p{(\manuscripttablewidth - 12\tabcolsep) * \real{0.1254}}
  >{\raggedright\arraybackslash}p{(\manuscripttablewidth - 12\tabcolsep) * \real{0.0930}}
  >{\raggedright\arraybackslash}p{(\manuscripttablewidth - 12\tabcolsep) * \real{0.0868}}@{}}
\caption{RUL prediction performance under the seven loss configurations.}\label{tab:13}\\
\toprule\noalign{}
\begin{minipage}[b]{\linewidth}\raggedright
ITs
\end{minipage} & \begin{minipage}[b]{\linewidth}\raggedright
Experiments
\end{minipage} & \begin{minipage}[b]{\linewidth}\raggedright
Loss function
\end{minipage} & \begin{minipage}[b]{\linewidth}\raggedright
Performance
\end{minipage} & \begin{minipage}[b]{\linewidth}\raggedright
\end{minipage} & \begin{minipage}[b]{\linewidth}\raggedright
\end{minipage} & \begin{minipage}[b]{\linewidth}\raggedright
\end{minipage} \\
& & & MAE & NRMSE & Score & NCR \\
\midrule\noalign{}
\endfirsthead

\toprule\noalign{}
\begin{minipage}[b]{\linewidth}\raggedright
ITs
\end{minipage} & \begin{minipage}[b]{\linewidth}\raggedright
Experiments
\end{minipage} & \begin{minipage}[b]{\linewidth}\raggedright
Loss function
\end{minipage} & \begin{minipage}[b]{\linewidth}\raggedright
Performance
\end{minipage} & \begin{minipage}[b]{\linewidth}\raggedright
\end{minipage} & \begin{minipage}[b]{\linewidth}\raggedright
\end{minipage} & \begin{minipage}[b]{\linewidth}\raggedright
\end{minipage} \\
& & & MAE & NRMSE & Score & NCR \\
\midrule\noalign{}
\endhead
\bottomrule\noalign{}
\endlastfoot

50 & LC1 & \(\mathcal{L}_{OS}\) & 43.7 & 0.882 & 0.053 & 0.0 \\
& LC2 & \(\mathcal{L}_{RO}\) & 169.8 & 5.002 & 0.026 & 0.3 \\
& LC3 & \(\mathcal{L}_{TG}\) & 19.9 & 0.458 & 0.346 & 0.0 \\
& LC4 & \(\mathcal{L}_{OS} + \mathcal{L}_{RO}\) & 90.5 & 3.109 & 0.169 & 0.1 \\
& LC5 & \(\mathcal{L}_{OS} + \mathcal{L}_{TG}\) & \secondbest{17.9} & \secondbest{0.426} & \secondbest{0.361} & 0.0 \\
& LC6 & \(\mathcal{L}_{RO} + \mathcal{L}_{TG}\) & 20.1 & 0.441 & 0.304 & 0.0 \\
& \textbf{LC7} & \(\mathcal{L}_{TG - RC}\) \textbf{(Proposed)} & \textbf{10.7} & \textbf{0.258} & \textbf{0.527} & 0.0 \\
\midrule\noalign{}
200 & LC1 & \(\mathcal{L}_{OS}\) & N/A & N/A & N/A & 1.0 \\
& LC2 & \(\mathcal{L}_{RO}\) & N/A & N/A & N/A & 1.0 \\
& LC3 & \(\mathcal{L}_{TG}\) & 84.9 & 0.433 & 0.241 & 0.0 \\
& LC4 & \(\mathcal{L}_{OS} + \mathcal{L}_{RO}\) & 291.6 & 1.461 & 0.001 & 0.2 \\
& LC5 & \(\mathcal{L}_{OS} + \mathcal{L}_{TG}\) & 50.8 & 0.259 & 0.420 & 0.0 \\
& LC6 & \(\mathcal{L}_{RO} + \mathcal{L}_{TG}\) & \secondbest{49.7} & \secondbest{0.251} & \secondbest{0.425} & 0.0 \\
& \textbf{LC7} & \(\mathcal{L}_{TG - RC}\) \textbf{(Proposed)} & \textbf{40.4} & \textbf{0.205} & \textbf{0.500} & 0.0 \\
\end{longtable}

Table \ref{tab:13} shows that the three single-term objectives produced markedly different results. LC1, which used only the one-shot MSE term, yielded an MAE of 43.7 at IT=50. At IT=200, none of its extrapolated trajectories reached the FT within the maximum forecasting horizon, and RUL could not be estimated. LC2, which used only the recursive-rollout MSE, performed poorly at IT=50 and also failed to produce an RUL estimate at IT=200. In contrast, LC3, which used only the trend-guided soft-DTW term, achieved MAEs of 19.9 and 84.9 at IT=50 and IT=200, respectively, and maintained an NCR of 0 at both ITs. Therefore, among the single-term configurations, the trend-guided term provided the most reliable directional guidance toward the FT.

Among the two-term configurations, LC5 and LC6, both of which included \(\mathcal{L}_{TG}\), outperformed LC4 at both ITs. At IT=50, LC5 achieved the lowest MAE among the two-term variants, at 17.9. At IT=200, LC6 achieved the lowest MAE of 49.7, compared with 50.8 for LC5. This small difference is consistent with the possible benefit of training directly on recursively accumulated errors. However, it is insufficient to establish a statistically meaningful advantage of LC6 over LC5. LC4, which combined the one-shot and rollout MSE terms without trend guidance, produced MAEs of 90.5 and 291.6 at IT=50 and IT=200, respectively. Its NCR values of 0.1 and 0.2 indicate that the predicted HI failed to reach the FT in one and two of the ten runs. Therefore, under the examined conditions, the two MSE-based terms alone were insufficient to maintain a reliable degradation direction during recursive extrapolation.

LC7, which combined all three terms, achieved the best overall performance, with MAEs of 10.7 at IT=50 and 40.4 at IT=200. It also attained the lowest NRMSE and highest Score at both ITs while maintaining an NCR of 0. These results support the complementary roles of the three terms: \(\mathcal{L}_{OS}\) promotes local prediction accuracy, \(\mathcal{L}_{RO}\) exposes the model to recursively generated inputs during training, and \(\mathcal{L}_{TG}\) provides directional guidance through alignment with the trend prior. All configurations containing \(\mathcal{L}_{TG}\)---LC3, LC5, LC6, and LC7---achieved an NCR of 0 at both ITs. This pattern indicates that the trend-guided soft-DTW term contributed substantially to reliable FT-reaching behavior in these experiments.

\subsubsection{Effect of TG-RC loss on RNN- and Transformer-based DP models}
\label{sec:ablation3.2}

This section examines the effect of the TG-RC loss on the RUL prediction performance of existing DP models. Training with the conventional MSE-based objective is denoted by ``w/o TG-RC,'' whereas training with the proposed objective is denoted by ``w/ TG-RC.'' The GDAU, DTGRU, CLSTM, and GTFDAU were originally developed as many-to-one models. Because the TG-RC loss contains sequence-level loss terms, these models were adapted to use a many-to-many output structure. To isolate the effect of the loss function, the same many-to-many architecture and experimental configuration were used for both training conditions.

\begingroup
\renewcommand{\manuscripttablefont}{\scriptsize}
\begin{longtable}[]{@{}
  >{\raggedright\arraybackslash}p{(\manuscripttablewidth - 18\tabcolsep) * \real{0.0852}}
  >{\raggedright\arraybackslash}p{(\manuscripttablewidth - 18\tabcolsep) * \real{0.1416}}
  >{\raggedright\arraybackslash}p{(\manuscripttablewidth - 18\tabcolsep) * \real{0.0892}}
  >{\raggedright\arraybackslash}p{(\manuscripttablewidth - 18\tabcolsep) * \real{0.1222}}
  >{\raggedright\arraybackslash}p{(\manuscripttablewidth - 18\tabcolsep) * \real{0.0906}}
  >{\raggedright\arraybackslash}p{(\manuscripttablewidth - 18\tabcolsep) * \real{0.0846}}
  >{\raggedright\arraybackslash}p{(\manuscripttablewidth - 18\tabcolsep) * \real{0.0892}}
  >{\raggedright\arraybackslash}p{(\manuscripttablewidth - 18\tabcolsep) * \real{0.1222}}
  >{\raggedright\arraybackslash}p{(\manuscripttablewidth - 18\tabcolsep) * \real{0.0906}}
  >{\raggedright\arraybackslash}p{(\manuscripttablewidth - 18\tabcolsep) * \real{0.0846}}@{}}
\caption{RUL prediction performance of the DP models with and without the TG-RC loss.}\label{tab:14}\\
\toprule\noalign{}
\begin{minipage}[b]{\linewidth}\raggedright
IT
\end{minipage} & \begin{minipage}[b]{\linewidth}\raggedright
Model
\end{minipage} & \multicolumn{4}{c}{w/o TG-RC} & \multicolumn{4}{c}{w/ TG-RC} \\
& & MAE & NRMSE & Score & NCR & MAE & NRMSE & Score & NCR \\
\midrule\noalign{}
\endfirsthead

\toprule\noalign{}
\begin{minipage}[b]{\linewidth}\raggedright
IT
\end{minipage} & \begin{minipage}[b]{\linewidth}\raggedright
Model
\end{minipage} & \multicolumn{4}{c}{w/o TG-RC} & \multicolumn{4}{c}{w/ TG-RC} \\
& & MAE & NRMSE & Score & NCR & MAE & NRMSE & Score & NCR \\
\midrule\noalign{}
\endhead
\bottomrule\noalign{}
\endlastfoot

\multirow{5}{*}{50} & GDAU & 41.5 & 0.832 & 0.058 & 0.0 & \textbf{7.0} & \textbf{0.158} & \textbf{0.286} & 0.0 \\
& DTGRU & \textbf{26.3} & \textbf{0.702} & 0.140 & 0.0 & 37.9 & 0.943 & \textbf{0.167} & 0.0 \\
& CLSTM & 33.6 & 0.701 & 0.133 & 0.0 & \textbf{9.0} & \textbf{0.245} & \textbf{0.376} & 0.0 \\
& GTFDAU & 40.4 & 0.811 & 0.062 & 0.0 & \textbf{36.6} & \textbf{0.749} & \textbf{0.078} & 0.0 \\
& LSTM-TF & 39.2 & \textbf{0.807} & 0.070 & 0.0 & \textbf{38.0} & 0.808 & \textbf{0.086} & 0.0 \\
\midrule\noalign{}
\multirow{5}{*}{200} & GDAU & N/A & N/A & N/A & 1.0 & \textbf{113.1} & \textbf{0.569} & \textbf{0.145} & \textbf{0.0} \\
& DTGRU & N/A & N/A & N/A & 1.0 & \textbf{74.7} & \textbf{0.522} & \textbf{0.113} & \textbf{0.0} \\
& CLSTM & N/A & N/A & N/A & 1.0 & \textbf{63.7} & \textbf{0.351} & \textbf{0.300} & \textbf{0.0} \\
& GTFDAU & N/A & N/A & N/A & 1.0 & \textbf{64.5} & \textbf{0.352} & \textbf{0.043} & \textbf{0.0} \\
& LSTM-TF & N/A & N/A & N/A & 1.0 & \textbf{211.5} & \textbf{1.153} & \textbf{0.116} & \textbf{0.3} \\
\end{longtable}
\endgroup

Table \ref{tab:14} summarizes ten comparisons involving the five models and two ITs. At IT=50, the TG-RC loss reduced the MAE of the GDAU, CLSTM, GTFDAU, and LSTM-TF by 34.5, 24.6, 3.8, and 1.2, respectively. The largest improvements were obtained for the GDAU and CLSTM. The DTGRU was the only model for which MAE increased, from 26.3 to 37.9, although its Score increased from 0.140 to 0.167. Because Score penalizes RUL overestimation more heavily than underestimation, this discrepancy indicates that the TG-RC loss changed the distribution or direction of the prediction errors. The effect of the TG-RC loss on FT-reaching behavior was more pronounced at IT=200. Without the TG-RC loss, none of the five models reached the FT within the maximum extrapolation horizon in any of the ten runs, resulting in an NCR of 1.0, thereby preventing RUL estimation. With the TG-RC loss, the GDAU, DTGRU, CLSTM, and GTFDAU achieved an NCR of 0, allowing an RUL estimate to be obtained in every run. The LSTM-TF also improved, although its NCR of 0.3 indicates that three of the ten runs still failed to reach the FT.

Overall, the TG-RC loss improved the MAE at IT=50 for four of the five models and substantially improved FT-reaching reliability at IT=200. These findings indicate that the loss can be incorporated into different DP architectures and improve recursive extrapolation stability under the studied conditions. Nevertheless, its effect on pointwise RUL accuracy was model-dependent, as shown by the DTGRU at IT=50 and the relatively large MAE of LSTM-TF at IT=200.

\section{Conclusion}
\label{sec:conclusion}

DP-based RUL estimation is a widely used approach to support predictive maintenance for rotating machinery. Because the estimated RUL is determined by the extrapolated HI trajectory and its first crossing of the FT, reliable DP modeling is central to prediction performance. Neural degradation models can learn complex trajectories without requiring a predefined degradation function. However, this flexibility permits recursively predicted trajectories that deviate from the irreversible direction expected by a DP or become unstable over multiple forecasting steps. Such behavior can prevent the predicted HI from reaching the FT and thereby undermine RUL estimation.

This study addressed these limitations by using two components. First, the LGFM models the HI sequence by combining its current level with local and global difference-based descriptors. Although its mappings are primarily linear, recursive application of the model and the use of nonlinear statistical DF descriptors allow it to represent nontrivial extrapolation trajectories at a low computational cost. Second, the TG-RC loss combines a one-shot MSE term, recursive-rollout MSE term, and trend-guided soft-DTW term. The rollout term reduces the mismatch between teacher-forced training and recursive inference, while the trend-guided term encourages alignment with the dominant degradation direction estimated from the observed HI history. The TG-RC loss does not mathematically enforce pointwise monotonicity; rather, it guides the extrapolated trajectory toward a more consistent global direction.

Experiments on two publicly available bearing datasets showed that several existing DP models became unstable under recursive extrapolation and, under some IT conditions, failed to produce trajectories that reached the FT. The proposed LGFM--TG-RC framework maintained more consistent FT-reaching behavior across the evaluated ITs and achieved competitive or superior RUL prediction performance in most settings. Applying the TG-RC loss to the existing RNN- and Transformer-based DP models also improved the extrapolation reliability, particularly at IT=200 for FB1-1. These results suggest that the TG-RC loss can serve as a model-independent training objective for a range of sequence-based DP models, although the magnitude of its benefit depends on the underlying architecture and inspection conditions.

Several directions remain for future research. First, the LGFM currently uses five statistical and physical DFs. Other degradation-sensitive features should be examined across a broader range of equipment and operating conditions. Second, the descriptor ablation showed that the most effective feature may vary with the observed degradation pattern. Therefore, a gating mechanism that adaptively weights the descriptors could improve the use of condition-dependent information. Third, the current trend prior assumes a manually selected linear functional form. An adaptive procedure that selects or estimates the trend family using only the observed HI prefix or validation data could reduce reliance on analyst-defined assumptions. Polynomial and exponential trend priors also produced viable trajectories in the present analysis; however, their applicability should be systematically evaluated. Mixtures of trend families or a learnable trend-prior module may provide greater flexibility, provided that the model selection does not use unobserved future observations. Finally, the soft-DTW term increases training time. Approximate or constrained soft-DTW formulations could reduce this cost while retaining the directional guidance of the trend prior. Such methods would be particularly relevant to the online setting considered in this study, where model training is repeated as additional target-bearing observations become available.

\clearpage
\bibliographystyle{unsrtnat}
\bibliography{references}

\clearpage
\appendix
\setcounter{table}{0}
\renewcommand{\thetable}{A\arabic{table}}
\renewcommand{\theHtable}{A\arabic{table}}
\section*{Appendix A. RUL prediction results for the remaining test sets}
\phantomsection
\label{sec:appendixA}

\begin{longtable}[]{@{}
  >{\raggedright\arraybackslash}p{(\manuscripttablewidth - 10\tabcolsep) * \real{0.1516}}
  >{\raggedright\arraybackslash}p{(\manuscripttablewidth - 10\tabcolsep) * \real{0.1748}}
  >{\raggedright\arraybackslash}p{(\manuscripttablewidth - 10\tabcolsep) * \real{0.1954}}
  >{\raggedright\arraybackslash}p{(\manuscripttablewidth - 10\tabcolsep) * \real{0.1704}}
  >{\raggedright\arraybackslash}p{(\manuscripttablewidth - 10\tabcolsep) * \real{0.1610}}
  >{\raggedright\arraybackslash}p{(\manuscripttablewidth - 10\tabcolsep) * \real{0.1468}}@{}}
\caption{RUL prediction performance on FB1-3 at different ITs.}\label{tab:A1}\\
\toprule\noalign{}
\begin{minipage}[b]{\linewidth}\raggedright
IT
\end{minipage} & \begin{minipage}[b]{\linewidth}\raggedright
Model
\end{minipage} & \begin{minipage}[b]{\linewidth}\raggedright
Performance
\end{minipage} & \begin{minipage}[b]{\linewidth}\raggedright
\end{minipage} & \begin{minipage}[b]{\linewidth}\raggedright
\end{minipage} & \begin{minipage}[b]{\linewidth}\raggedright
\end{minipage} \\
& & MAE & NRMSE & Score & NCR \\
\midrule\noalign{}
\endfirsthead

\toprule\noalign{}
\begin{minipage}[b]{\linewidth}\raggedright
IT
\end{minipage} & \begin{minipage}[b]{\linewidth}\raggedright
Model
\end{minipage} & \begin{minipage}[b]{\linewidth}\raggedright
Performance
\end{minipage} & \begin{minipage}[b]{\linewidth}\raggedright
\end{minipage} & \begin{minipage}[b]{\linewidth}\raggedright
\end{minipage} & \begin{minipage}[b]{\linewidth}\raggedright
\end{minipage} \\
& & MAE & NRMSE & Score & NCR \\
\midrule\noalign{}
\endhead
\bottomrule\noalign{}
\endlastfoot

50 & GDAU & 253.6 & 6.186 & 0.017 & 0.2 \\
& DTGRU & 50.7 & 1.311 & \secondbest{0.184} & 0.0 \\
& CLSTM & 206.2 & 5.519 & 0.067 & 0.2 \\
& \textbf{GTFDAU} & \textbf{22.8} & \textbf{0.568} & \textbf{0.337} & 0.0 \\
& \secondbest{LSTM-TF} & \secondbest{29.2} & 0.718 & 0.126 & 0.0 \\
& LGFM & 43.8 & \secondbest{0.569} & 0.161 & 0.0 \\
\midrule\noalign{}
100 & GDAU & 162.0 & 1.783 & 0.002 & 0.0 \\
& \textbf{DTGRU} & \textbf{3.5} & \textbf{0.040} & \secondbest{0.649} & 0.0 \\
& CLSTM & 261.2 & 2.884 & 0.001 & 0.4 \\
& \secondbest{GTFDAU} & \secondbest{4.3} & \secondbest{0.059} & \textbf{0.691} & 0.0 \\
& LSTM-TF & 57.6 & 0.588 & 0.267 & 0.0 \\
& LGFM & 49.7 & 0.517 & 0.003 & 0.0 \\
\midrule\noalign{}
150 & GDAU & 78.4 & 0.667 & 0.076 & 0.0 \\
& \secondbest{DTGRU} & \secondbest{60.2} & \secondbest{0.460} & 0.049 & 0.0 \\
& CLSTM & 246.6 & 1.801 & 0.001 & 0.5 \\
& GTFDAU & 87.0 & 0.639 & 0.007 & 0.0 \\
& LSTM-TF & 114.0 & 0.507 & \secondbest{0.236} & 0.0 \\
& \textbf{LGFM} & \textbf{14.8} & \textbf{0.131} & \textbf{0.522} & 0.0 \\
\midrule\noalign{}
200 & GDAU & 265.5 & 1.371 & 0.011 & 0.7 \\
& DTGRU & 239.9 & 1.259 & 0.001 & 0.7 \\
& \secondbest{CLSTM} & \secondbest{189.2} & 1.042 & \secondbest{0.083} & 0.7 \\
& GTFDAU & 257.4 & 1.330 & 0.001 & 0.7 \\
& LSTM-TF & 200.3 & \secondbest{0.527} & \textbf{0.194} & 0.2 \\
& \textbf{LGFM} & \textbf{74.1} & \textbf{0.649} & 0.033 & 0.0 \\
\end{longtable}

\begin{longtable}[]{@{}
  >{\raggedright\arraybackslash}p{(\manuscripttablewidth - 10\tabcolsep) * \real{0.1516}}
  >{\raggedright\arraybackslash}p{(\manuscripttablewidth - 10\tabcolsep) * \real{0.1748}}
  >{\raggedright\arraybackslash}p{(\manuscripttablewidth - 10\tabcolsep) * \real{0.1954}}
  >{\raggedright\arraybackslash}p{(\manuscripttablewidth - 10\tabcolsep) * \real{0.1704}}
  >{\raggedright\arraybackslash}p{(\manuscripttablewidth - 10\tabcolsep) * \real{0.1610}}
  >{\raggedright\arraybackslash}p{(\manuscripttablewidth - 10\tabcolsep) * \real{0.1468}}@{}}
\caption{RUL prediction performance on FB2-4 at different ITs.}\label{tab:A2}\\
\toprule\noalign{}
\begin{minipage}[b]{\linewidth}\raggedright
IT
\end{minipage} & \begin{minipage}[b]{\linewidth}\raggedright
Model
\end{minipage} & \begin{minipage}[b]{\linewidth}\raggedright
Performance
\end{minipage} & \begin{minipage}[b]{\linewidth}\raggedright
\end{minipage} & \begin{minipage}[b]{\linewidth}\raggedright
\end{minipage} & \begin{minipage}[b]{\linewidth}\raggedright
\end{minipage} \\
& & MAE & NRMSE & Score & NCR \\
\midrule\noalign{}
\endfirsthead

\toprule\noalign{}
\begin{minipage}[b]{\linewidth}\raggedright
IT
\end{minipage} & \begin{minipage}[b]{\linewidth}\raggedright
Model
\end{minipage} & \begin{minipage}[b]{\linewidth}\raggedright
Performance
\end{minipage} & \begin{minipage}[b]{\linewidth}\raggedright
\end{minipage} & \begin{minipage}[b]{\linewidth}\raggedright
\end{minipage} & \begin{minipage}[b]{\linewidth}\raggedright
\end{minipage} \\
& & MAE & NRMSE & Score & NCR \\
\midrule\noalign{}
\endhead
\bottomrule\noalign{}
\endlastfoot

50 & GDAU & 46.1 & 0.922 & 0.041 & 0.0 \\
& DTGRU & 20.4 & 0.416 & 0.253 & 0.0 \\
& CLSTM & 25.9 & 0.587 & 0.196 & 0.0 \\
& \secondbest{GTFDAU} & \secondbest{19.4} & \secondbest{0.398} & \secondbest{0.273} & 0.0 \\
& LSTM-TF & 40.0 & 0.840 & 0.125 & 0.0 \\
& \textbf{LGFM} & \textbf{7.5} & \textbf{0.201} & \textbf{0.651} & 0.0 \\
\midrule\noalign{}
100 & GDAU & 61.1 & 0.869 & \secondbest{0.228} & 0.0 \\
& DTGRU & 130.2 & 1.326 & 0.001 & 0.0 \\
& CLSTM & 76.9 & 0.902 & 0.008 & 0.0 \\
& GTFDAU & 119.2 & 1.207 & 0.001 & 0.0 \\
& LSTM-TF & \secondbest{59.2} & \secondbest{0.631} & 0.169 & 0.0 \\
& \textbf{LGFM} & \textbf{17.3} & \textbf{0.196} & \textbf{0.577} & 0.0 \\
\midrule\noalign{}
150 & GDAU & 329.4 & 2.219 & 0.001 & 0.8 \\
& DTGRU & \secondbest{106.7} & \secondbest{0.742} & 0.007 & 0.0 \\
& CLSTM & 163.1 & 1.445 & \secondbest{0.156} & 0.4 \\
& GTFDAU & 147.6 & 1.027 & 0.001 & 0.4 \\
& LSTM-TF & 209.5 & 1.667 & 0.064 & 0.5 \\
& \textbf{LGFM} & \textbf{46.1} & \textbf{0.308} & \textbf{0.346} & 0.0 \\
\midrule\noalign{}
200 & GDAU & N/A & N/A & N/A & 1.0 \\
& DTGRU & N/A & N/A & N/A & 1.0 \\
& CLSTM & N/A & N/A & N/A & 1.0 \\
& GTFDAU & N/A & N/A & N/A & 1.0 \\
& LSTM-TF & \secondbest{102.3} & \secondbest{0.593} & \secondbest{0.219} & 0.1 \\
& \textbf{LGFM} & \textbf{57.5} & \textbf{0.292} & \textbf{0.376} & 0.0 \\
\end{longtable}

\begin{longtable}[]{@{}
  >{\raggedright\arraybackslash}p{(\manuscripttablewidth - 10\tabcolsep) * \real{0.1516}}
  >{\raggedright\arraybackslash}p{(\manuscripttablewidth - 10\tabcolsep) * \real{0.1748}}
  >{\raggedright\arraybackslash}p{(\manuscripttablewidth - 10\tabcolsep) * \real{0.1954}}
  >{\raggedright\arraybackslash}p{(\manuscripttablewidth - 10\tabcolsep) * \real{0.1704}}
  >{\raggedright\arraybackslash}p{(\manuscripttablewidth - 10\tabcolsep) * \real{0.1610}}
  >{\raggedright\arraybackslash}p{(\manuscripttablewidth - 10\tabcolsep) * \real{0.1468}}@{}}
\caption{RUL prediction performance on FB2-7 at different ITs.}\label{tab:A3}\\
\toprule\noalign{}
\begin{minipage}[b]{\linewidth}\raggedright
IT
\end{minipage} & \begin{minipage}[b]{\linewidth}\raggedright
Model
\end{minipage} & \begin{minipage}[b]{\linewidth}\raggedright
Performance
\end{minipage} & \begin{minipage}[b]{\linewidth}\raggedright
\end{minipage} & \begin{minipage}[b]{\linewidth}\raggedright
\end{minipage} & \begin{minipage}[b]{\linewidth}\raggedright
\end{minipage} \\
& & MAE & NRMSE & Score & NCR \\
\midrule\noalign{}
\endfirsthead

\toprule\noalign{}
\begin{minipage}[b]{\linewidth}\raggedright
IT
\end{minipage} & \begin{minipage}[b]{\linewidth}\raggedright
Model
\end{minipage} & \begin{minipage}[b]{\linewidth}\raggedright
Performance
\end{minipage} & \begin{minipage}[b]{\linewidth}\raggedright
\end{minipage} & \begin{minipage}[b]{\linewidth}\raggedright
\end{minipage} & \begin{minipage}[b]{\linewidth}\raggedright
\end{minipage} \\
& & MAE & NRMSE & Score & NCR \\
\midrule\noalign{}
\endhead
\bottomrule\noalign{}
\endlastfoot

15 & \textbf{GDAU} & \textbf{3.7} & \textbf{0.282} & \textbf{0.434} & 0.0 \\
& DTGRU & 13.1 & \secondbest{0.887} & 0.001 & 0.0 \\
& CLSTM & \secondbest{13.0} & 0.967 & 0.003 & 0.0 \\
& GTFDAU & 13.4 & 0.906 & 0.001 & 0.0 \\
& LSTM-TF & 23.3 & 2.113 & \secondbest{0.262} & 0.0 \\
& LGFM & 17.8 & 1.190 & 0.001 & 0.0 \\
\midrule\noalign{}
30 & \textbf{GDAU} & \textbf{8.0} & \secondbest{0.392} & \textbf{0.324} & \textbf{0.0} \\
& \secondbest{DTGRU} & \secondbest{8.1} & \textbf{0.289} & 0.052 & 0.0 \\
& CLSTM & 184.1 & 8.771 & 0.001 & 0.7 \\
& GTFDAU & 15.9 & 0.563 & 0.004 & 0.0 \\
& LSTM-TF & 84.3 & 5.537 & \secondbest{0.185} & 0.1 \\
& LGFM & 43.9 & 1.482 & 0.001 & 0.0 \\
\midrule\noalign{}
45 & GDAU & N/A & N/A & N/A & 1.0 \\
& \secondbest{DTGRU} & \secondbest{302.1} & \secondbest{7.415} & 0.001 & \secondbest{0.7} \\
& CLSTM & N/A & N/A & N/A & 1.0 \\
& GTFDAU & N/A & N/A & N/A & 1.0 \\
& LSTM-TF & N/A & N/A & N/A & 1.0 \\
& \textbf{LGFM} & \textbf{78.9} & \textbf{1.771} & 0.001 & \textbf{0.0} \\
\midrule\noalign{}
60 & GDAU & N/A & N/A & N/A & 1.0 \\
& DTGRU & N/A & N/A & N/A & 1.0 \\
& CLSTM & N/A & N/A & N/A & 1.0 \\
& GTFDAU & N/A & N/A & N/A & 1.0 \\
& LSTM-TF & N/A & N/A & N/A & 1.0 \\
& \textbf{LGFM} & \textbf{93.7} & \textbf{1.675} & \textbf{0.001} & \textbf{0.0} \\
\end{longtable}

\begin{longtable}[]{@{}
  >{\raggedright\arraybackslash}p{(\manuscripttablewidth - 10\tabcolsep) * \real{0.1516}}
  >{\raggedright\arraybackslash}p{(\manuscripttablewidth - 10\tabcolsep) * \real{0.1748}}
  >{\raggedright\arraybackslash}p{(\manuscripttablewidth - 10\tabcolsep) * \real{0.1954}}
  >{\raggedright\arraybackslash}p{(\manuscripttablewidth - 10\tabcolsep) * \real{0.1704}}
  >{\raggedright\arraybackslash}p{(\manuscripttablewidth - 10\tabcolsep) * \real{0.1610}}
  >{\raggedright\arraybackslash}p{(\manuscripttablewidth - 10\tabcolsep) * \real{0.1468}}@{}}
\caption{RUL prediction performance on FB3-3 at different ITs.}\label{tab:A4}\\
\toprule\noalign{}
\begin{minipage}[b]{\linewidth}\raggedright
IT
\end{minipage} & \begin{minipage}[b]{\linewidth}\raggedright
Model
\end{minipage} & \begin{minipage}[b]{\linewidth}\raggedright
Performance
\end{minipage} & \begin{minipage}[b]{\linewidth}\raggedright
\end{minipage} & \begin{minipage}[b]{\linewidth}\raggedright
\end{minipage} & \begin{minipage}[b]{\linewidth}\raggedright
\end{minipage} \\
& & MAE & NRMSE & Score & NCR \\
\midrule\noalign{}
\endfirsthead

\toprule\noalign{}
\begin{minipage}[b]{\linewidth}\raggedright
IT
\end{minipage} & \begin{minipage}[b]{\linewidth}\raggedright
Model
\end{minipage} & \begin{minipage}[b]{\linewidth}\raggedright
Performance
\end{minipage} & \begin{minipage}[b]{\linewidth}\raggedright
\end{minipage} & \begin{minipage}[b]{\linewidth}\raggedright
\end{minipage} & \begin{minipage}[b]{\linewidth}\raggedright
\end{minipage} \\
& & MAE & NRMSE & Score & NCR \\
\midrule\noalign{}
\endhead
\bottomrule\noalign{}
\endlastfoot

15 & GDAU & 128.1 & 14.72 & 0.075 & 0.3 \\
& DTGRU & 35.7 & 1.167 & 0.043 & 0.0 \\
& CLSTM & 22.3 & 2.657 & \textbf{0.167} & 0.0 \\
& GTFDAU & 35.5 & 0.816 & 0.124 & 0.0 \\
& \secondbest{LSTM-TF} & \secondbest{11.5} & \secondbest{0.786} & 0.086 & 0.0 \\
& \textbf{LGFM} & \textbf{7.6} & \textbf{0.552} & \secondbest{0.127} & 0.0 \\
\midrule\noalign{}
30 & GDAU & 19.5 & 0.659 & \secondbest{0.113} & 0.0 \\
& DTGRU & 32.9 & 1.328 & 0.001 & 0.0 \\
& \secondbest{CLSTM} & \secondbest{19.1} & \secondbest{0.645} & 0.102 & 0.0 \\
& GTFDAU & 59.0 & 2.195 & 0.001 & 0.1 \\
& LSTM-TF & 26.4 & 0.886 & 0.051 & 0.0 \\
& \textbf{LGFM} & \textbf{11.5} & \textbf{0.407} & \textbf{0.293} & \textbf{0.0} \\
\midrule\noalign{}
45 & GDAU & 33.2 & 0.741 & 0.079 & 0.0 \\
& DTGRU & 71.8 & 1.821 & 0.001 & 0.0 \\
& \secondbest{CLSTM} & \secondbest{31.7} & \secondbest{0.714} & \secondbest{0.095} & 0.0 \\
& GTFDAU & 103.1 & 3.512 & 0.001 & 0.1 \\
& LSTM-TF & 80.0 & 3.301 & 0.060 & 0.1 \\
& \textbf{LGFM} & \textbf{8.3} & \textbf{0.198} & \textbf{0.544} & \textbf{0.0} \\
\midrule\noalign{}
60 & \secondbest{GDAU} & \secondbest{36.4} & \secondbest{0.639} & 0.156 & \secondbest{0.0} \\
& DTGRU & 96.4 & 1.800 & 0.025 & 0.0 \\
& CLSTM & 78.2 & 2.406 & \secondbest{0.160} & 0.0 \\
& GTFDAU & 118.1 & 3.365 & 0.050 & 0.0 \\
& LSTM-TF & 56.2 & 0.937 & 0.039 & 0.0 \\
& \textbf{LGFM} & \textbf{10.0} & \textbf{0.1904} & \textbf{0.288} & \textbf{0.0} \\
\end{longtable}

\begin{longtable}[]{@{}
  >{\raggedright\arraybackslash}p{(\manuscripttablewidth - 10\tabcolsep) * \real{0.1516}}
  >{\raggedright\arraybackslash}p{(\manuscripttablewidth - 10\tabcolsep) * \real{0.1748}}
  >{\raggedright\arraybackslash}p{(\manuscripttablewidth - 10\tabcolsep) * \real{0.1954}}
  >{\raggedright\arraybackslash}p{(\manuscripttablewidth - 10\tabcolsep) * \real{0.1704}}
  >{\raggedright\arraybackslash}p{(\manuscripttablewidth - 10\tabcolsep) * \real{0.1610}}
  >{\raggedright\arraybackslash}p{(\manuscripttablewidth - 10\tabcolsep) * \real{0.1468}}@{}}
\caption{RUL prediction performance on XB1-5 at different ITs.}\label{tab:A5}\\
\toprule\noalign{}
\begin{minipage}[b]{\linewidth}\raggedright
IT
\end{minipage} & \begin{minipage}[b]{\linewidth}\raggedright
Model
\end{minipage} & \begin{minipage}[b]{\linewidth}\raggedright
Performance
\end{minipage} & \begin{minipage}[b]{\linewidth}\raggedright
\end{minipage} & \begin{minipage}[b]{\linewidth}\raggedright
\end{minipage} & \begin{minipage}[b]{\linewidth}\raggedright
\end{minipage} \\
& & MAE & NRMSE & Score & NCR \\
\midrule\noalign{}
\endfirsthead

\toprule\noalign{}
\begin{minipage}[b]{\linewidth}\raggedright
IT
\end{minipage} & \begin{minipage}[b]{\linewidth}\raggedright
Model
\end{minipage} & \begin{minipage}[b]{\linewidth}\raggedright
Performance
\end{minipage} & \begin{minipage}[b]{\linewidth}\raggedright
\end{minipage} & \begin{minipage}[b]{\linewidth}\raggedright
\end{minipage} & \begin{minipage}[b]{\linewidth}\raggedright
\end{minipage} \\
& & MAE & NRMSE & Score & NCR \\
\midrule\noalign{}
\endhead
\bottomrule\noalign{}
\endlastfoot

3 & \secondbest{GDAU} & \secondbest{1.7} & \secondbest{0.586} & 0.163 & 0.0 \\
& DTGRU & 2.0 & 0.666 & 0.099 & 0.0 \\
& \textbf{CLSTM} & \textbf{1.2} & \textbf{0.471} & \textbf{0.365} & \textbf{0.0} \\
& GTFDAU & 2.0 & 0.666 & 0.099 & 0.0 \\
& LSTM-TF & 2.3 & 0.610 & \secondbest{0.185} & 0.0 \\
& LGFM & 5.5 & 5.254 & 0.031 & 0.0 \\
\midrule\noalign{}
6 & \secondbest{GDAU} & \secondbest{4.0} & \secondbest{0.666} & 0.099 & 0.0 \\
& DTGRU & 5.0 & 0.833 & 0.055 & 0.0 \\
& \textbf{CLSTM} & \textbf{2.6} & \textbf{0.532} & \textbf{0.291} & \textbf{0.0} \\
& GTFDAU & 5.0 & 0.833 & 0.055 & 0.0 \\
& LSTM-TF & 5.0 & 0.833 & 0.055 & 0.0 \\
& LGFM & 15.6 & 3.248 & \secondbest{0.100} & 0.0 \\
\midrule\noalign{}
9 & GDAU & N/A & N/A & N/A & 1.0 \\
& DTGRU & N/A & N/A & N/A & 1.0 \\
& CLSTM & N/A & N/A & N/A & 1.0 \\
& GTFDAU & N/A & N/A & N/A & 1.0 \\
& LSTM-TF & N/A & N/A & N/A & 1.0 \\
& \textbf{LGFM} & \textbf{56.5} & \textbf{10.7} & \textbf{0.001} & \textbf{0.0} \\
\midrule\noalign{}
12 & GDAU & N/A & N/A & N/A & 1.0 \\
& DTGRU & N/A & N/A & N/A & 1.0 \\
& CLSTM & N/A & N/A & N/A & 1.0 \\
& GTFDAU & N/A & N/A & N/A & 1.0 \\
& LSTM-TF & N/A & N/A & N/A & 1.0 \\
& LGFM & N/A & N/A & N/A & 1.0 \\
\end{longtable}

\begin{longtable}[]{@{}
  >{\raggedright\arraybackslash}p{(\manuscripttablewidth - 10\tabcolsep) * \real{0.1516}}
  >{\raggedright\arraybackslash}p{(\manuscripttablewidth - 10\tabcolsep) * \real{0.1748}}
  >{\raggedright\arraybackslash}p{(\manuscripttablewidth - 10\tabcolsep) * \real{0.1954}}
  >{\raggedright\arraybackslash}p{(\manuscripttablewidth - 10\tabcolsep) * \real{0.1704}}
  >{\raggedright\arraybackslash}p{(\manuscripttablewidth - 10\tabcolsep) * \real{0.1610}}
  >{\raggedright\arraybackslash}p{(\manuscripttablewidth - 10\tabcolsep) * \real{0.1468}}@{}}
\caption{RUL prediction performance on XB2-5 at different ITs.}\label{tab:A6}\\
\toprule\noalign{}
\begin{minipage}[b]{\linewidth}\raggedright
IT
\end{minipage} & \begin{minipage}[b]{\linewidth}\raggedright
Model
\end{minipage} & \begin{minipage}[b]{\linewidth}\raggedright
Performance
\end{minipage} & \begin{minipage}[b]{\linewidth}\raggedright
\end{minipage} & \begin{minipage}[b]{\linewidth}\raggedright
\end{minipage} & \begin{minipage}[b]{\linewidth}\raggedright
\end{minipage} \\
& & MAE & NRMSE & Score & NCR \\
\midrule\noalign{}
\endfirsthead

\toprule\noalign{}
\begin{minipage}[b]{\linewidth}\raggedright
IT
\end{minipage} & \begin{minipage}[b]{\linewidth}\raggedright
Model
\end{minipage} & \begin{minipage}[b]{\linewidth}\raggedright
Performance
\end{minipage} & \begin{minipage}[b]{\linewidth}\raggedright
\end{minipage} & \begin{minipage}[b]{\linewidth}\raggedright
\end{minipage} & \begin{minipage}[b]{\linewidth}\raggedright
\end{minipage} \\
& & MAE & NRMSE & Score & NCR \\
\midrule\noalign{}
\endhead
\bottomrule\noalign{}
\endlastfoot

50 & \secondbest{GDAU} & \secondbest{7.1} & \secondbest{0.177} & \textbf{0.633} & 0.0 \\
& DTGRU & 17.6 & 0.359 & 0.305 & 0.0 \\
& CLSTM & 12.3 & 0.270 & 0.458 & 0.0 \\
& GTFDAU & 10.4 & 0.226 & 0.511 & 0.0 \\
& LSTM-TF & 26.5 & 0.574 & 0.132 & 0.0 \\
& \textbf{LGFM} & \textbf{4.9} & \textbf{0.112} & \secondbest{0.534} & 0.0 \\
\midrule\noalign{}
100 & GDAU & N/A & N/A & N/A & 1.0 \\
& \textbf{DTGRU} & \textbf{24.1} & \textbf{0.257} & \secondbest{0.357} & 0.0 \\
& CLSTM & 334.1 & 3.603 & 0.084 & 0.8 \\
& GTFDAU & 70.3 & 1.371 & \textbf{0.372} & 0.1 \\
& LSTM-TF & 81.9 & 0.972 & 0.094 & 0.0 \\
& \secondbest{LGFM} & \secondbest{42.1} & \secondbest{0.432} & 0.246 & 0.0 \\
\midrule\noalign{}
150 & GDAU & N/A & N/A & N/A & 1.0 \\
& DTGRU & N/A & N/A & N/A & 1.0 \\
& CLSTM & N/A & N/A & N/A & 1.0 \\
& GTFDAU & N/A & N/A & N/A & 1.0 \\
& LSTM-TF & N/A & N/A & N/A & 1.0 \\
& \textbf{LGFM} & \textbf{61.1} & \textbf{0.446} & \textbf{0.298} & \textbf{0.0} \\
\midrule\noalign{}
200 & GDAU & N/A & N/A & N/A & 1.0 \\
& DTGRU & N/A & N/A & N/A & 1.0 \\
& CLSTM & N/A & N/A & N/A & 1.0 \\
& GTFDAU & N/A & N/A & N/A & 1.0 \\
& LSTM-TF & N/A & N/A & N/A & 1.0 \\
& \textbf{LGFM} & \textbf{170.5} & \textbf{1.073} & \textbf{0.183} & \textbf{0.0} \\
\end{longtable}

\begin{longtable}[]{@{}
  >{\raggedright\arraybackslash}p{(\manuscripttablewidth - 10\tabcolsep) * \real{0.1516}}
  >{\raggedright\arraybackslash}p{(\manuscripttablewidth - 10\tabcolsep) * \real{0.1748}}
  >{\raggedright\arraybackslash}p{(\manuscripttablewidth - 10\tabcolsep) * \real{0.1954}}
  >{\raggedright\arraybackslash}p{(\manuscripttablewidth - 10\tabcolsep) * \real{0.1704}}
  >{\raggedright\arraybackslash}p{(\manuscripttablewidth - 10\tabcolsep) * \real{0.1610}}
  >{\raggedright\arraybackslash}p{(\manuscripttablewidth - 10\tabcolsep) * \real{0.1468}}@{}}
\caption{RUL prediction performance on XB3-3 at different ITs.}\label{tab:A7}\\
\toprule\noalign{}
\begin{minipage}[b]{\linewidth}\raggedright
IT
\end{minipage} & \begin{minipage}[b]{\linewidth}\raggedright
Model
\end{minipage} & \begin{minipage}[b]{\linewidth}\raggedright
Performance
\end{minipage} & \begin{minipage}[b]{\linewidth}\raggedright
\end{minipage} & \begin{minipage}[b]{\linewidth}\raggedright
\end{minipage} & \begin{minipage}[b]{\linewidth}\raggedright
\end{minipage} \\
& & MAE & NRMSE & Score & NCR \\
\midrule\noalign{}
\endfirsthead

\toprule\noalign{}
\begin{minipage}[b]{\linewidth}\raggedright
IT
\end{minipage} & \begin{minipage}[b]{\linewidth}\raggedright
Model
\end{minipage} & \begin{minipage}[b]{\linewidth}\raggedright
Performance
\end{minipage} & \begin{minipage}[b]{\linewidth}\raggedright
\end{minipage} & \begin{minipage}[b]{\linewidth}\raggedright
\end{minipage} & \begin{minipage}[b]{\linewidth}\raggedright
\end{minipage} \\
& & MAE & NRMSE & Score & NCR \\
\midrule\noalign{}
\endhead
\bottomrule\noalign{}
\endlastfoot

6 & GDAU & \secondbest{20.1} & \secondbest{3.371} & 0.001 & 0.2 \\
& DTGRU & 138.3 & 37.86 & 0.001 & 0.5 \\
& CLSTM & 100.9 & 36.82 & \textbf{0.201} & 0.5 \\
& GTFDAU & 123.7 & 37.11 & 0.001 & 0.5 \\
& LSTM-TF & 57.4 & 14.02 & \secondbest{0.089} & 0.0 \\
& \textbf{LGFM} & \textbf{5.0} & \textbf{0.833} & 0.055 & 0.0 \\
\midrule\noalign{}
12 & \secondbest{GDAU} & \secondbest{10.4} & \textbf{0.867} & 0.050 & 0.0 \\
& DTGRU & 13.5 & 1.185 & 0.001 & 0.0 \\
& CLSTM & 10.9 & 0.959 & 0.055 & 0.0 \\
& GTFDAU & 156.7 & 22.3 & 0.031 & 0.4 \\
& LSTM-TF & 26.5 & 2.971 & \textbf{0.113} & 0.0 \\
& \textbf{LGFM} & \textbf{8.3} & \textbf{0.711} & \secondbest{0.106} & 0.0 \\
\midrule\noalign{}
18 & \secondbest{GDAU} & \secondbest{16.0} & \secondbest{0.888} & 0.045 & 0.0 \\
& DTGRU & 152.8 & 14.67 & 0.047 & 0.2 \\
& \textbf{CLSTM} & \textbf{15.0} & \textbf{0.834} & \secondbest{0.056} & 0.0 \\
& GTFDAU & 200.1 & 16.94 & 0.026 & 0.4 \\
& LSTM-TF & 21.7 & 1.567 & \textbf{0.120} & 0.0 \\
& LGFM & 25.5 & 2.241 & 0.001 & 0.0 \\
\midrule\noalign{}
24 & GDAU & N/A & N/A & N/A & 1.0 \\
& DTGRU & N/A & N/A & N/A & 1.0 \\
& CLSTM & N/A & N/A & N/A & 1.0 \\
& GTFDAU & N/A & N/A & N/A & 1.0 \\
& LSTM-TF & N/A & N/A & N/A & 1.0 \\
& \textbf{LGFM} & \textbf{34.6} & \textbf{1.505} & \textbf{0.001} & \textbf{0.0} \\
\end{longtable}

\clearpage
\setcounter{figure}{0}
\renewcommand{\thefigure}{B\arabic{figure}}
\renewcommand{\theHfigure}{B\arabic{figure}}
\section*{Appendix B. RtF trajectories and FPTs of the XB test sets}
\phantomsection
\label{sec:appendixB}

\begin{figure}[!htbp]
\centering
\includegraphics[width=\textwidth,height=0.78\textheight,keepaspectratio]{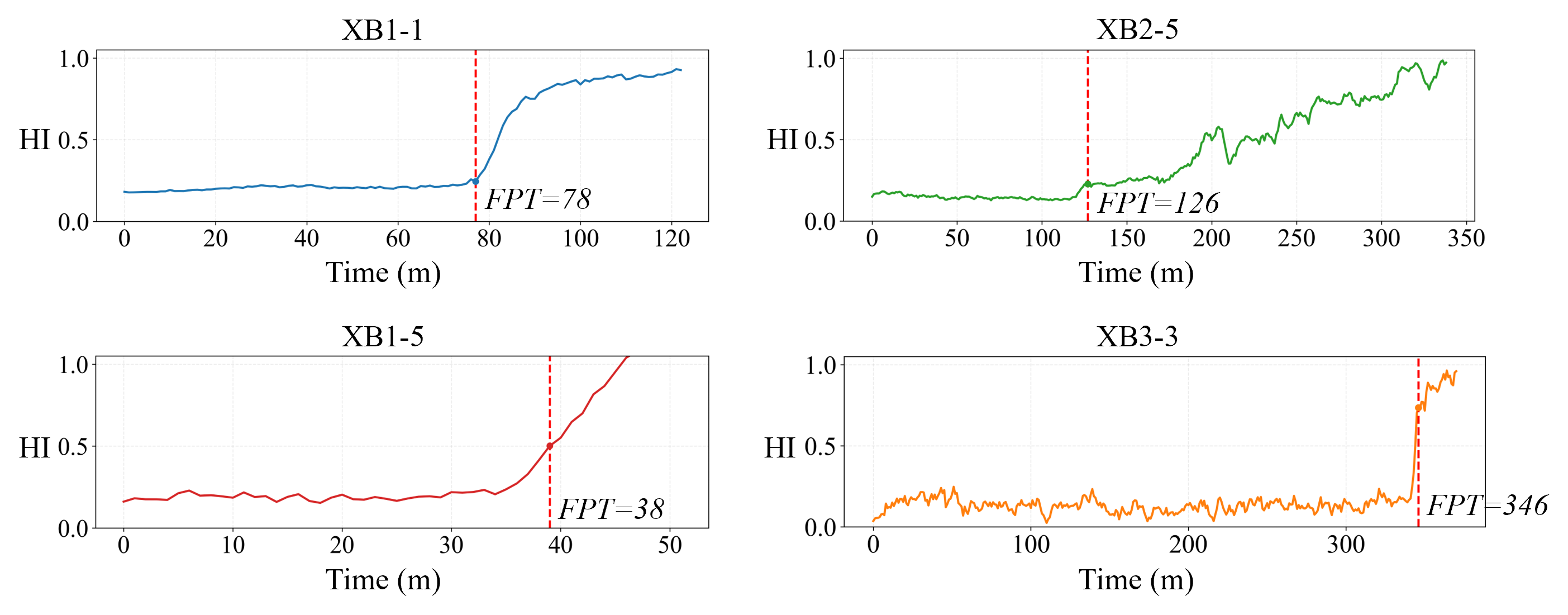}
\caption{RtF HI trajectories and corresponding FPTs of the XB test bearings.}
\label{fig:B1}
\end{figure}

Figure \ref{fig:B1} shows the HI trajectories of the XB test bearings over their complete RtF periods. The FPT identified for each bearing was used as the reference point for selecting its ITs.

\clearpage
\setcounter{figure}{0}
\renewcommand{\thefigure}{C\arabic{figure}}
\renewcommand{\theHfigure}{C\arabic{figure}}
\section*{Appendix C. Mathematical background of the soft-DTW loss}
\phantomsection
\label{sec:appendixC}

Consider two univariate sequences, \(\mathbf{\alpha} = \left\{ \alpha_{1},\ldots,\alpha_{N} \right\} \in \mathbb{R}^{N}\) and \(\mathbf{\beta} = \left\{ \beta_{1},\ldots,\beta_{M} \right\} \in \mathbb{R}^{M}\). DTW is defined as \citep{Cuturi2017SoftDTW}

\begin{tabularx}{\textwidth}{@{}>{\raggedright\arraybackslash}X>{\raggedleft\arraybackslash}p{0.08\textwidth}@{}}

\begin{minipage}[b]{\linewidth}\raggedright
\[DTW\left( \mathbf{\alpha},\mathbf{\beta} \right) = \min_{\mathbf{A} \in \mathbf{A}_{N.M}}\left\langle \mathbf{A},\mathrm{\Delta}\left( \mathbf{\alpha},\mathbf{\beta} \right) \right\rangle\]
\end{minipage} & \begin{minipage}[b]{\linewidth}\raggedright
(C1)
\end{minipage} \\

\end{tabularx}

Here, \(\mathbf{A}_{N.M}\) denotes the set of admissible alignment matrices, and \(\mathbf{A} \in \left\{ 0,1 \right\}^{N \times M}\) is a binary matrix representing an admissible alignment path between the two sequences. The pairwise cost matrix is defined as \(\left\lbrack \mathrm{\Delta}\left( \mathbf{\alpha},\mathbf{\beta} \right) \right\rbrack_{i,j} = \delta\left( \alpha_{i},\beta_{j} \right) \in \mathbb{R}^{N \times M}\), where \(\delta\) is a local cost function, commonly the squared Euclidean distance \(\delta\left( \alpha_{i},\beta_{j} \right) = \left( \alpha_{i} - \beta_{j} \right)^{2}\). The notation \(\left\langle \mathbf{A},\mathrm{\Delta} \right\rangle = \sum_{i = 1}^{N}{\sum_{j = 1}^{M}{\mathbf{A}_{i,j}\mathrm{\Delta}_{i,j}}}\) denotes the Frobenius inner product and gives the total cost associated with alignment \(\mathbf{A}\). DTW therefore selects the admissible alignment path with the lowest cumulative cost.

Because DTW uses a hard minimum over all admissible paths, it is not differentiable at points where the optimal path changes. Even where a gradient exists, it depends only on the selected optimal path and may change discontinuously after a small perturbation of either sequence. These properties make conventional DTW difficult to use directly as a stable training objective.

\begin{figure}[!htbp]
\centering
\includegraphics[scale=0.5]{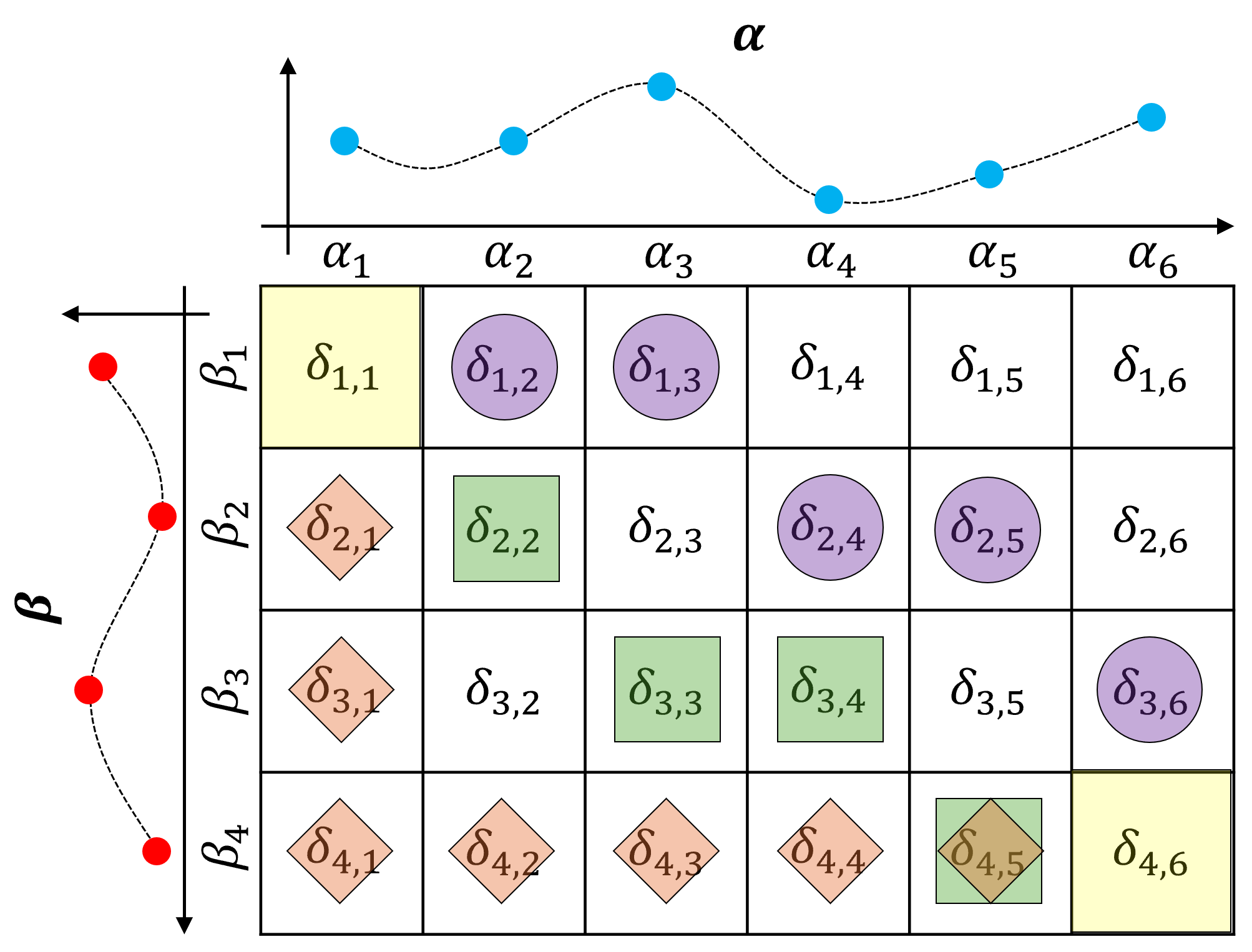}
\caption{Soft-DTW computation between \(\alpha\) and \(\beta\) for \(N = 6\) and \(M = 4\). The cost of an alignment is the sum of the entries visited along its path. The purple circles indicate the optimal DTW alignment, corresponding to the limiting case \(\gamma = 0\). Conventional DTW considers only the minimum-cost alignment, whereas soft-DTW aggregates the costs of all admissible alignments through a smooth minimum.}
\label{fig:C1}
\end{figure}

For continuous-valued sequences without exact cost ties, conventional DTW generally selects a single optimal alignment, as illustrated by the purple path in Figure \ref{fig:C1}. Nevertheless, a small change in either sequence may alter the selected path and produce a discontinuous change in the gradient. Soft-DTW replaces the hard minimum with a differentiable soft minimum over all admissible alignments. This smoothing produces more stable gradients and makes the alignment cost more suitable for gradient-based learning \citep{Chen2024SoftDTWPeak}.

Soft-DTW is defined as \citep{Cuturi2017SoftDTW,Chen2024SoftDTWPeak}

\begin{tabularx}{\textwidth}{@{}>{\raggedright\arraybackslash}X>{\raggedleft\arraybackslash}p{0.08\textwidth}@{}}

\begin{minipage}[b]{\linewidth}\raggedright
\[{sDTW}_{\gamma}\left( \mathbf{\alpha},\mathbf{\beta} \right) = \left\{ \begin{array}{r}
 - \gamma log\sum_{\mathbf{A} \in \mathbf{A}_{N.M}}^{}{\exp{\left( - \frac{\left\langle \mathbf{A},\mathrm{\Delta}\left( \mathbf{\alpha},\mathbf{\beta} \right) \right\rangle}{\gamma} \right),\ \ \gamma > 0}} \\
DTW\left( \mathbf{\alpha},\mathbf{\beta} \right),\ \ \gamma = 0
\end{array} \right.\ \]
\end{minipage} & \begin{minipage}[b]{\linewidth}\raggedright
(C2)
\end{minipage} \\

\end{tabularx}

The parameter \(\gamma \geq 0\) controls the degree of smoothing. When \(\gamma = 0\), soft-DTW is reduced to conventional DTW. For \(\gamma > 0\), soft-DTW is differentiable with respect to both input sequences provided that the local cost function \(\delta\) is differentiable \citep{Cuturi2017SoftDTW,Chen2024SoftDTWPeak}. Similar values of \(\gamma\) yield a closer approximation to the hard minimum, whereas larger values distribute greater influence across alternative alignment paths.

The normalized soft-DTW divergence is defined as

\begin{equation}
	\begin{aligned}
		\operatorname{sDTW}_{\gamma}^{\mathrm{norm}}
		\left(\boldsymbol{\alpha},\boldsymbol{\beta}\right)
		&=
		\operatorname{sDTW}_{\gamma}
		\left(\boldsymbol{\alpha},\boldsymbol{\beta}\right)
		\\
		&\quad
		-\frac{1}{2}
		\operatorname{sDTW}_{\gamma}
		\left(\boldsymbol{\alpha},\boldsymbol{\alpha}\right)
		-\frac{1}{2}
		\operatorname{sDTW}_{\gamma}
		\left(\boldsymbol{\beta},\boldsymbol{\beta}\right).
	\end{aligned}
	\tag{C3}
	\label{eq:c3}
\end{equation}

Equation~\eqref{eq:c3} corrects the entropic bias introduced by the soft minimum operation. The unnormalized values \({sDTW}_{\gamma}\left( \mathbf{\alpha},\mathbf{\beta} \right)\) may be negative and need not attain its minimum when the two sequences are identical. Subtracting the two self-similarity terms yields the normalized soft-DTW divergence used in this study.

\end{document}